\documentclass[onecolumn]{aastex61}
\pdfoutput=1
\usepackage{natbib}
\usepackage{gensymb}
\usepackage[section]{placeins}
\usepackage{graphicx}
\usepackage{amsmath}
\usepackage[colorinlistoftodos]{todonotes}
\usepackage{float}
\usepackage{chngcntr}
\counterwithin*{footnote}{section}
\usepackage[english]{babel}
\usepackage[utf8x]{inputenc}

\begin{document}

\title{The VLA Nascent Disk And Multiplicity Survey of Perseus Protostars (VANDAM). \\
III. Extended Radio Emission from Protostars in Perseus}
\author{Łukasz Tychoniec}
\affiliation{Leiden Observatory, Leiden University, P.O. Box 9513, NL-2300RA Leiden, The Netherlands} 
\affiliation{Astronomical Observatory, Faculty of Physics, Adam Mickiewicz University, Słoneczna 36, PL-60268 Poznań, Poland}

\author{John J. Tobin}
\affiliation{Leiden Observatory, Leiden University, P.O. Box 9513, NL-2300RA Leiden, The Netherlands} 
\affiliation{Homer L. Dodge Department of Physics and Astronomy, University of Oklahoma, 440 W. Brooks Street, Norman, Oklahoma 73019, USA}

\author{Agata Karska}
\affiliation{Centre for Astronomy, Nicolaus Copernicus University in Toruń, Faculty of Physics, Astronomy and Informatics, Grudziadzka 5, PL-87100 Toruń, Poland} 

\author{Claire Chandler}
\affiliation{National Radio Astronomy Observatory, P.O. Box O, 1003 Lopezville Road, Socorro, NM 87801-0387, USA}

\author{Michael M. Dunham}
\affiliation{Harvard-Smithsonian Center for Astrophysics, 60 Garden St., Cambridge, MA, USA}
\affiliation{Department of Physics, State University of New York Fredonia, Fredonia, New York 14063, USA}

\author{Zhi-Yun Li}
\affiliation{Department of Astronomy, University of Virginia, Charlottesville, VA 22903, USA}

\author{Leslie W. Looney} 
\affiliation{Department of Astronomy, University of Illinois, Urbana, IL 61801, USA}

\author{Dominique Segura-Cox}
\affiliation{Department of Astronomy, University of Illinois, Urbana, IL 61801, USA}

\author{Robert J. Harris}
\affiliation{Department of Astronomy, University of Illinois, Urbana, IL 61801, USA}

\author{Carl Melis}
\affiliation{Center for Astrophysics and Space Sciences, University of California, San Diego, CA 92093, USA}

\author{Sarah I. Sadavoy}
\affiliation{Max-Planck-Institut f{\"u}r Astronomie, K{\"o}nigstuhl 17, D-69117 Heidelberg, Germany}

\correspondingauthor{Łukasz Tychoniec}
\email{tychoniec@strw.leidenuniv.nl}

\begin{abstract}
Centimeter continuum emission from protostars offers insight into the innermost part of the outflows, as shock-ionized gas produces free-free emission. We observed a complete population of Class 0 and I protostars in the Perseus molecular cloud at 4.1 cm and 6.4 cm with resolution and sensitivity superior to previous surveys. From a total of 71 detections, 8 sources exhibit resolved emission at 4.1 cm and/or 6.4 cm. In this paper we focus on this sub-sample, analyzing their spectral indices along the jet, and their alignment with respect to the large-scale molecular outflow. Spectral indices for fluxes integrated toward the position of the protostar are consistent with free-free thermal emission. The value of the spectral index along a radio jet decreases with distance from the protostar. For six sources, emission is well aligned with the outflow central axis, showing that we observe the ionized base of the jet. This is not the case for two sources, where we note misalignment of the emission with respect to the large-scale outflow. This might indicate that the emission does not originate in the radio jet, but rather in an ionized outflow cavity wall or disk surface. For five of the sources, the spectral indices along the jet decrease well below the thermal free-free limit of -0.1 with $>2\sigma$ significance. This is indicative of synchrotron emission, meaning that high energy electrons are being produced in the outflows close to the disk. This result can have far-reaching implications for the chemical composition of the embedded disks.

\end{abstract}

\section{Introduction}

Outflows and jets are ubiquitous phenomena in star formation. The youngest and most embedded Class 0 and Class I protostars are known to drive some of the most  powerful outflows \citep{Bontemps1996,CarattioGaratti2012}.
Outflows can be observed across the electro-magnetic spectrum: as a Herbig-Haro jets in the
visible regime (e.g., \citealt{Reipurth1995}; \citealt{Bally1996}), hot molecular and atomic gas in the near and mid-infrared
 (e.g., \citealt{Nisini2002}) and cold entrained gas in the millimeter (e.g., \citealt{Plunkett2013}; \citealt{Lee2015}).
At centimeter wavelengths, we are able to trace the free-free emission arising from the ionized gas in the outflows
 (\citealt{Reynolds1986}; \citealt{Anglada1998}). Centimeter emission from protostars, if resolved, often appears extended,  
matching well with directions of the large-scale outflow (e.g., \citealt{Rodriguez1986, Marti1993, Rodriguez1994a, Anglada1995}).
Thus, it is inferred that radio jets trace regions that constitute the base of the outflow, and the correlation of radio emission with outflow force provides strong evidence for this link \citep{Cabrit1992, Anglada1995, Wu2004, Shirley2007} 
 
With the spectral index at centimeter wavelengths, we can discriminate between mechanisms responsible for the observed emission.
 Thermal free-free emission is expected to have flat or positive spectral index ($-0.1\leq \alpha < 2.0$; \citealt{Panagia1975, Rodriguez1993}) varying with the optical depth.
Similar or steeper indices are expected for thermal dust emission but it does not contribute significantly to the radio emission above 4 cm, because for typical dust masses and opacities in the low-mass star forming regions, especially for the youngest protostars, the emission is $\sim$10$\times$ below the sensitivity limit \citep{Tobin2015a}. That emission becomes significant as the protostellar disk evolves and the grain growth proceeds \citep[e.g.,][]{Wilner2005}.  Negative spectral indices ($\alpha < -0.1$) are a  
manifestation of a non-thermal emission. 

Positive spectral indices are most frequently observed from young Class 0/I protostars, while for more evolved Class II/III young stellar objects, coronal activity commonly produces non-thermal radiation \citep[e.g.,][]{Dzib2013, Pech2016}. However, there are a growing number of sources with observed negative spectral indices in the outflow positions of young protostars. Since first observed in the protostellar outflows, negative spectral indices were explained by synchrotron emission, with shocks as a mechanism to accelerate electrons \citep{Rodriguez1989b}.
\cite{Curiel1993} observed significant variations of the spectral index along the jet of the Serpens SMM1 protostar, also known as a Serpens triple radio source \citep[e.g.,][]{Snell1986}, and suggested that both free-free and synchrotron emission mechanisms can simultaneously operate in outflows.   
Direct confirmation of synchrotron emission contributing to at least a part of the radio flux from protostars was provided  
by detecting linear polarization in the HH80-81 outflows of a massive protostar IRAS 18162-2048 \citep{Carrasco-Gonzalez2010}.
\cite{Ainsworth2014} also observed synchrotron emission at the bow shock position in the jet from DG Tau, a solar-mass pre-main
sequence star, suggesting that synchrotron emission phenomenon can be observed even in the outflows of low-mass stars.

Resolved radio jets were observed in many protostars; however, the sample of the low-mass sources with luminosities  $\sim$1~L$_{\sun}$  is still sparse. One of the rare cases is SVS 13C \citep{Rodriguez1997} with bolometric luminosity of 1.5 L$_{\sun}$, which we also present in this paper.
Centimeter radio emission is correlated with the bolometric luminosity \citep{Cabrit1992, Anglada1995, Shirley2007}; thus, $\sim$1~L$_{\sun}$ sources require higher sensitivity than massive protostars to have their radio emission detected. Also, if we want to observe the region closest to the protostar, we must resolve where the jet is being collimated, so resolution well below arcsecond is necessary.
Most previous radio observations found extended radio jets only toward higher luminosity sources, and there have only been a few surveys with high sensitivity, but not sub-arcsecond resolution, or probing a selected sample of protostars in different regions, instead of surveying complete population in one cloud. \cite{Rodriguez1999} achieved sensitivity up to 10 $\mu$Jy, with observations at 3.6 cm and 6 cm but their sample was limited to the central part of the NGC1333 region in Perseus and the resolution was around 5\arcsec. The \cite{Scaife2011a, Scaife2011, Scaife2012, Scaife2012a} observed selected protostars in Perseus, Serpens, Taurus and several isolated cores at low resolution (30\arcsec) and sensitivity up to 15 $\mu$Jy at 1.8 cm. The Gould's Belt VLA Survey targeted large sample of protostars in selected star forming regions with 16 $\mu$Jy sensitivity and resolution up to 0\farcs4 \citep{Dzib2013, Dzib2015, Kounkel2014, Ortiz-Leon2015, Pech2016} in C-band (4.1 and 6.4 cm). \cite{Reipurth2002, Reipurth2004} surveyed a sample of protostars in different star forming regions with high-resolution (0\farcs35) and sensitivity ($\sim$ 10 $\mu$Jy) at 3.6 cm and found evidence for a few modestly extended jets.
 
To further investigate the properties of the protostellar jets and to overcome limitations of the previous surveys we use VLA Nascent Disk and Multiplicity (VANDAM) survey C-band observations of all known (84) Class 0/I protostars in the Perseus molecular cloud, one of the most active stellar nurseries in the solar neighborhood (d $\sim$ 230 pc, \citealt{Hirota2008}). VANDAM is the largest radio survey of the youngest (Class 0 \& I) protostars in a single cloud ever undertaken, with high resolution up to $\sim 80$ AU (0\farcs3) at the distance to Perseus and remarkable sensitivity of $\sim 5\  \mu $Jy in the C-band, both superior to previous surveys. Protostars targeted by the survey span the low-mass regime with luminosities between 0.1 L$_\odot$ and 30 L$_\odot$. A comprehensive study of the C-band observations will be the scope of a future paper (Tychoniec et al. in prep.). Here,  we present a subset of sources with resolved radio jets. We will investigate their properties and spectral index distributions along the jet and discuss their relation to the larger scale outflows.

\section{Observations}
Observations were a part of the VANDAM Survey\footnote{FITS files from both Ka and C-band observations will be available online at: https://dataverse.harvard.edu/dataverse/VANDAM}. Observations were taken in A configuration between 2014 February 28 and
2014 April 12, with 3C48 as an absolute flux and bandpass calibrator and J0336+3218 as the complex gain calibrator.  
The observations were performed in 8-bit mode, resulting in 2 GHz of bandwidth divided into sixteen 128 MHz sub-bands with 2 MHz channels and full polarization products recorded; the polarization results will be published in a future paper.  
The 2 GHz is divided into two 1~GHz base bands that were centered at 4.7~GHz (6.4~cm) and 7.3 GHz (4.1~cm).  
The basebands were centered away from the band edges to avoid persistent radio frequency interference. The absolute flux uncertainty is estimated to be $\sim$ 5\%. With the simultaneous observations at the two ends of C-band, the limiting factor in our ability to determine the spectral index is the accuracy of the flux density model of 3C48  which is known to better than $\sim 2\%$ \citep{Perley2017}.
 
The data were reduced and calibrated using CASA 4.1.0 \citep{McMullin2007} and version 1.2.2 of the VLA Pipeline. Further flagging was conducted after the pipeline run and is detailed further in \citet{Tobin2015a}. We imaged the data using the \textit{clean} task in CASA version 4.1.0 with both natural weighting and Briggs weighting with the robust parameters set to 0.25. In all cases, we imaged the full field of the dataset using images with dimensions either 8192$^2$ or 16384$^2$ pixels and pixel widths of 0\farcs05. The nominal angular resolution at 4.1~cm was $\sim$0\farcs30 and 0\farcs25 for
natural and robust~$=0.25$ weighting, respectively; the angular resolution at 6.4~cm was $\sim$0\farcs50 and 0\farcs35.
For some images with bright extragalactic sources, we performed two iterations of phase-only self-calibration to achieve better dynamic range. Self-calibration  was not necessary for any fields containing resolved jets.

\section{Results}
From the 71 protostars detected in the C-band, we identified 8 that had clearly extended emission, listed in Table \ref{tab:sourcesinfo}.
The integrated flux densities were obtained by fitting the 2D Gaussian function to the sources with CASA \textit{imfit} task. In most cases, a multiple Gaussian fit was necessary to obtain as small residual as possible. The integrated flux value is then the sum of all Gaussians. Uncertainty of the integrated flux was obtained from \textit{imfit} task. Peak flux density and the rms were measured with CASA \textit{imstat} task. Both values and their uncertainties were extracted from the images before primary beam correction and subsequently corrected for primary beam response measured at the position of the source. For each source, we calculated spectral index values of both integrated and peak flux using the following equation:
\begin{equation}
\alpha_{\ 4.1/6.4}=\frac{\log (F_{4.1}/F_{6.4})}{\log (6.4/4.1)}
\label{eq:alpha}
\end{equation}
where $F_{4.1}$ and $F_{6.4}$ are integrated or peak fluxes at 4.1 cm and 6.4 cm respectively. Uncertainty of the spectral index was obtained using
standard error propagation  \citep{Chiang2012}. Our analysis was done primarily on robust~$=0.25$ images, except for Per-emb-33 and SVS 13C which have the most extended radio emission and hence the natural weighting
was more appropriate. A summary of the measured properties of the protostars with resolved jets
is presented in Table \ref{tab:cband_res}.

We measured spectral indices at both central and off-source positions along the resolved jets.
This was done by taking a median of 4.1 cm and 6.4 cm flux from area of the size of synthesized beam using CASA \textit{imstat} task, and then calculating the spectral index in the way described above with the rms of the image used as a flux uncertainty. Positions where the spectral index was measured are showed in the appendix. Using the area of the synthesized beam takes into account the resolution of the observations, but as the emission fades steeply with distance to the protostar, we use median as less sensitive to those outliers. While some of the measured beams are overlapping, making measurements not entirely independent, we note that the trends seen in the spectral index maps are consistent with the obtained values.
Protostellar positions are adopted from the Ka-band survey results that identify the protostar positions with the highest accuracy \citep{Tobin2016}.
To measure the direction of the radio jet, we obtain the position angle of a
single 2D Gaussian fit to the source and summarize the results in Table \ref{tab:indextab}.
For all sources, we produced a spectral index map. To obtain the same angular resolution at both wavelengths, the 4.1 cm frames were convolved with the same restoring beam as 6.4 cm frames in the \textit{clean} procedure in CASA.
Pixels with values below $3\times$ rms were masked.

We don't expect the dust emission to significantly contribute to the C-band flux. The values of the spectral indices at Ka-band for all sources presented here, are found below the value expected from the thermal dust emission \citep{Tobin2016} which shows that free-free emission is dominating the C-band flux, and also singificantly affects the Ka-band fluxes. We also note that any contribution of the thermal dust emission would be limited the the central source position. Regarding the detected negative spectral indices, any thermal dust contribution would increase the value of spectral index between 4.1 and 6.4 cm, thus any potential contribution does not cast a doubt on the tentative detection of the synchrotron emission.

We detected 8 protostars with resolved radio jets extended on scales between 80 AU and 900 AU. For all the objects, the spectral
indices at the position of the protostar are consistent with free-free emission. For 5 sources we detect
negative spectral indices along the radio jet/outflow direction, possibly indicative of synchrotron emission.
We also find that the integrated spectral indices of extended jet sources are lower than the median observed for the whole sample ($\alpha_{\textrm{med}}=0.51$; Tychoniec et al. in prep.). The significant contribution of radio flux from the more optically thin extended jets in the form of free-free or synchrotron emission would lower the value of the overall spectral index.

Extended emission is observed from protostars with bolometric luminosities between 1 L$_\odot$ and
9.2 L$_\odot$. The sample includes both Class 0 and Class I protostars with low bolometric temperatures of  $T_{\rm bol} < 100 $ K, meaning that the protostars are all relatively young \citep[e.g.,][]{Chen1995, Enoch2009}. Other than this upper limit on $T_{\rm bol}$ there are no specific trends between protostar properties and extended radio emission. We will discuss results for each source individually.
\subsection{Per-emb-36}
Per-emb-36 (also known as NGC1333 IRAS 2B) is a Class I system comprised of 2 protostars separated by $\sim$ 70 AU \citep{Tobin2016}.  Centimeter emission is dominated by the bipolar jet from Per-emb-36-A. 
Per-emb-36-B is not clearly resolved from Per-emb-36-A  at 4.1 cm or 6.4 cm, 
but a small asymmetry in the 4.1 cm emission is observed
toward the position of Per-emb-36-B as identified at Ka-band. There is a clear decrease in a spectral index along the outflow direction, from $0.67\pm 0.18$ at the protostellar position to $-0.64\pm 0.20$ in the northern outflow. The  position angle of the extended emission is $24\pm 2\degree$, closely matching the position angle
of the CO outflow observed by \cite{Plunkett2013}. This is a strong indication that the jet from the protostar is producing the extended radio emission. The outflow is asymmetric, showing stronger emission from the northern part and only on this side we
 detect a steep spectral index which likely indicates the presence of the 
synchrotron emission. It is noteworthy that the 6.4 cm emission peaks significantly 
away from the source at 0.4\arcsec (93 AU) which is not the case for any other 
sources we observed. Shift is also clearly seen in full bandwidth image (Fig. \ref{fig:per36zoom} in the appendix). We estimate the error of the position measurement at 6.4 cm, following Equation 1 in \cite{Condon1998}, and the resulting error is below 0\farcs01 in each coordinate, showing that this displacement is robustly detected. It is possible that this offset in the peak emission at 6.4~cm is due to a strong bow shock producing synchrotron emission, which may have a counterpart in X-rays similar
to DG Tau \citep{Ainsworth2014}. We examined the literature for possible X-ray detections and 
find an X-ray source \citep{Preibisch1997} near Per-emb-36. However, the X-ray emission may be from a Class III (BD +30 547)
source with a projected separation of $\sim$4\arcsec\ from Per-emb-36; the source(s) were detected toward the edge of the X-ray observations field where the positional uncertainty is large.

\begin{figure}[H]
  \centering
  \includegraphics[width=0.36\linewidth]{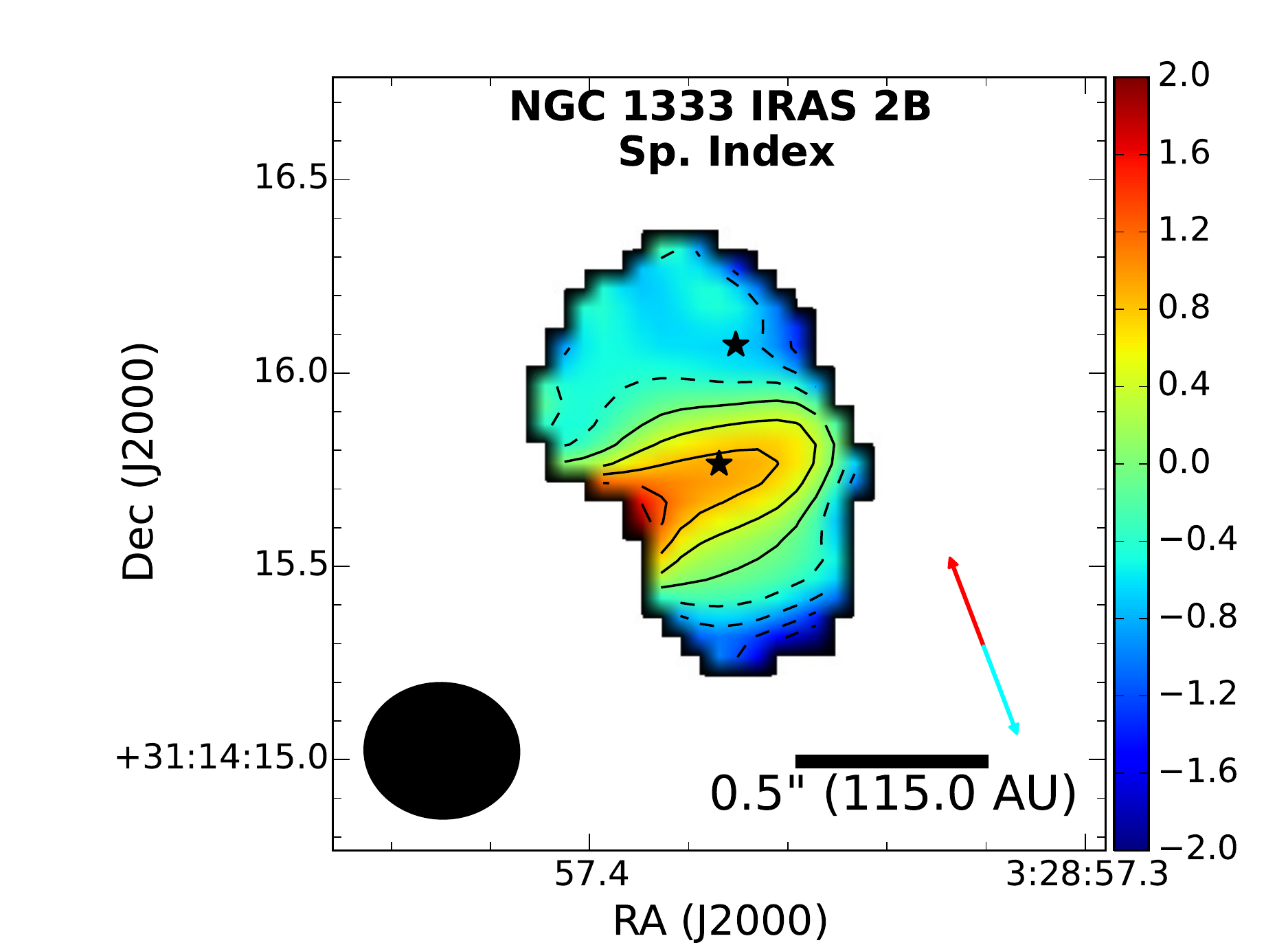}
  \label{fig:lowhist}
 \centering
  \includegraphics[width=0.29\linewidth]{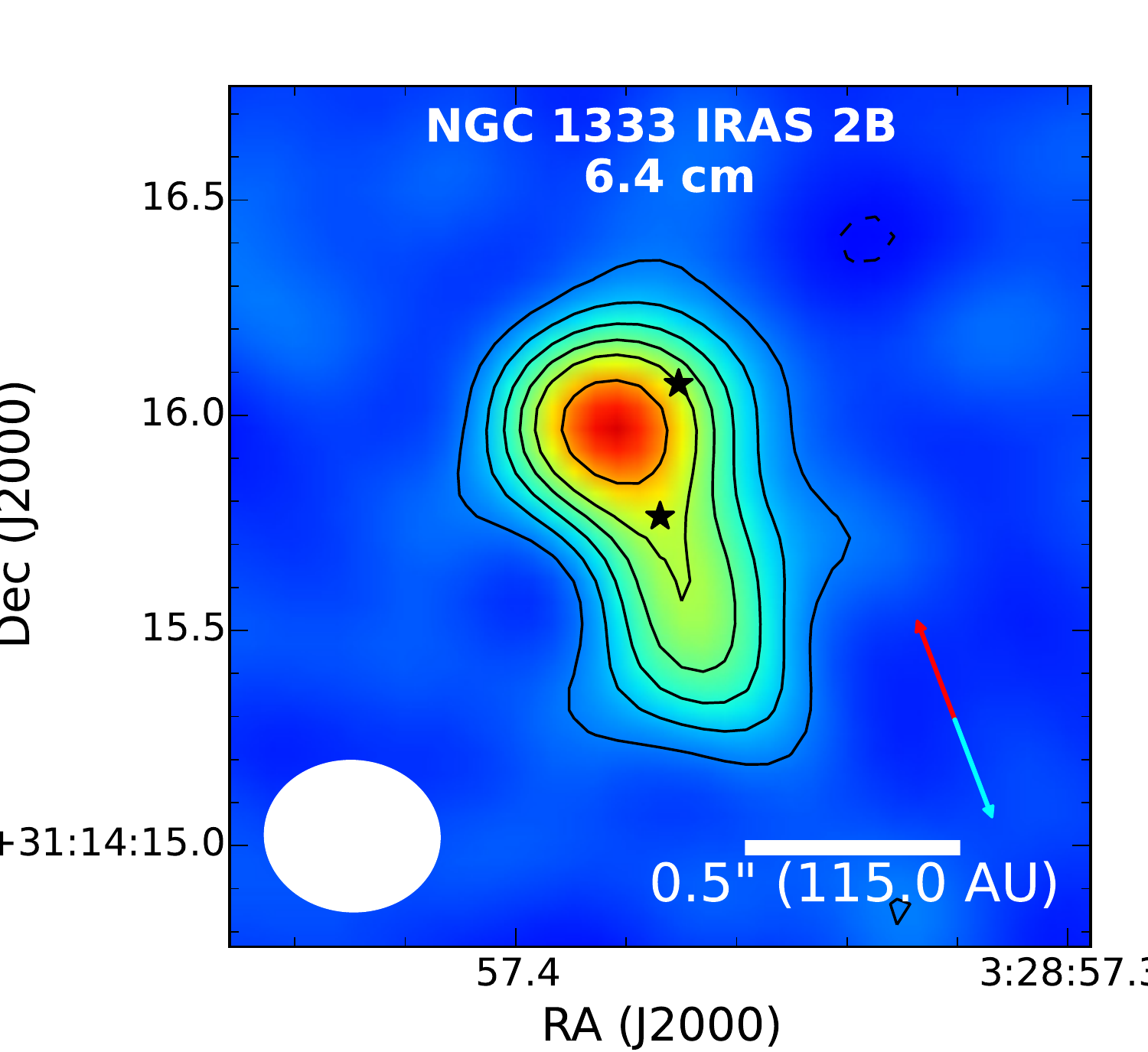}
  \label{fig:lowhist}
  \centering
  \includegraphics[width=0.29\linewidth]{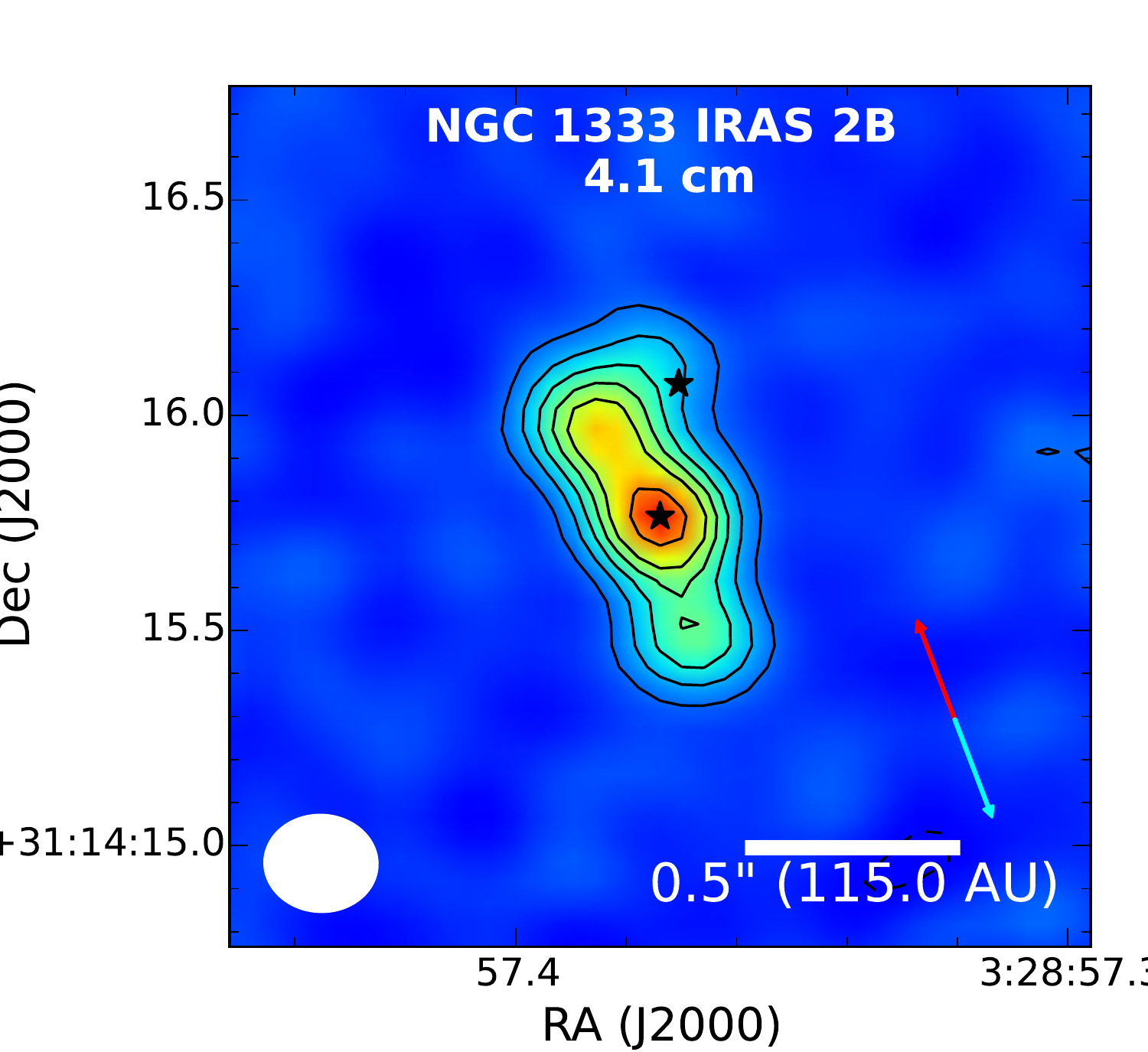}

\caption{ Images of Per-emb-36. Spectral index map with contours: [-2, -1.6, -1.2, -0.8, -0.4, 0.0, 0.4, 0.8, 1.2,
 1.6, 2.0]. 6.4 cm and 4.1 cm maps with contours [-3, 3, 6, 9, 12, 15, 20] x $\sigma$ where $\sigma _{6.4\ cm}=4.83\ \mu $Jy and $\sigma _{4.1\ cm}=3.90\ \mu $Jy.
Synthesized beam is shown in the left bottom corner (Sp. Index and 6.4 cm: 0\farcs41$\times$0\farcs35, 4.1~cm: 0\farcs26$\times$0\farcs22). The stars mark the position of the protostars based on Ka-band observations
\citep{Tobin2016} and the red and blue arrows indicate outflow direction from \cite{Plunkett2013} with a position angle $24\degree$. 6.4 cm and 4.1 cm maps are not corrected for a primary beam response (spectral index derived from PB corrected map).}
\label{fig:per36}
\end{figure}

\subsection{SVS 13C}
SVS 13C is a Class 0 object, as classified by \cite{sadavoy2014}, located in the NGC1333 region in Perseus.
This source was also detected at millimeter wavelengths by \cite{Looney2000}. It produces the most prominent
extended radio jet in our sample. The position angle for its centimeter emission is $9\pm 1\degree$. It extends $\sim$4\arcsec\ (900 AU) in both the north and south direction. The
extended centimeter emission toward this source had been previously reported by \cite{Rodriguez1997} (their source VLA2) and
later by \cite{Reipurth2002} and \cite{Carrasco-Gonzalez2008}. However, our observations are the highest fidelity maps to date taken toward this source. We find irregular, clumpy structure along the jet, with the spectral index significantly decreasing outwards. At 2\arcsec~(460 AU) distance in the northern direction, we can observe a source of emission detached from the main component, suggesting that there is a clump of denser matter in the area.  \cite{Plunkett2013} tentatively identified an outflow with $8\degree$ position angle and \cite{Lee2016} found a similar position angle of $0\degree$, both
of which are consistent with an extended radio jet. The outflow also appeared to be in the plane of the sky in both the
observation from \cite{Plunkett2013} and \cite{Lee2016}. This is consistent with the large proper motions
of $\sim$100 km\ s$^{-1}$ found by \citet{Raga2013}.

\begin{figure}[H]
 \centering
  \includegraphics[width=0.36\linewidth]{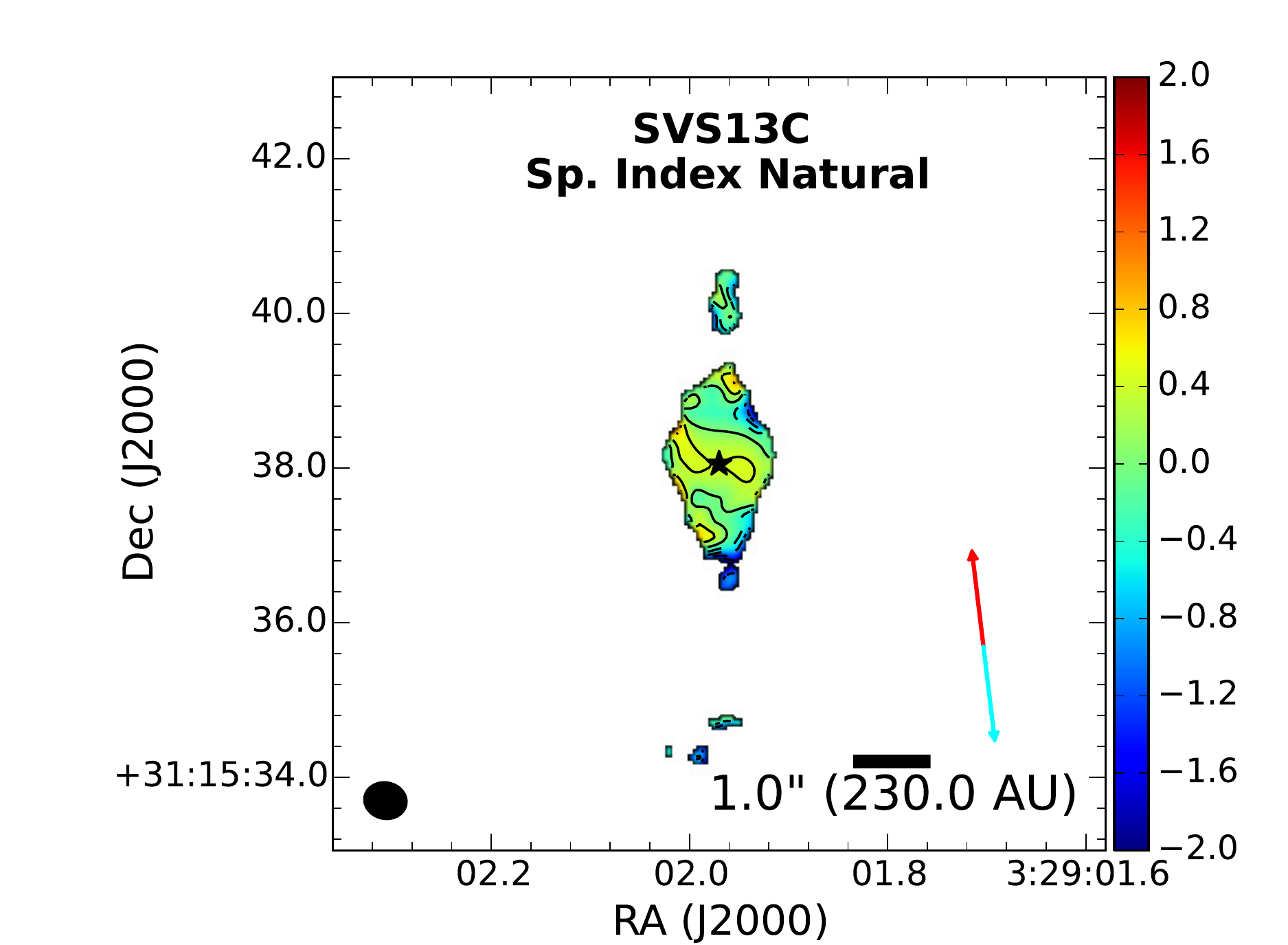}
  \centering
  \includegraphics[width=0.29\linewidth]{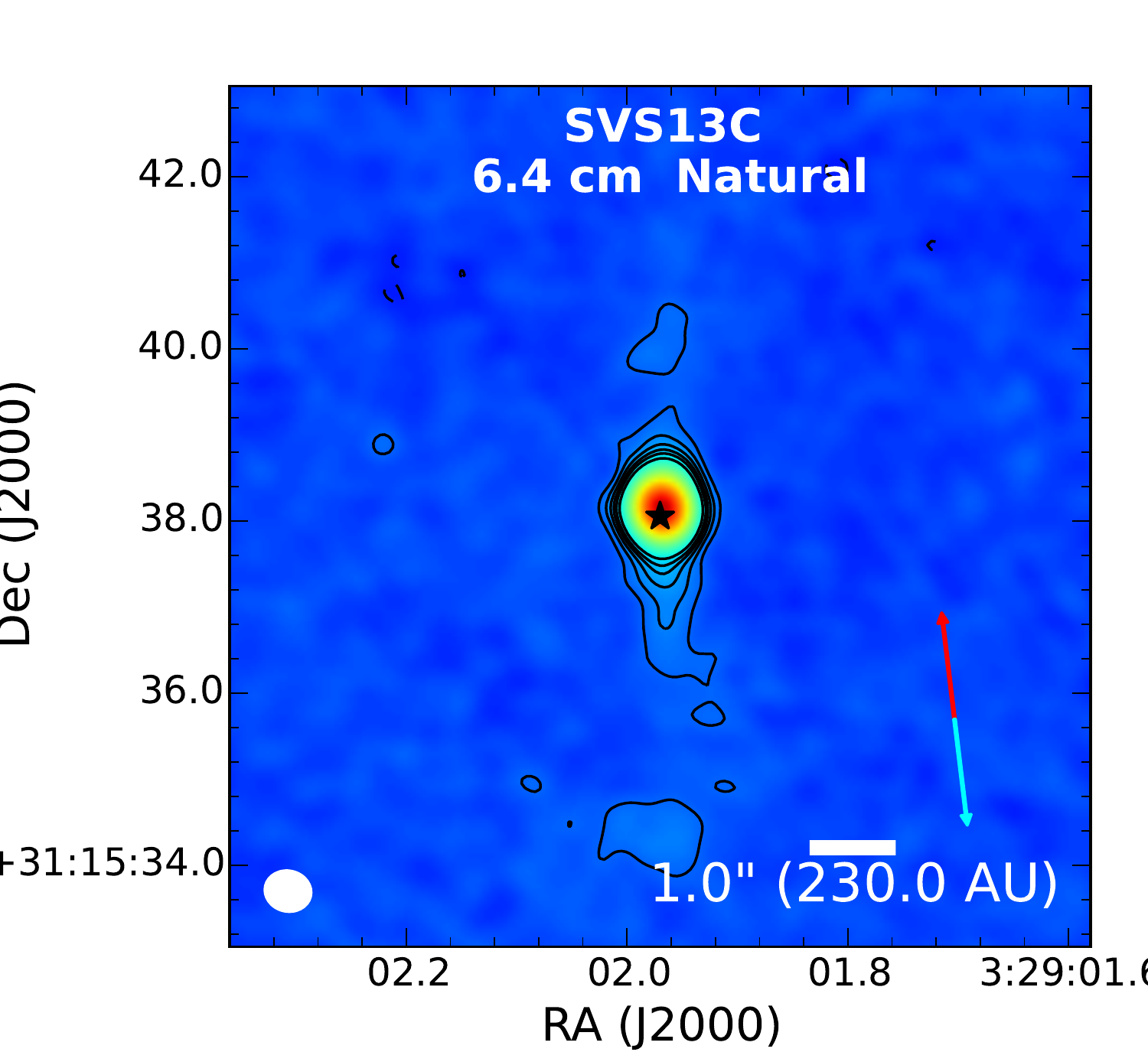}
  \centering
  \includegraphics[width=0.29\linewidth]{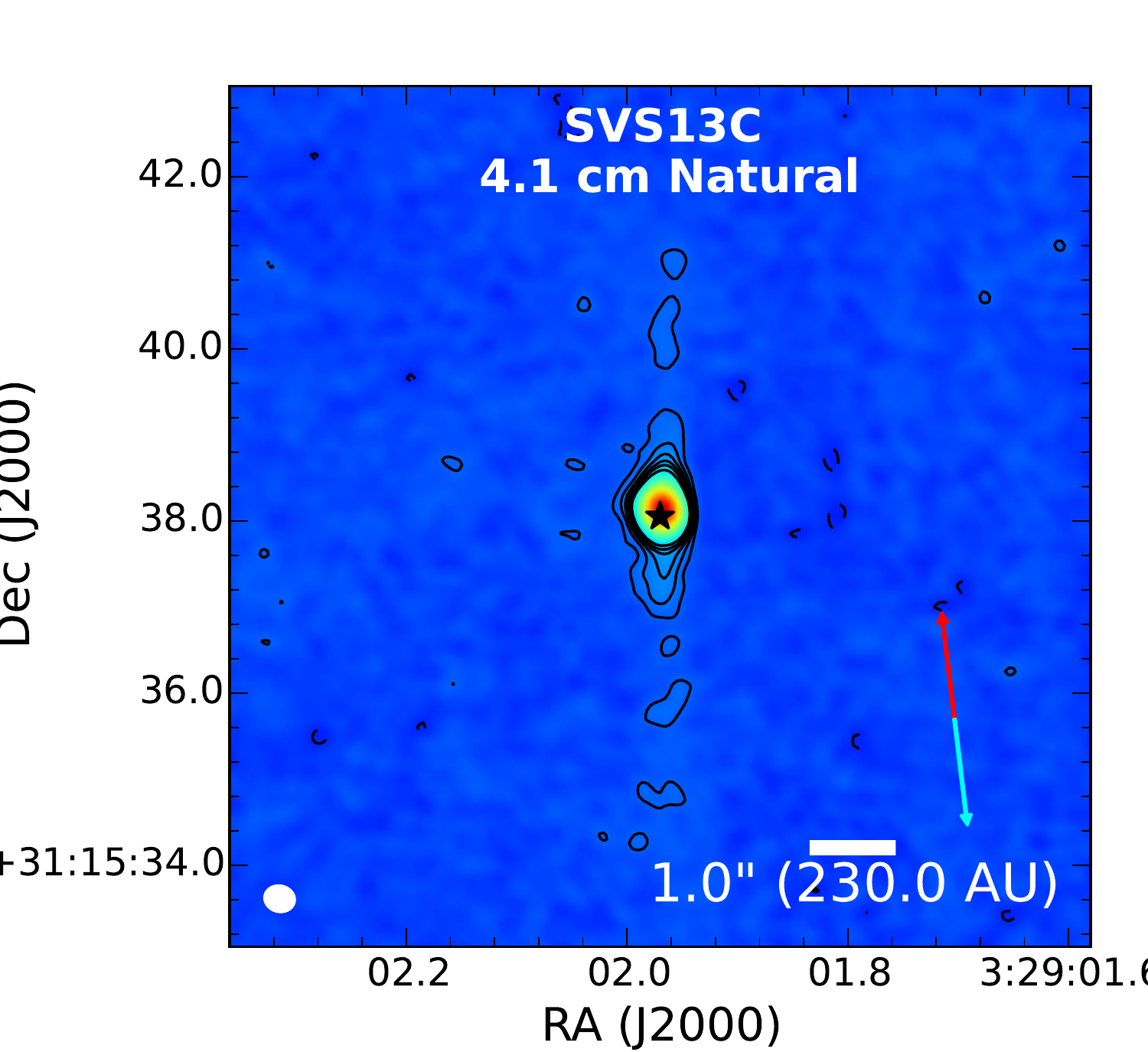}

\caption{
Naturally-weighted map of SVS 13C with contours as in Figure \ref{fig:per36} ($\sigma _{6.4\ cm}=4.83\ \mu $Jy and $\sigma _{4.1\ cm}=3.90\ \mu $Jy). Synthesized beam is shown in the left bottom corner (Sp. Index and 6.4 cm::
0\farcs41$\times$0\farcs35, 4.1~cm: 0\farcs26$\times$0\farcs22). The star marks the position of the protostar based on Ka-band observations \citep{Tobin2016} and the red and blue arrows indicate outflow direction from \cite{Plunkett2013} and \citet{Lee2016}.}
\label{fig:per109}
\end{figure}

\subsection{Per-emb-30}
Per-emb-30 is a Class 0 source located in the Barnard 1 region.
We find radio emission at 6.4~cm extending $\sim$1\arcsec\ (230 AU) from the source and more compact emission at 4.1~cm. The more compact emission at 4.1~cm results in a strongly negative spectral index ($-1.10\ \pm\ 0.47$) along the jet direction. The direction of extended radio emission is consistent with the H\raisebox{-.4ex}{\scriptsize 2} feature (HH790) originating from this source \citep{Davis2008}, having a position angle of $\sim$109\degr\ which is consistent with our fit of $117 \pm 4\degree$. \cite{Storm2014} also finds evidence for an HCO$^+$ outflow toward Per-emb-30 with a similar position angle. Interestingly, the H\raisebox{-.4ex}{\scriptsize 2}, radio, and HCO$^+$ outflow appear mono-polar; upcoming CO observations from the MASSES survey (e.g., \citealt{Lee2014}) will be more definitive. The mono-polar appearance may reflect the spatial distribution of dense gas around the protostar; south-eastern portion of the outflow is interacting with ambient medium while the north-western portion is not.

\begin{figure}[H]

  \centering
  \includegraphics[width=0.36\linewidth]{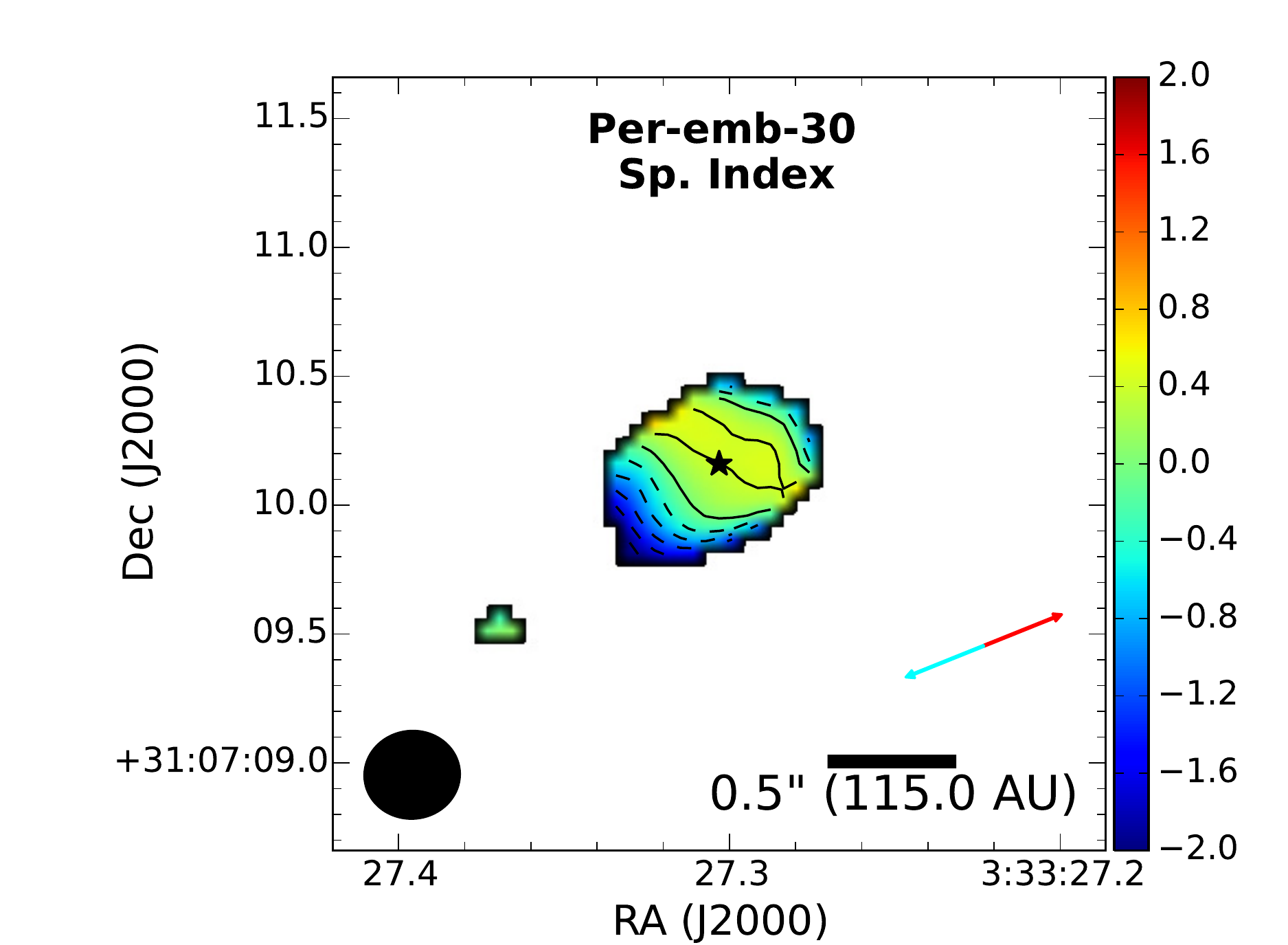}
 \centering
  \includegraphics[width=0.29\linewidth]{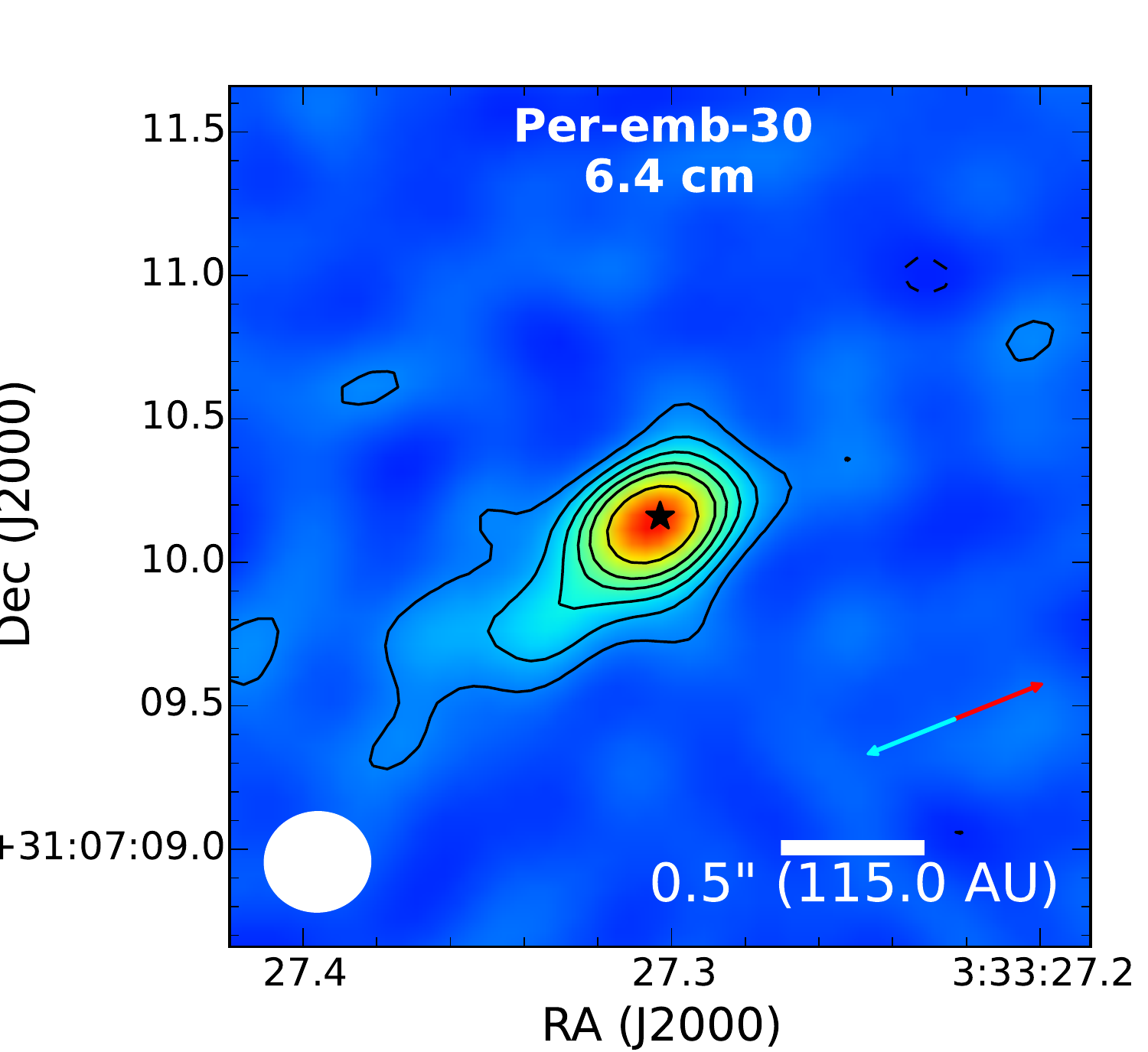}
 \centering
  \includegraphics[width=0.29\linewidth]{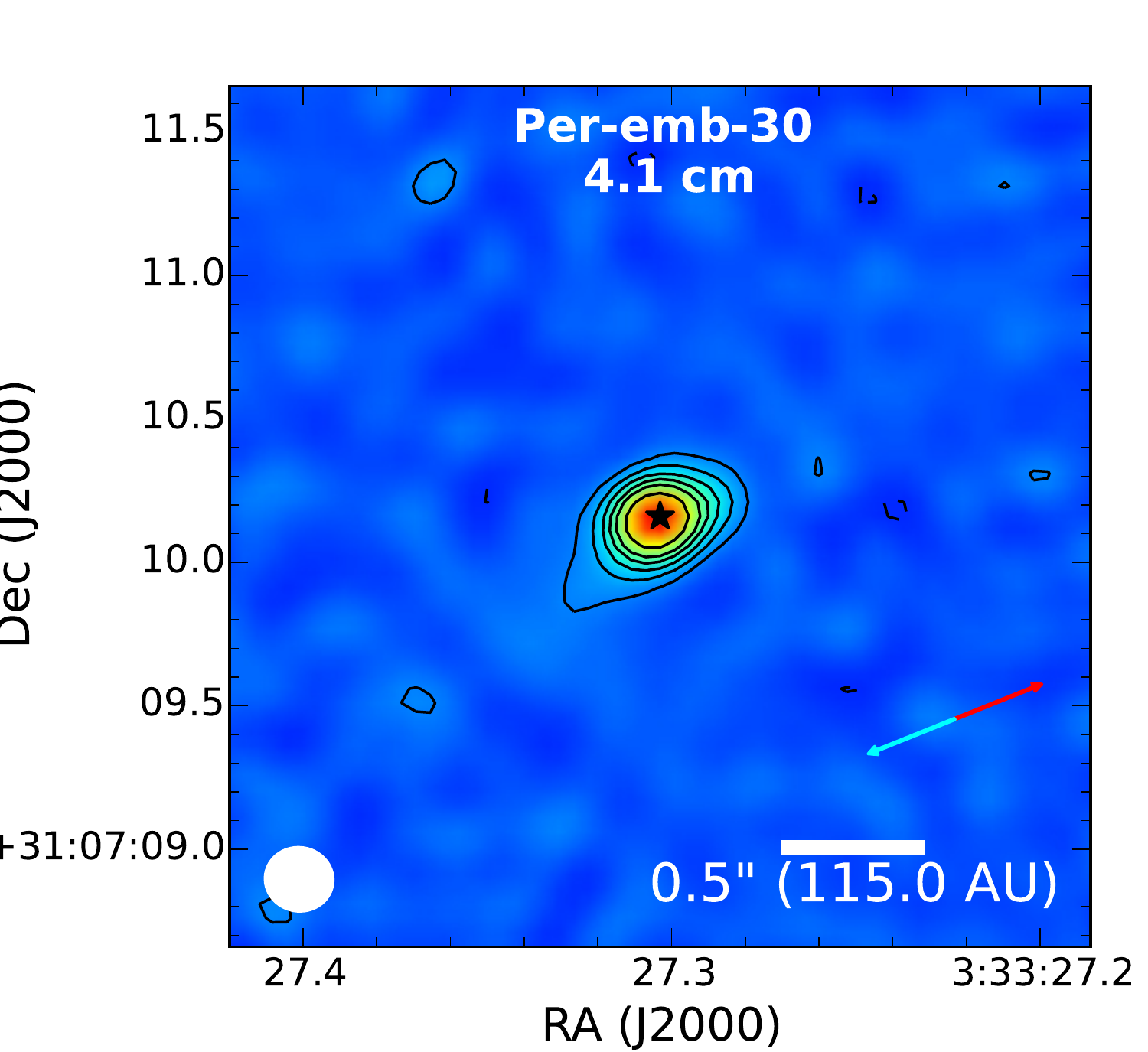}

\caption{Images of Per-emb-30 with contours as in Figure \ref{fig:per36} ($\sigma _{6.4\ cm}=4.87\ \mu $Jy and $\sigma _{4.1\ cm}=4.01\ \mu $Jy).
Synthesized beam is shown in the left bottom corner (Sp. Index and 6.4 cm: 0\farcs34 x 0\farcs37, 4.1 cm: 0\farcs24 x 0\farcs22).
 The star marks the position of the protostar based on Ka-band observations \citep{Tobin2016} and the
red and blue arrows indicate the outflow direction from \cite{Davis2008} with a position angle $109\degree$.}
\label{fig:per30}

\end{figure}

\subsection{Per-emb-33}
Per-emb-33 is a triple system of Class 0 protostars, also known as L1448~IRS3B \citep{Looney2000} and L1448 N(B) \citep{Curiel1990}. Per-emb-33-A is dominating the emission in the C-band and Per-emb-33-C is marginally detected, in contrast to the 9~mm observations, where source C has the highest flux density \citep{Tobin2016}. The resolution was not sufficient to resolve Per-emb-33-B from source A. The radio emission is also extended along the jet outflow direction measured by \citet{Lee2015}, making it difficult to determine whether or not Per-emb-33-C is actually detected or if this is just the extended jet emission. Along the radio jet/outflow direction to the west, there is a clump of emission separated from the source position by a distance 2\farcs4 (540 AU), which also appears  to be due to the radio jet. The spectral index along the outflow is consistent with optically thin free-free emission, varying from $-0.14 \pm 0.14$ in the A source position to $-0.30 \pm 0.74$ in the north-western clump and $-0.53 \pm 0.50$ in the position of source C. Position angle of the extended jet $107 \pm 5\degree$ is consistent with the CO outflow position angle obtained by \cite{Kwon2006} and \citep{Lee2015}. \cite{Tobin2016a} find that the rotational center of the system is located closest to Per-emb-33-A, consistent with it driving the radio jet. They also showed that Per-emb-33-C drives a distinct, collimated outflow.

\begin{figure}[H]
 \centering
  \centering
  \includegraphics[width=0.36\linewidth]{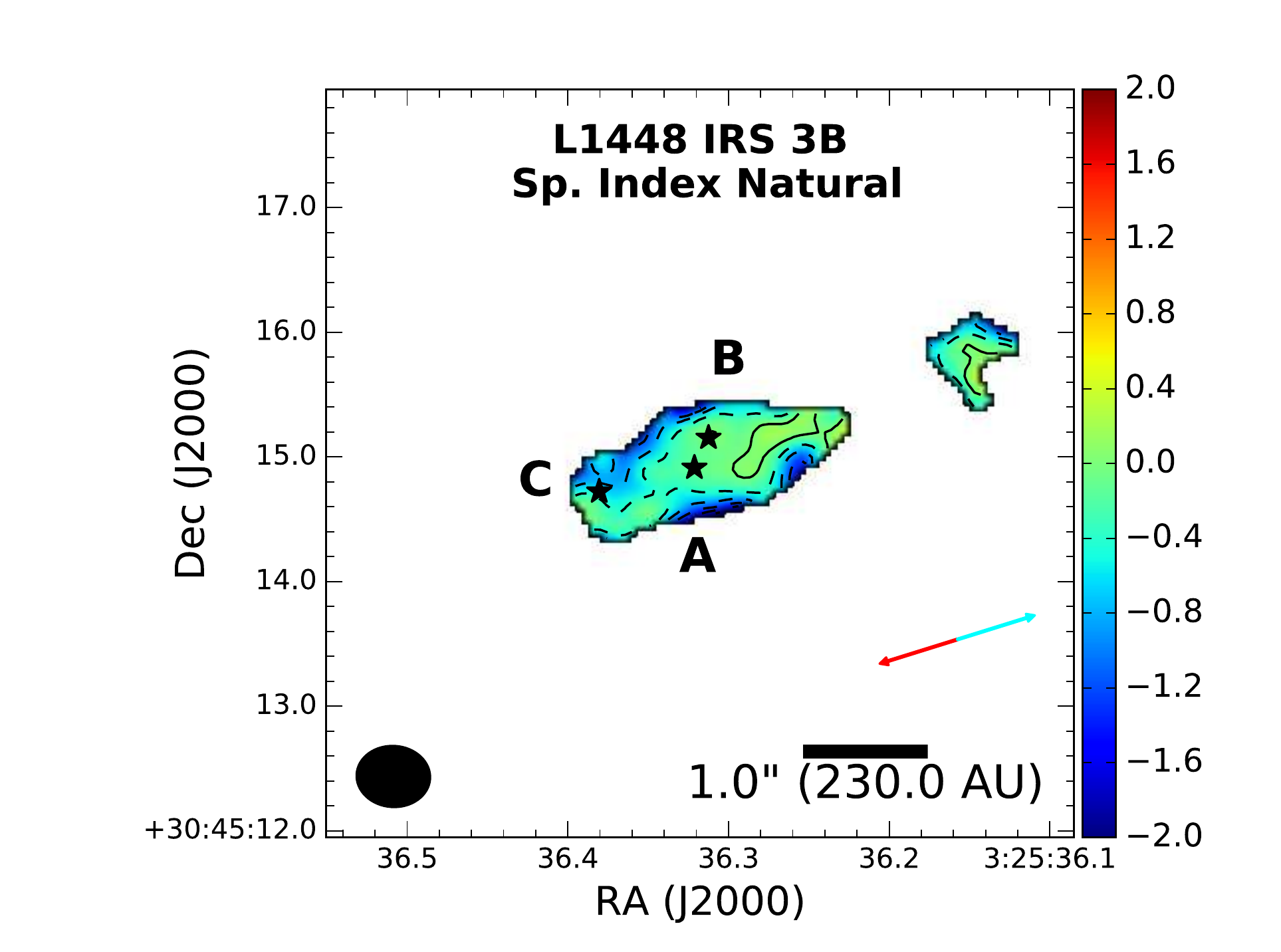}
  \label{fig:lowhist}
  \includegraphics[width=0.29\linewidth]{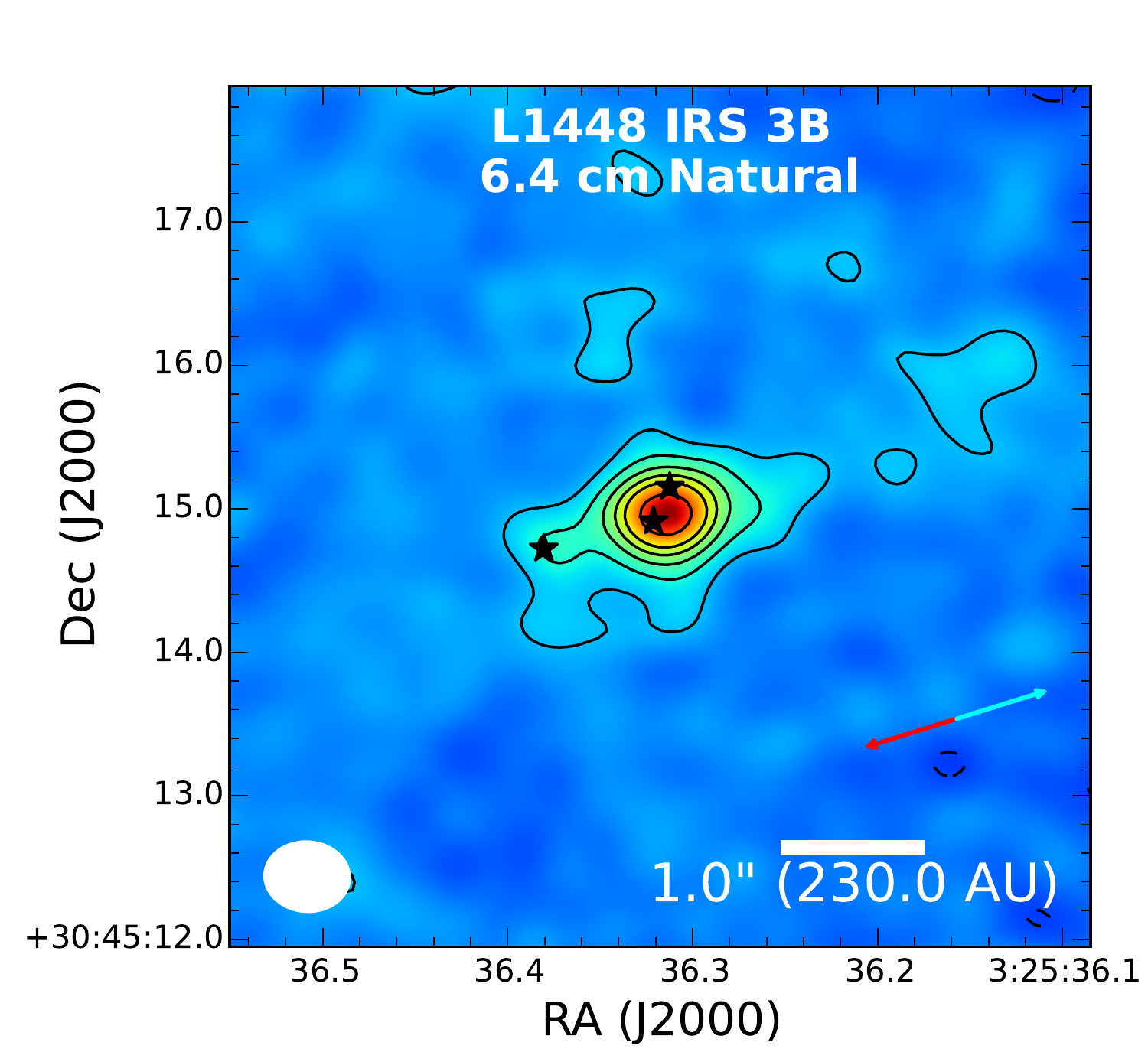}
  \label{fig:lowhist}
  \centering
  \includegraphics[width=0.29\linewidth]{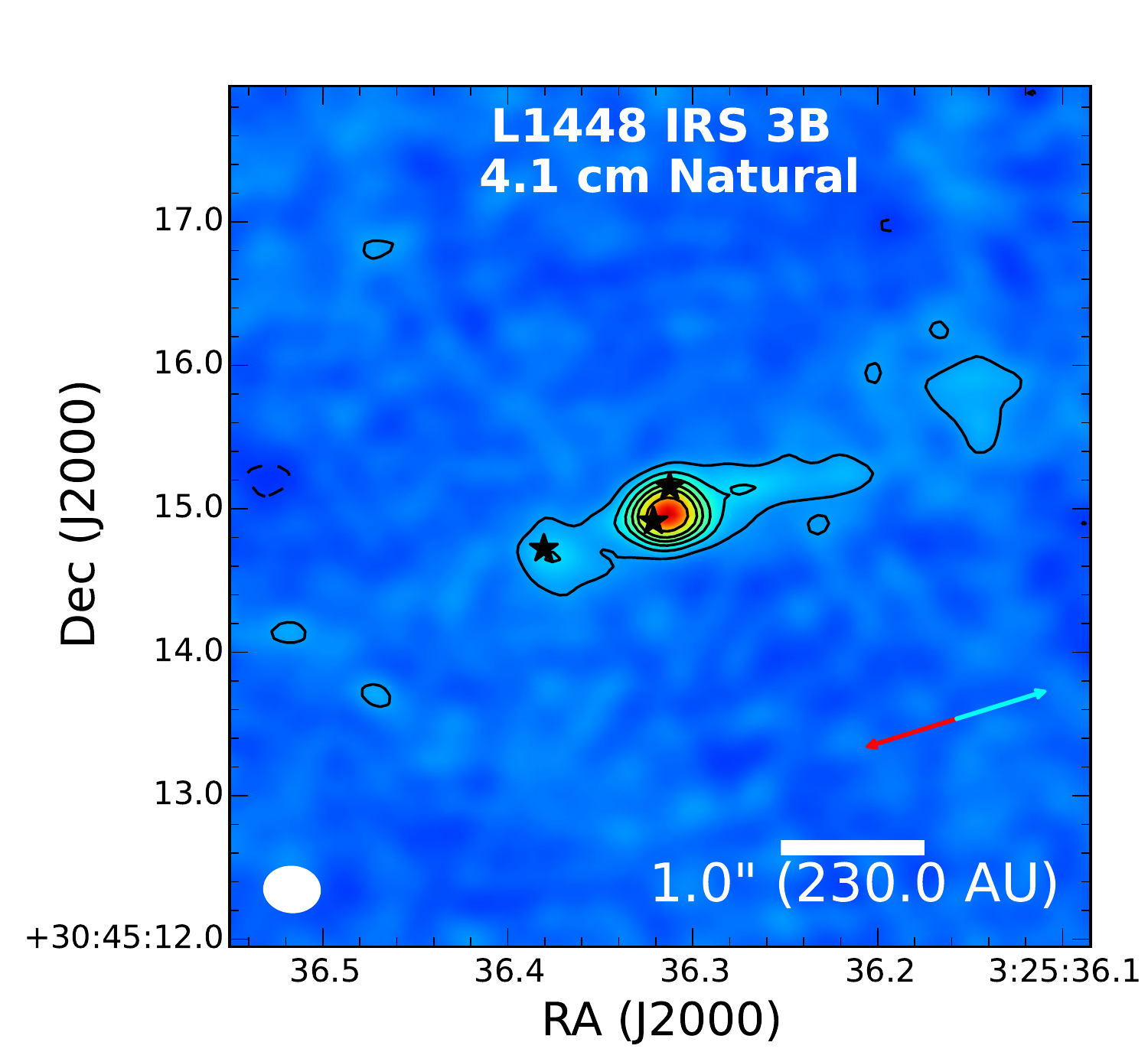}
  \label{fig:highhist}

\caption{ 
Naturally-weighted images of Per-emb-33 with contours as in Figure \ref{fig:per36}. ($\sigma _{6.4\ cm}=4.87\ \mu $Jy and $\sigma _{4.1\ cm}=4.01\ \mu $Jy).
Synthesized beam is shown in the left bottom corner (Sp. Index and 6.4 cm: 0\farcs34 x 0\farcs37, 4.1 cm: 0\farcs24 x 0\farcs22).
Stars mark the positions of the protostars based on Ka-band observations \citep{Tobin2016}. The source on the east is 33-C, while A (south) and B (north) form a tight binary.
The red and blue arrows indicate outflow direction from \cite{Kwon2006} with a position angle of $105\degree$.}
\label{fig:per33}
\end{figure}

\subsection{L1448 IRS3A}
L1448 IRS3A is a Class I protostar and a wide companion of Per-emb-33, separated by 7\farcs3 \citep[$\sim$1700 AU;][]{Looney2000, Tobin2016}.
It has extended centimeter emission in the western direction at both 4.1~cm and 6.4~cm. The spectral index
northeast of the protostar position is $-0.61 \pm 0.24$. This indicates that the emission could
be produced by synchrotron emission, but the fact that emission arises on the edge of the source casts a doubt on this detection. The portion more extended to the west has a spectral index of $-0.13 \pm 0.12$ and is consistent with free-free emission. Measured position angle ($79\pm 1\degree$) of the 4.1~cm and 6.4~cm emission
is notably different from the CO outflow position angle of $38\degree$ measured by \cite{Lee2015}. Recent ALMA observations show a disk in dust emission aligned perpendicular to the
CO outflow (Tobin et al in prep.); thus the difference in PA between the centimeter radio and
the molecular outflow could indicate that the free-free emission is tracing a portion of an ionized outflow cavity. This may be similar to what was observed toward a more luminous protostar
in Serpens \citep{hull2016}. The 9~mm observations of L1448~IRS3A are also extended in the same direction and the spectral index was found to be relatively flat \citep{Tobin2016}. Thus, the free-free emission is likely contributing also at shorter wavelengths.

\begin{figure}[H]
 \centering
  \includegraphics[width=0.36\linewidth]{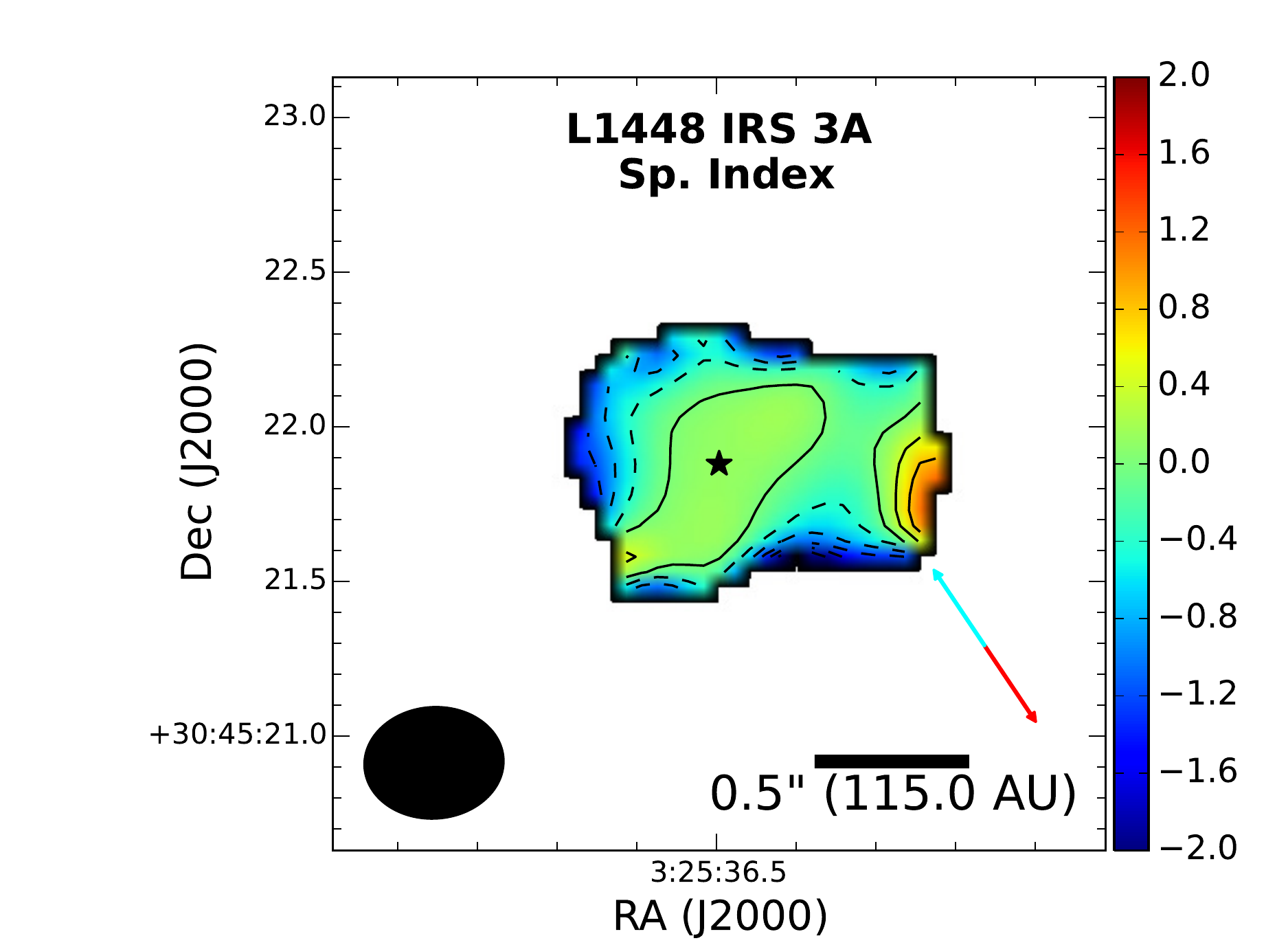}
  \label{fig:lowhist}
  \centering
  \includegraphics[width=0.29\linewidth]{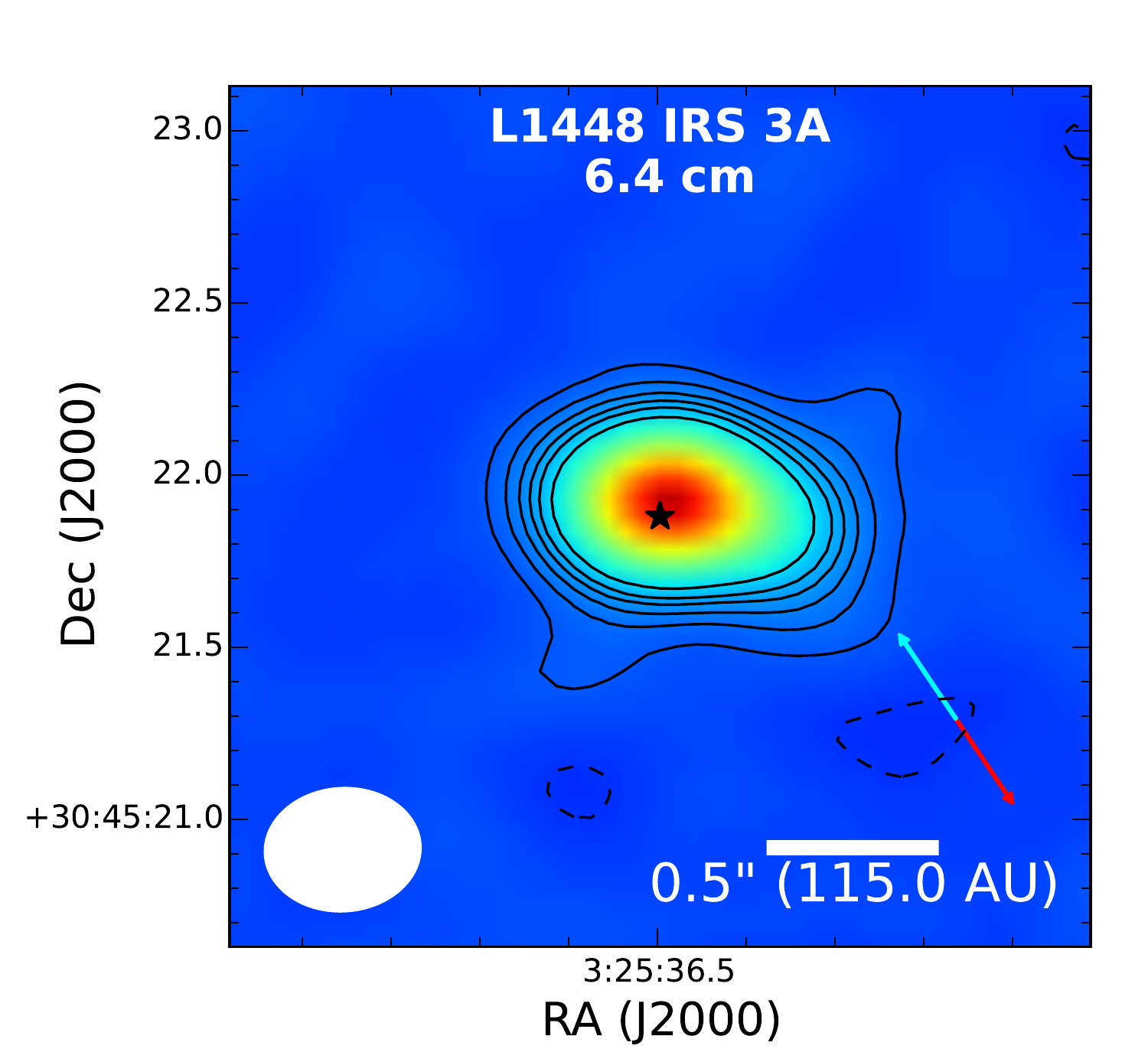}
  \label{fig:highhist}
  \centering
  \includegraphics[width=0.29\linewidth]{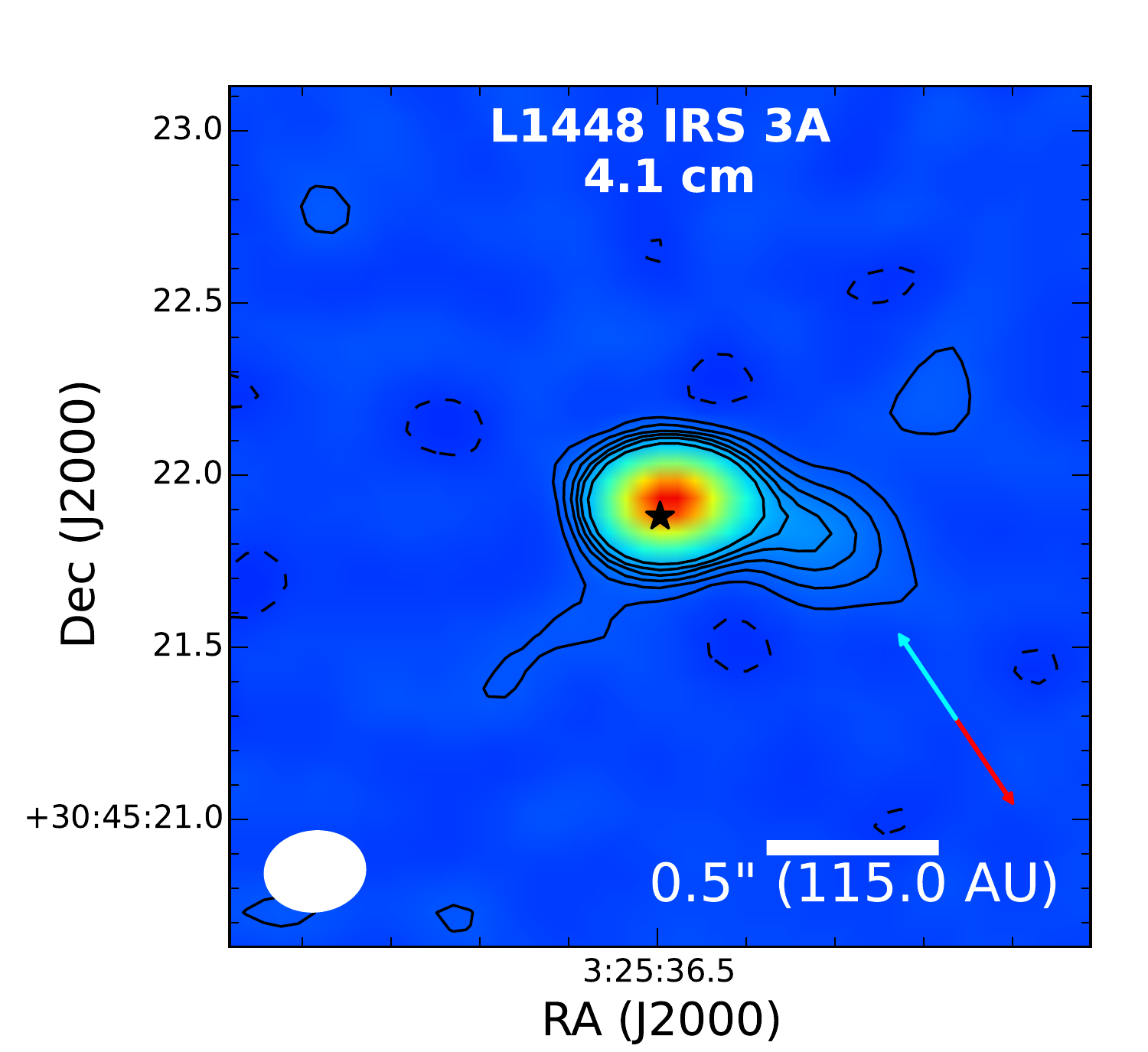}
  \label{fig:lowhist}
\caption{ Images of L1448 IRS3A with contours as in Figure \ref{fig:per36} ($\sigma _{6.4\ cm}=5.26\ \mu $Jy and $\sigma _{4.1\ cm}=4.20\ \mu $Jy).
Synthesized beam is shown in the left bottom corner (Sp. Index and 6.4 cm: 0\farcs45 x 0\farcs36, 4.1 cm: 0\farcs29 x 0\farcs23).
 The star marks the position of the protostar based on Ka-band observations \citep{Tobin2016}. 
Red and blue arrows indicate outflow direction from \cite{Lee2015} with a position angle of $38\degree$.} 
\label{fig:per8}
\end{figure}

\subsection{Per-emb-8}
Per-emb-8 is a Class 0 protostar in the IC 348 region, and it is located 9\farcs6 ($\sim$2200 AU) from Per-emb-55 which is a close multiple itself \citep{Tobin2016}. Per-emb-8 shows
4.1~cm and 6.4~cm emission extended north and slightly east, the position angle is $13\pm 2\degr$.
The spectral index smoothly decreases further away from the protostar, with a value of $-0.25 \pm 0.16$ along
the extension, and the spectral index remains consistent with free-free emission despite being marginally
steeper than optically thin free-free emission. This source did not have previously
published CO or H\raisebox{-.4ex}{\scriptsize 2} observations, which would indicate outflow direction and the \textit{Spitzer}
data are not well-enough resolved to be definitive. However, recent ALMA observations resolved a disk around Per-emb-8 (Tobin et al. in prep.) and observed the outflow in $^{12}$CO.
The observed disk has a position angle of 45\degr\ and the CO outflow is orthogonal to this.
The extended 4.1~cm and 6.4~cm emission has a position angle of about $\sim$45\degr\ different from the {CO outflow
direction. Thus, like L1448 IRS3A, the extended radio emission might trace the edge of an ionized outflow cavity
or the ionized surface of the disk.

\begin{figure}[H]
 \centering
  \includegraphics[width=0.36\linewidth]{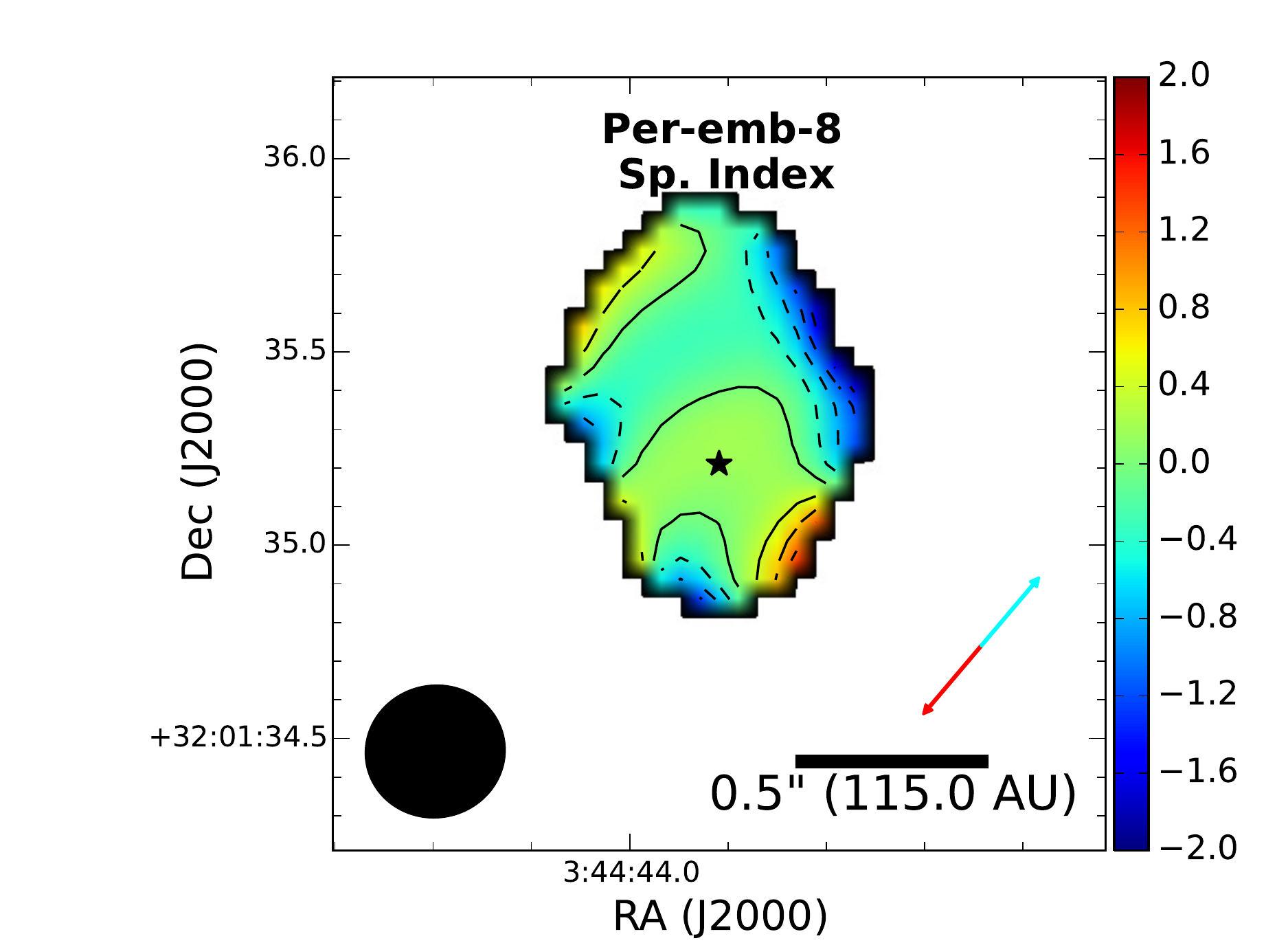}
  \label{fig:lowhist}
  \centering
  \includegraphics[width=0.29\linewidth]{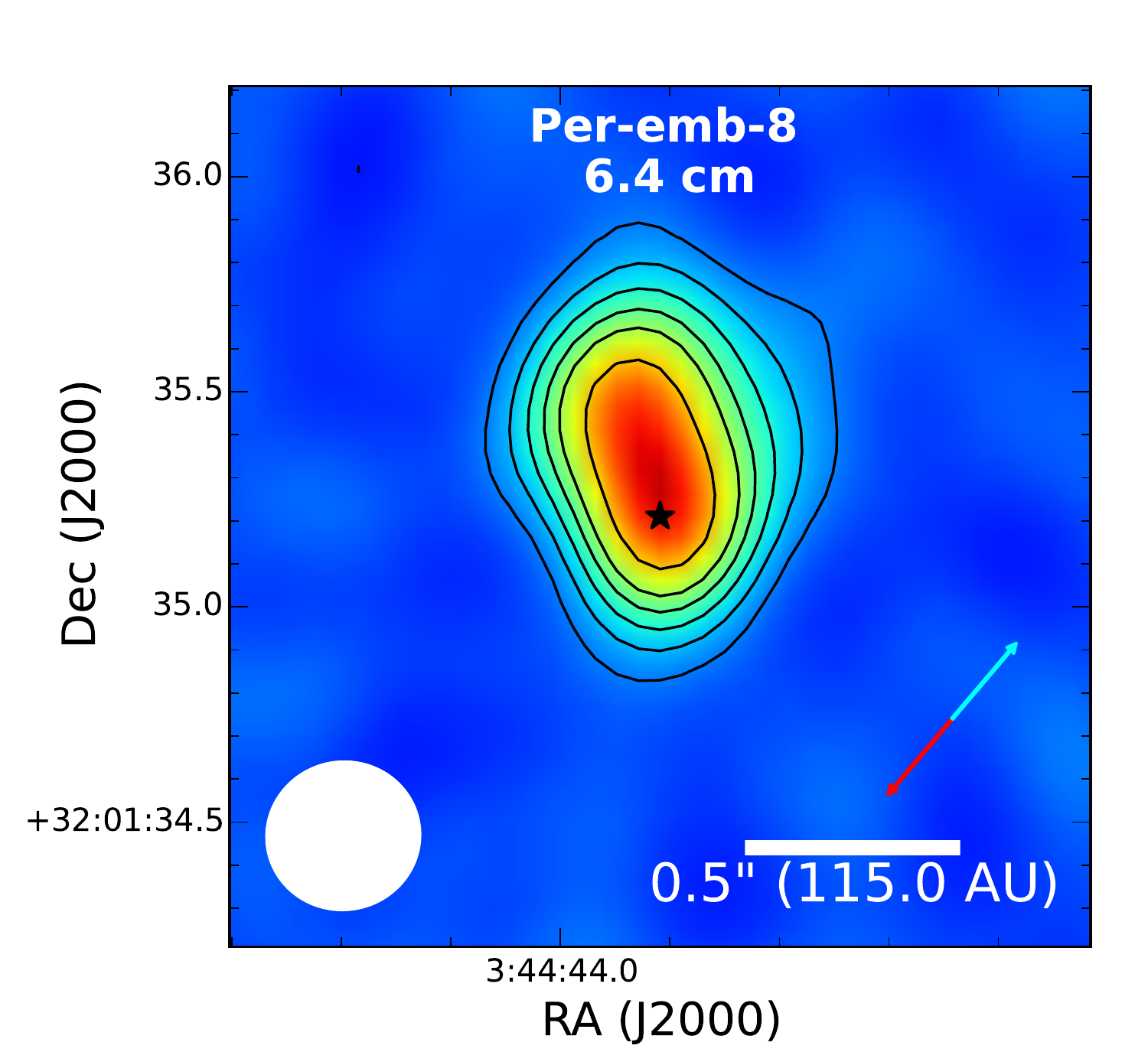}
  \label{fig:highhist}
  \centering
  \includegraphics[width=0.29\linewidth]{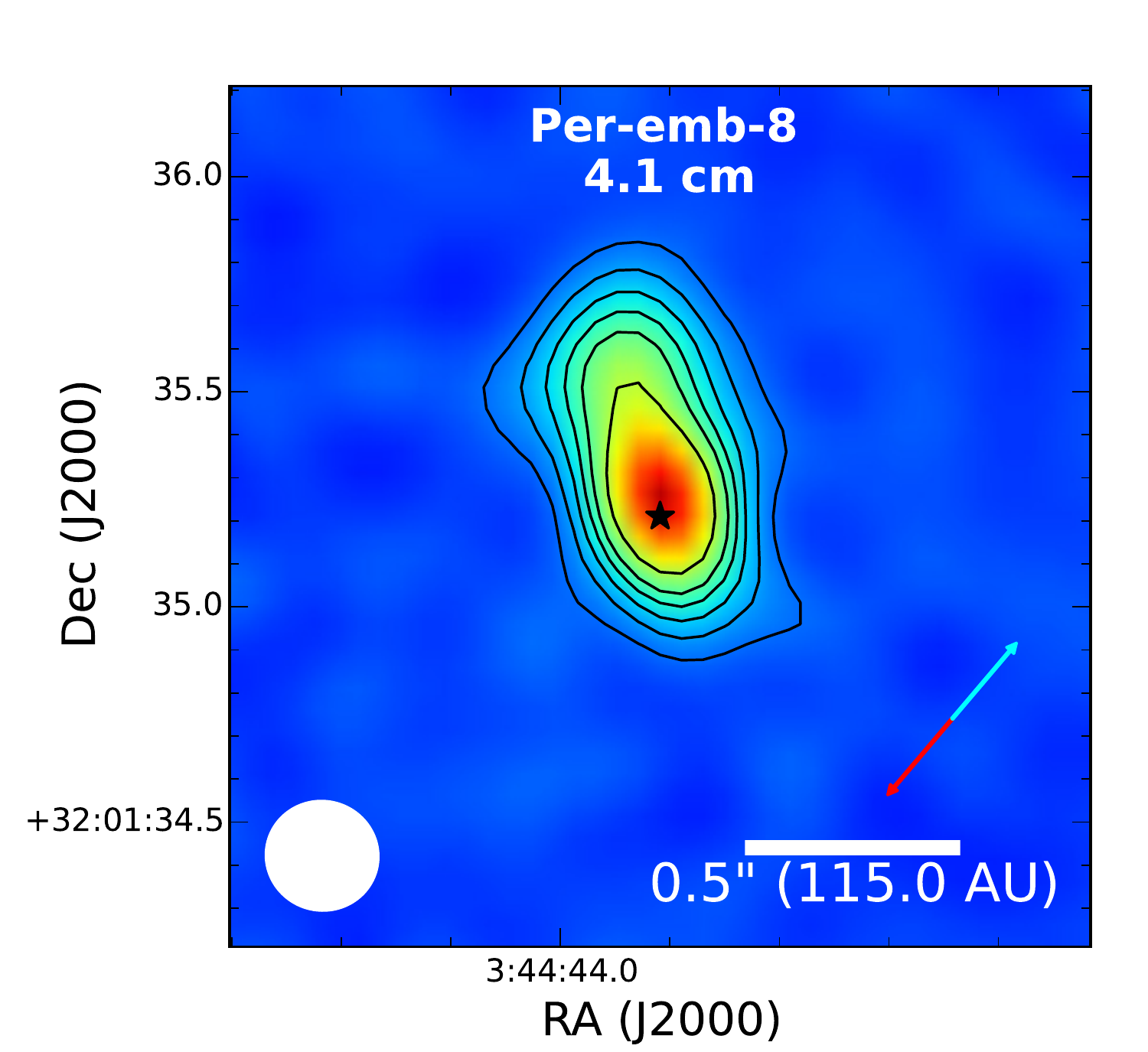}
  \label{fig:lowhist}
\caption{ Images of Per-emb-8 with contours as in Figure \ref{fig:per36} ($\sigma _{6.4\ cm}=4.90\ \mu $Jy and $\sigma _{4.1\ cm}=3.70\ \mu $Jy).
Synthesized beam is shown in the left bottom corner (Sp. Index and 6.4 cm: 0\farcs36$\times$0\farcs34, 4.1~cm: 0\farcs26$\times$0\farcs25).
 The star marks the position of the protostar based on Ka-band observations \citep{Tobin2016}. Red and blue arrows are orthogonal to the major axis of the disk based on ALMA observations (Tobin et al. in prep.).} 
\label{fig:per8}
\end{figure}

\subsection{Per-emb-18}

Per-emb-18 is a Class 0 system in the NGC1333 IRAS7 region consisting of two protostars separated by only 20 AU \citep{Tobin2016}. Thus, C-band observations were
unable to resolve the system. The radio emission is asymmetric and the spectral index decreases steeply from $0.37\pm 0.14$ at the protostellar position to $-1.13\pm 0.42$ in the southern outflow. Measured position angle of $164\pm 16\degree$ is consistent
with the value of the H$_2$ outflow ($159\degree$) observed by \citet{Davis2008}. The steep negative spectral
index in this source is also partly due to the lack of emission detected at 4.1~cm where extended 6.4~cm emission is present.

\begin{figure}[H]
 \centering
  \includegraphics[width=0.36\linewidth]{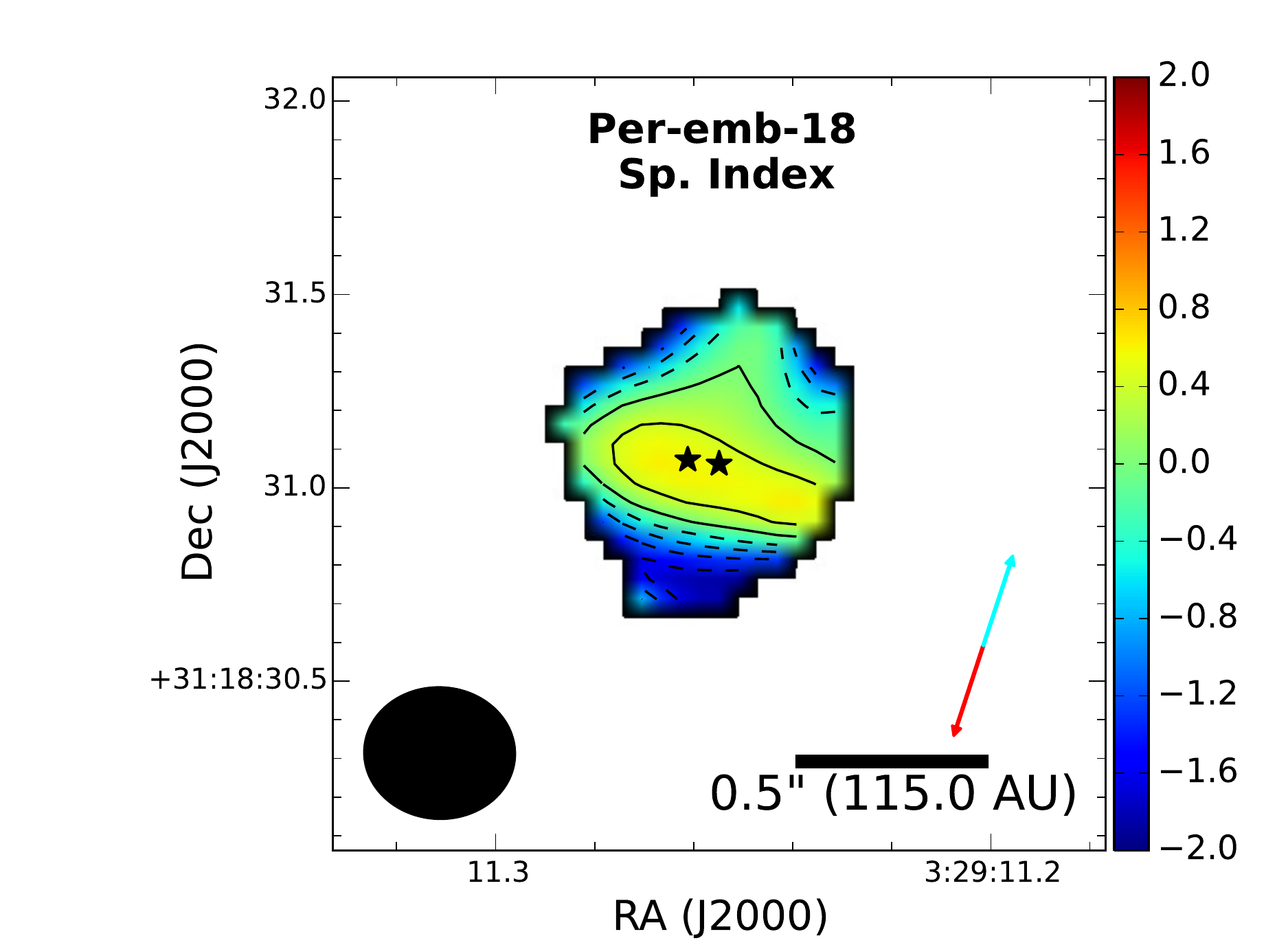}
  \label{fig:lowhist}
  \centering
  \includegraphics[width=0.29\linewidth]{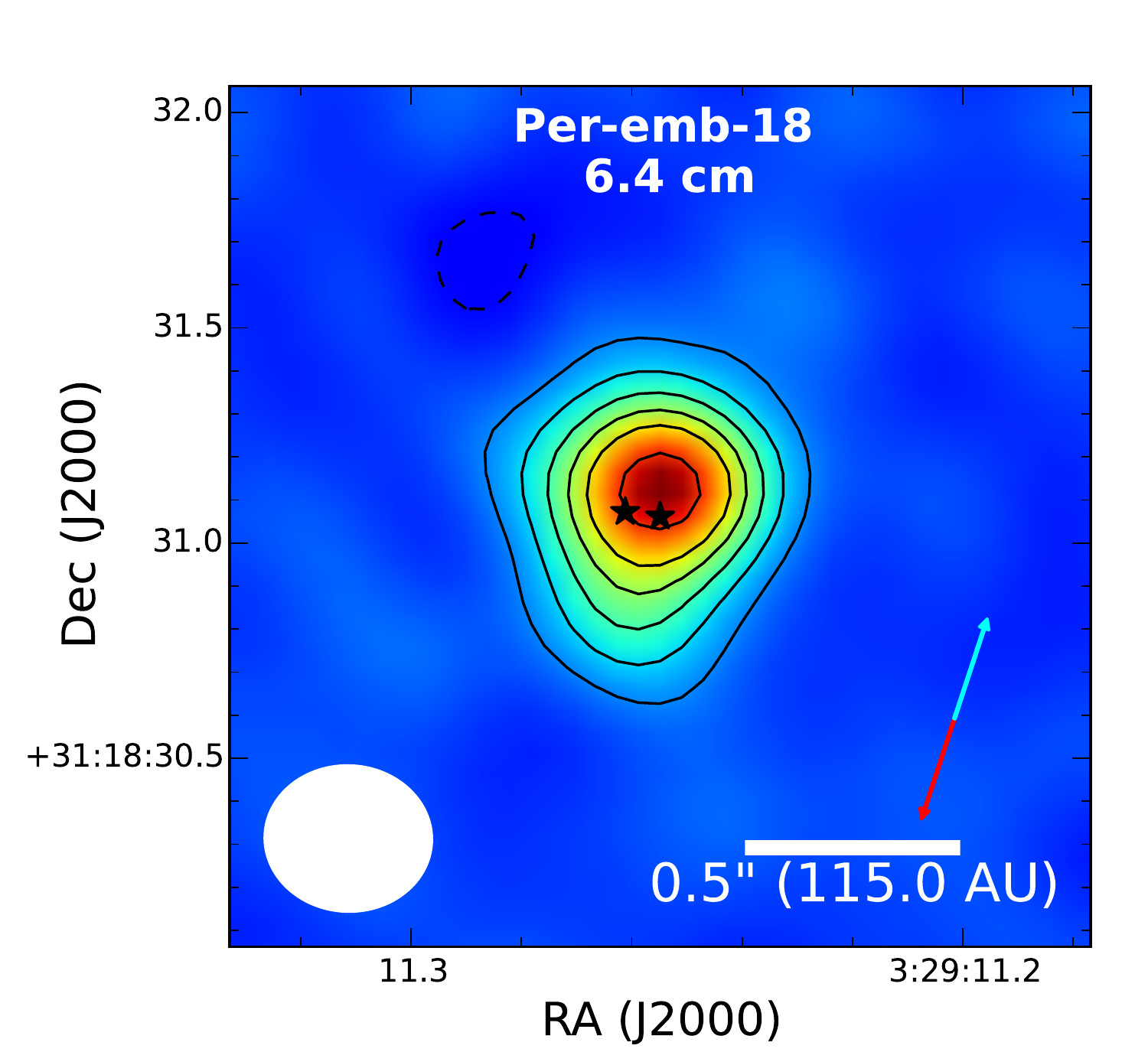}
  \label{fig:highhist}
  \centering
  \includegraphics[width=0.29\linewidth]{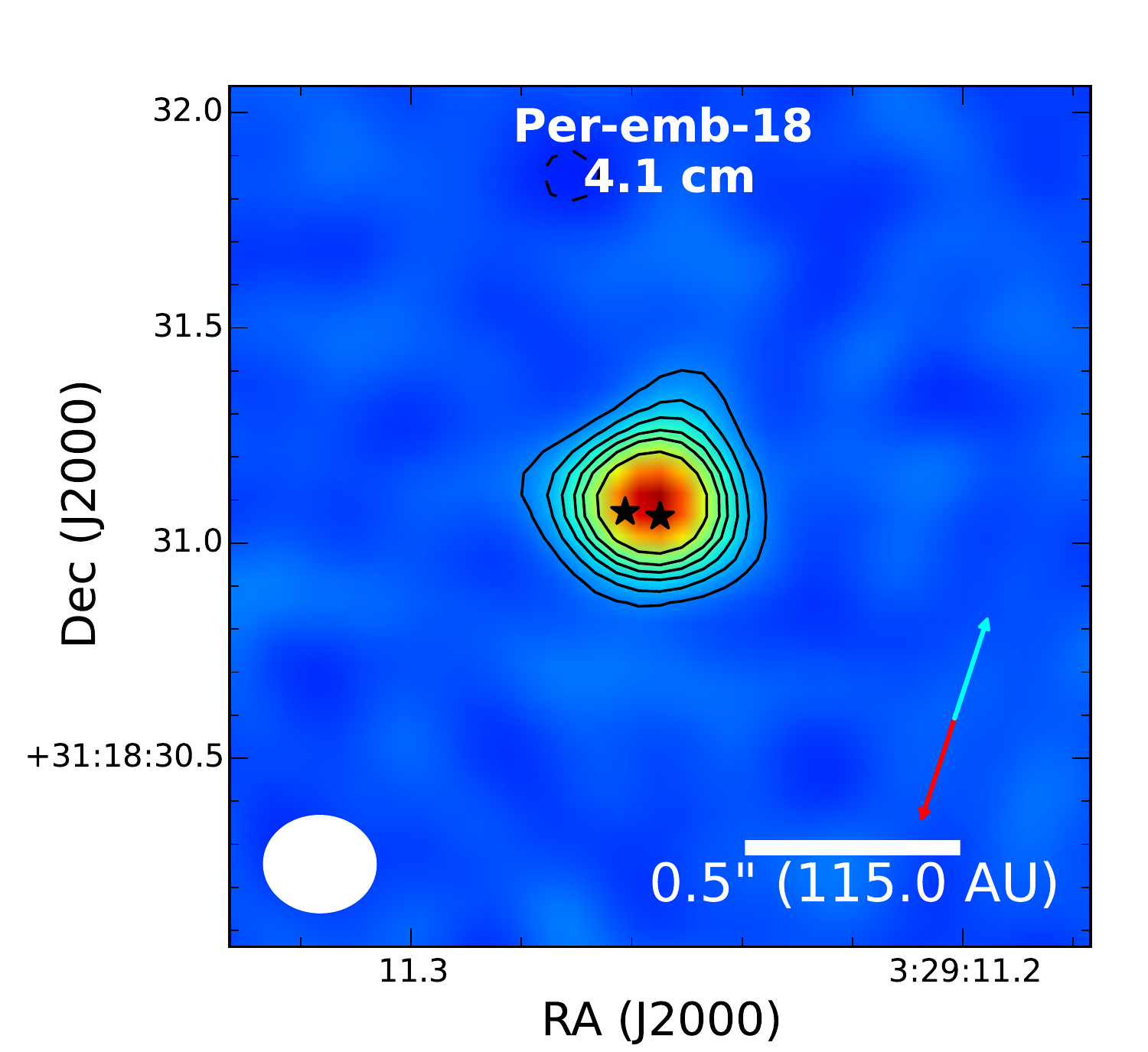}
  \label{fig:lowhist}
\caption{ Images of Per-emb-18 with contours as in Figure \ref{fig:per36} 
($\sigma _{6.4\ cm}=6.77\ \mu $Jy and $\sigma _{4.1\ cm}=5.54\ \mu $Jy).
Synthesized beam is shown in the left bottom corner (Sp. Index and 6.4 cm: 0\farcs56$\times$0\farcs38, 4.1~cm: 0\farcs37$\times$0\farcs25).
 The stars mark the position of the protostars based on Ka-band observations \citep{Tobin2016}. 
Red and blue arrows indicate outflow direction from \cite{Davis2008} with a position angle of $159\degree$.} 
\label{fig:per18}
\end{figure}

\subsection{Per-emb-20}
Per-emb-20 is a Class 0 protostar known also as L1455 IRS4. An extension of the 4.1~cm and 6.4~cm is observed with a position angle of $137 \pm 8\degr$. This position angle is reasonably consistent with the 115\degr\ position angle found by \citet{Davis2008} from H$_2$ observations. The extension is small, but is present at both wavelengths, and other sources in the map do not show a similar extension. The spectral index is flat in the central part (0.00$\pm$0.18) and is decreasing toward the northwest along the extension. The spectral index measured along the extension is -1.38$\pm$0.51, but
the extension is only marginally resolved at 6.4~cm.
\begin{figure}[H]
 \centering
  \includegraphics[width=0.36\linewidth]{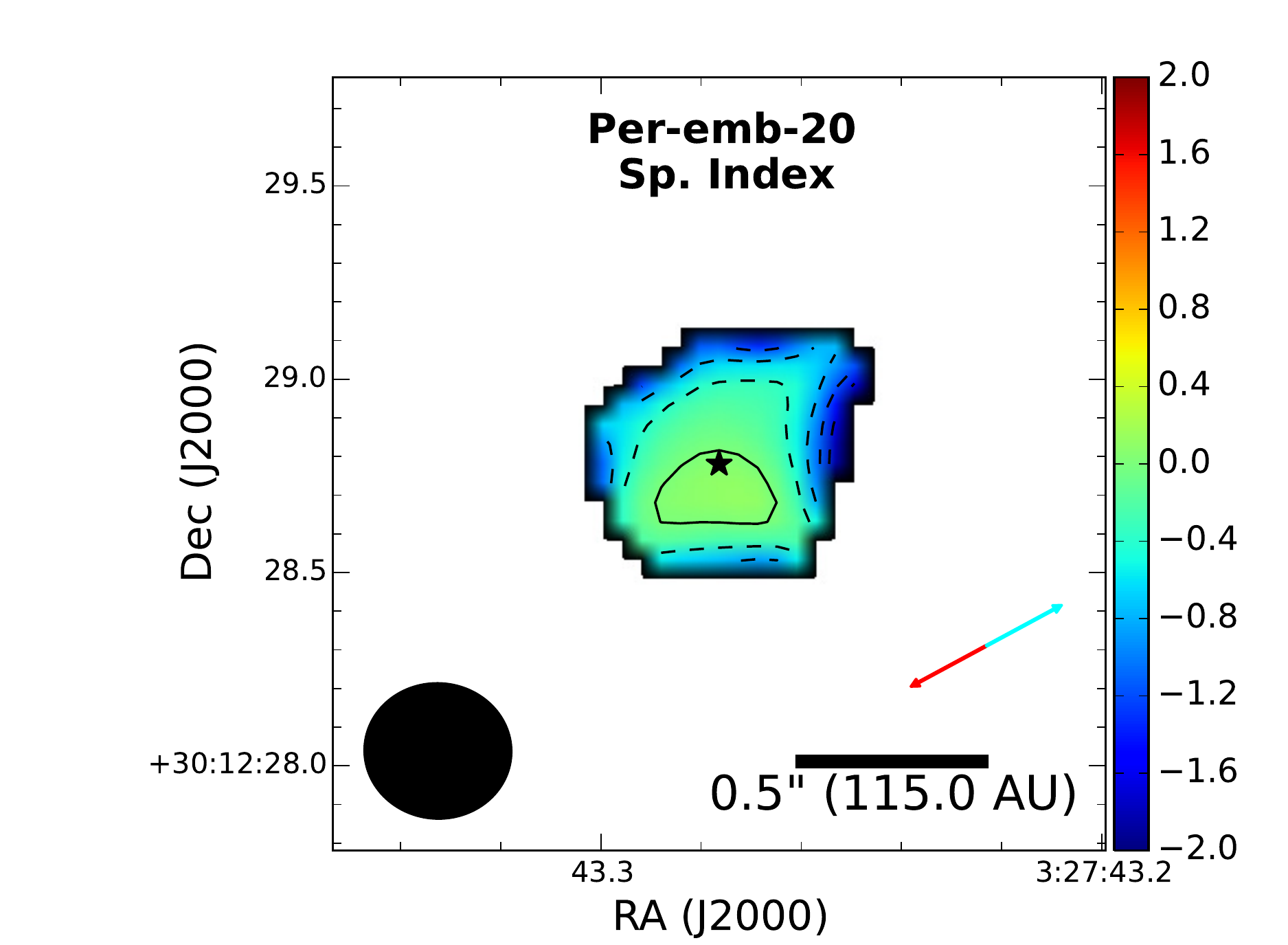}
  \includegraphics[width=0.29\linewidth]{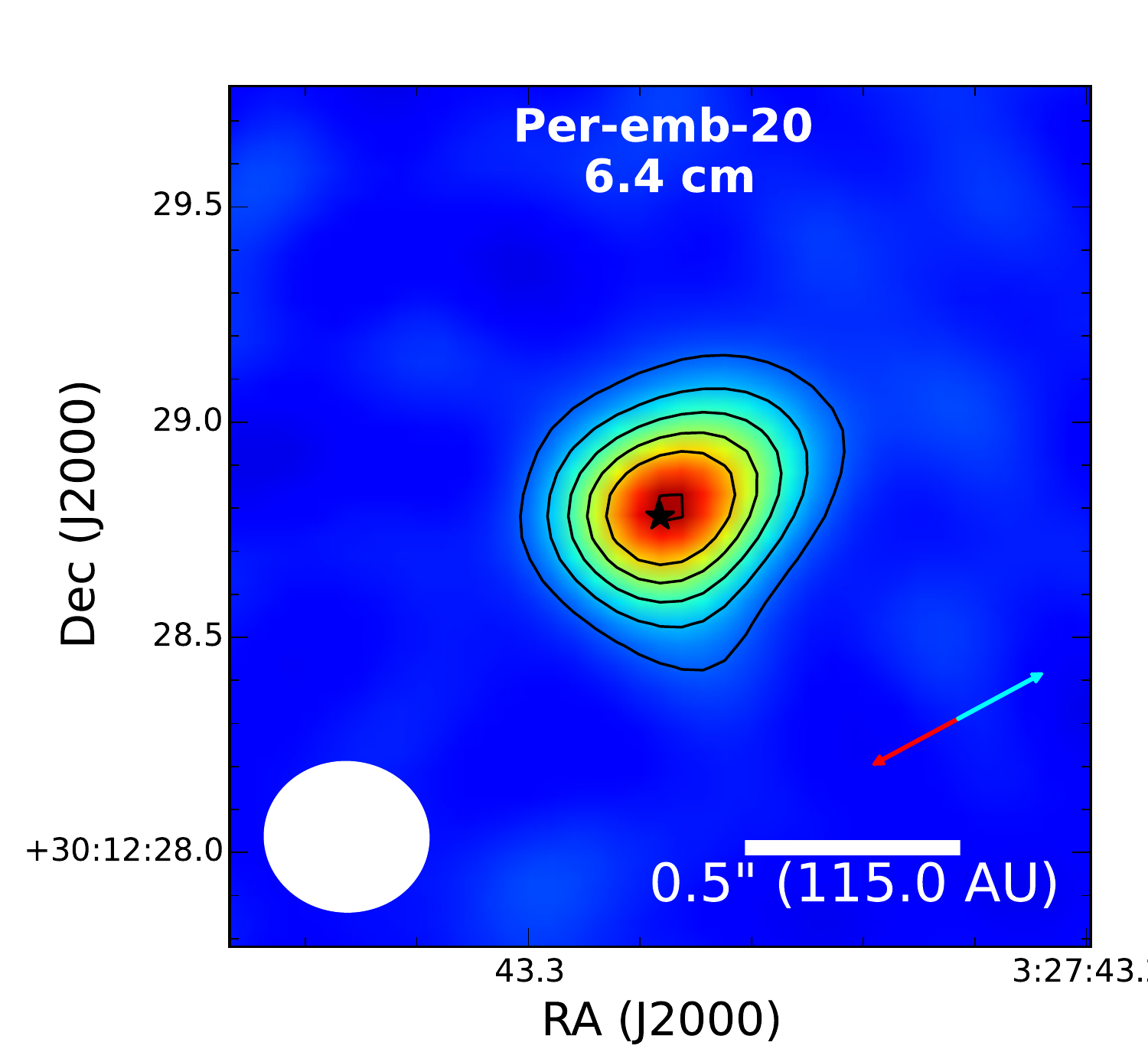}
  \includegraphics[width=0.29\linewidth]{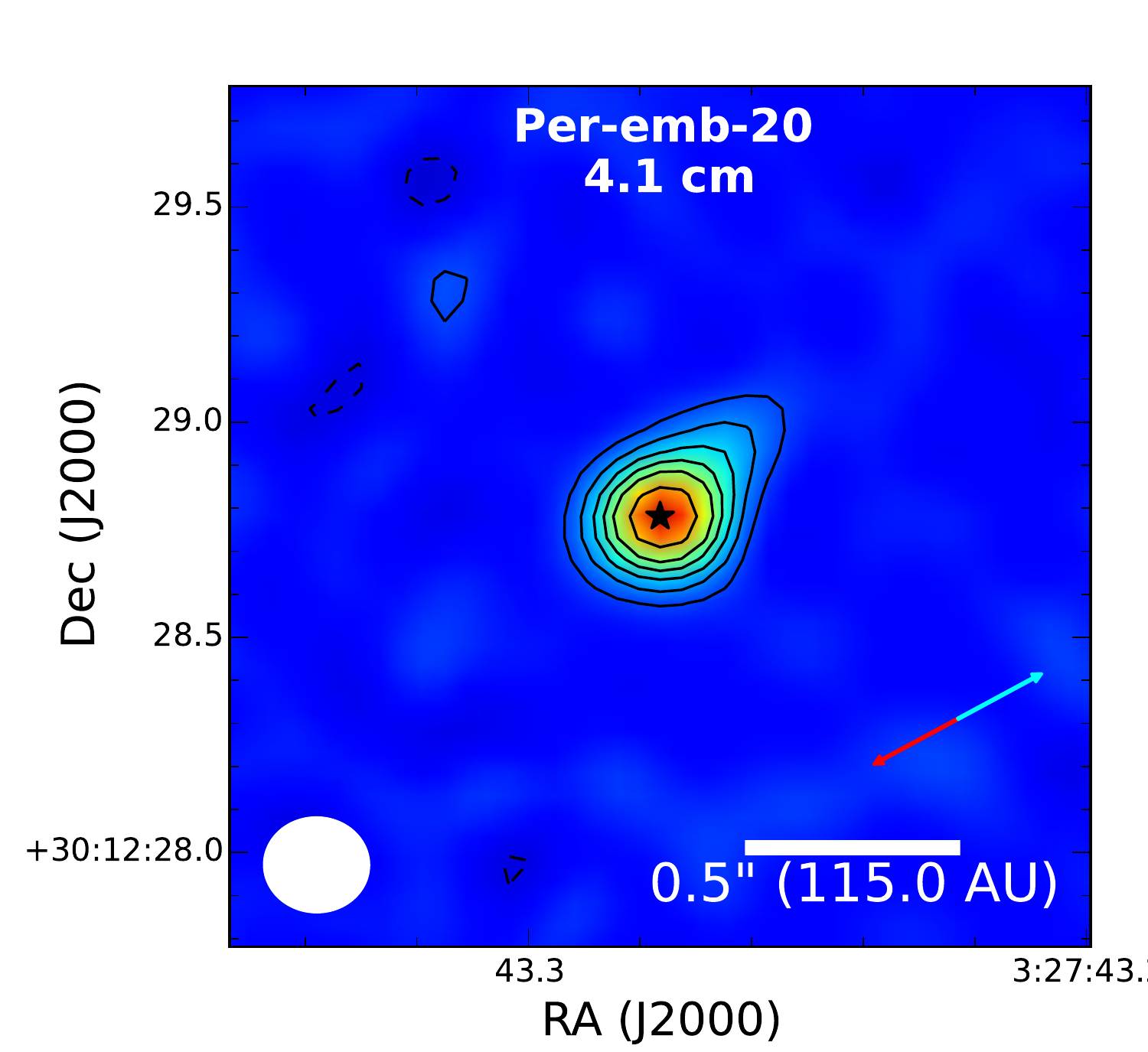}

\caption{ Images of Per-emb-20 with contours as in Figure \ref{fig:per36} 
($\sigma _{6.4\ cm}=5.30\ \mu $Jy and $\sigma _{4.1\ cm}=4.05\ \mu $Jy).
Synthesized beam is shown in the left bottom corner (Sp. Index and 6.4 cm: 
0\farcs38$\times$0\farcs35, 4.1~cm: 0\farcs24$\times$0\farcs22).
The star marks the position of the protostar based on Ka-band observations \citep{Tobin2016}. 
Red and blue arrows indicate outflow direction from \cite{Davis2008} with a position angle of $115\degree$.} 
\label{fig:per20}
\end{figure}

\section{Discussion}
Resolved jets at centimeter wavelengths have typically been associated with massive protostars (e.g., HH 80-81; see \citealt{Marti1995}) and with a few remarkable low-mass protostars that have exceptionally powerful jets (e.g., L1527, HH 1-2 VLA 1, SVS 13C; see \citealt{Rodriguez1997,Rodriguez2000,Reipurth2004}). However, analysis of the frequency of extended radio jets toward an unbiased sample of low-mass protostars has only recently become possible with the advent of the upgraded Jansky VLA. While some of the previous studies looked at a reasonably large number of sources \citep[e.g.,][]{Anglada1998}, the observations were often conducted at resolutions of $\sim$1\arcsec~or greater and the observations were not very sensitive due to the bandwidth limitations of the old VLA. Thus, while some of the protostars
observed may have had extended radio jets, the surface brightness sensitivities of the previous observations may have been too low even if the resolution was high enough. Thus, our sensitive, unbiased survey has enabled the detection of 8 extended radio jets toward low-mass stars in Perseus, where only one extended radio jet had been previously found \citep{Rodriguez1997}. The observations presented here provide a more complete picture of extended radio jets in low-mass protostellar systems.

In the study of radio emission toward selected regions with prominent Herbig-Haro objects, \cite{Reipurth2004} showed that almost half of the detected sources have radio jets. Only 11\% of protostars detected in our unbiased survey have resolved emission. It is worth noting that sources of some well-known HH objects in Perseus (e.g., HH 7-11) appear compact in our observations  (Tychoniec et al. in prep.).

\subsection{Why Are Some Jets Extended?}

With a larger sample of extended jets, we are able to investigate if there are any evolutionary or morphological trends behind the detection of extended jets. The majority of extended jets are found toward Class 0 protostars, Per-emb-8, Per-emb-18, Per-emb-20, Per-emb-33, SVS13C, and Per-emb-30 which is a borderline Class 0/I source; the others are Class I protostars (Per-emb-36, L1448 IRS3A). Taking into account a similar number of Class 0 (40) and Class I (38) protostars in the sample, and additional 6 borderline sources, at first glance, it appears that youth tends to play a role in the presence of extended jet emission. Furthermore, the two Class I protostars with extended jets are found within dense regions with Class 0 protostars and dense molecular gas in close proximity. Per-emb-20 is the most isolated protostar found with an extended jet.  It should be noted that protostellar ages, based on bolometric temperature, are subject to large uncertainties, due to inclination and reddening.
The sources have bolometric luminosities that range between 1 to 10 L$_\odot$, whereas the most luminous sources in the VANDAM sample (e.g., 4 sources with L$_{\textrm{bol}}\ >\ $10 L$_\odot$) do not have extended jets. Thus, extended jets are not directly linked with protostellar luminosity. Geometric orientation of the sources could play a role; for instance, the pole-on alignment would make extended jet
emission difficult to observe, but this would prevent the detection of the extended emission of only about 10 to 20\% of sources based on the probabilities of random orientations. Edge-on orientation, on the other hand, can make sources appear under-luminous \citep{Whitney2003}.

It is also possible that extended radio emission could be a transient phenomenon. Strong variability of jets from protostars is supported both with numerical modelling \citep{Machida2014} and observations \citep{Rodriguez2000}. Much less is known about the origin of the variability. Regarding the close relation between accretion and ejection, resolved jets could be induced by recent accretion burst.  However, the association of enhanced free-free emission with an accretion burst has not been firmly established given the recent study of \cite{galvan-madrid2015} toward the outbursting source HOPS-383 where an increase in the centimeter flux density was not observed post-outburst. It is possible that the enhanced free-free emission is visible later post-outburst when additional outflow material propagates to greater distances from the protostar.

\subsection{Why Are Some Jets Not Extended?}
While the detection of these extended radio jets is remarkable, it is also remarkable that many sources
driving powerful outflows do not have extended radio jets. Famous outflow driving
sources such as HH 211 \citep[e.g.,][]{Gueth1999}, L1448-C \citep[e.g.,][]{Bachiller1990}, SVS 13A, NGC 1333 IRAS 2A, NGC 1333 IRAS 4A/B \citep[e.g.,][]{Plunkett2013} only show compact free-free emission.
These sources are at a variety of inclination angles, so the pole-on viewing should not be a systematic problem. They are all also very young sources and are located in regions with high protostellar density,
high surrounding gas density, and/or nearby sources driving extended jets. Furthermore, some of these sources
have high-velocity molecular emission features located near the protostar position, indicating that energies are
high enough for shocks to produce extended free-free emission on these scales. Thus, it remains unclear why some of
these powerful outflow sources do not have extended jets, while some less energetic sources do. Our observations are sensitive to structures as large as $\sim 9\arcsec$, thus any structure in the crucial region close to the protostar should be detectable up to the sensitivity limit.

It is possible that those sources have high density material surrounding them, as indicated by vast amount of entrained gas in their molecular outflows. Higher densities could result in effective molecular cooling and rapid recombination that will keep the level of ionization low, hampering the detection of the free-free emission. Rapid cooling in the protostellar envelope also implicates that for extended jets, heating occurs in-situ by internal shocks, thus it is likely that more variable jets, by producing more internal shocks, are the sources of extended radio emission. It would explain the lack of relation between the molecular outflow strength and the presence of extended radio jet, as the former is expected to be less sensitive to short-time variations.

Recent observations with \textit{Herschel} provided new insights into the properties of jets. \cite{Nisini2015} used observations of [OI] to probe the atomic component of outflows. 
Such observations allowed to describe a process of the jet evolution. In this scheme, the molecular jet would be progressively dissociated, becoming mostly atomic, and eventually appear as hot and ionized. Radio observations present a unique tool to study the base of the protostellar jet during the most embedded phase. It appears that the ionized component of the outflow can be present at very early stages of star formation.
\cite{Lahuis2010} observed signatures of the ionization with \textit{Spitzer} but without the ability to disentangle the processes spatially. From our observations, we can show that ionization can frequently happen away from the protostar, although we are not able to say whether this happens in the cavity walls or in the spot shocks.

\subsection{Resolved Emission Not Aligned With the Outflow}
Per-emb-8 and L1448 IRS3A uniquely have misaligned radio emission relative to their molecular outflows. While jets are known to precess \citep[e.g., L1157;][]{Gueth1996, Looney2007}, precession angles are typically small for low-mass protostars \citep{Frank2014} while a massive protostar has shown changes in the jet direction of up to $\sim$ 45\degr~\citep{Cunningham2009}.
Thus, it is possible, but perhaps not likely, that jet precession is the cause of the misalignment between the outflow and radio emission for these two sources. We note that a water maser had been previously identified toward the L1448 IRS3 region
with the 37m Haystack Telescope \citep{Anglada1989}, but the maser faded before
follow-up with the VLA, so exact positional relation with the extended radio jet cannot be determined.

An additional possibility is that we are tracing an ionized portion of an outflow cavity or disk surface, where there is a low-extinction pathway from the protostar emitting UV radiation to the cavity or disk surface. Both Per-emb-8 and L1448 IRS3A have well-developed disks observed in the dust continuum and molecular lines (Tobin et al. in prep.). Moreover, toward more
massive protostar Serpens SMM1, \cite{hull2016} recently discovered extended radio emission from
the outflow cavity walls. While SMM1 is a much more luminous source ($\sim$100 L$_\odot$; \citealt{Kristensen2012}) and the radio emission is extended over a much larger area, we suggest that we are observing a similar phenomenon on much smaller scales toward these two low-mass protostars.

\subsection{Significance of Synchrotron Emission}

Toward 5 of 8 protostars in our sample of extended radio jets, we detect negative spectral index in the positions away from the protostar. To date, this phenomenon has only been observed within outflows of massive protostars (e.g., Serpens SMM1, HH 80-81; see \citealt{Curiel1993} and \citealt{Marti1995}), one solar-mass Class II protostar (DG Tau; \citealt{Ainsworth2014}), and a single example of Class I low-mass protostar \citep[L778 VLA5;][]{Girart2002}. In our observation, we see the tentative detection of synchrotron emission in the outflow of low-mass Class 0/I protostars. Of the entire sample, the most robust detection of synchrotron emission offset from the protostar position is found toward Per-emb-36. It has visibly shifted peak emission at 6.4 cm, presumably associated with the intensely shocked material. This shock may be strong enough to produce cosmic rays. 

In the study of Serpens SMM1 - intermediate-mass Class 0 protostar powering radio jet with negative spectral indices, \citep{Rodriguez-Kamenetzky2016} showed that conditions in this jet allows the production of the cosmic rays by diffuse shock acceleration \citep[DSA; e.g.,][]{Drury1991}. They suggested that this process might be episodic, and for example close binary encounter might trigger jet velocity enhancement, creating favorable conditions for the cosmic rays production.
 
The presence of synchrotron emission being generated within the star-forming core could have significant implications on the chemistry of the gas associated with disk and planet formation.  Recent work has shown that high-energy cosmic rays may be attenuated by the magnetic field of the young stars \citep{Cleeves2013}, leaving only
radioactive decay as a source of ionization inside proto-planetary disks. While that model was
specific to more evolved pre-main sequence stars with disks, \cite{Padovani2013,Padovani2014} show that magnetic fields in collapsing protostellar clouds can also shield cosmic rays from penetrating too deeply into clouds.

Thus, the potential lack of ionization in these clouds will strongly affect the chemical composition of the gas in the disk forming region of the clouds. However, \cite{Padovani2016} demonstrated that cosmic rays can be accelerated to relativistic energies through first-order Fermi acceleration (DSA) even at jet velocities as low as 100 km/s, typical for low-mass outflows. Thus, if there is a component of synchrotron emission in some of these extended jets, it could be acting as a local source of cosmic rays, enabling richer chemistry \citep[e.g.,][]{Aikawa1997, Padovani2013, Eistrup2016, Rodgers-Lee2017}.

It should be noted that for some of the sources the uncertainty of the calculated spectral index value is high ($\sim0.5$), so we cannot completely rule out the possibility that the emission is optically thin free-free. \cite{Rodriguez1993} showed that indices below -0.1 are virtually impossible for thermal processes, which suggests that at least some of our sources are strong candidates for synchrotron emission.

In the positions of the shocks, we may also find associated X-ray emission \citep{Pravdo2004}, such as was found toward DG Tau \citep{Guedel2008}. The only protostar within our sample with sensitive X-ray data is SVS13C, and it was undetected. Per-emb-36 lies at the outskirts of the X-ray map taken towards NGC 1333, but the X-ray emission toward
its position cannot be positively associated with it due to the nearby, more-evolved star BD~+30~547. Non-detections have also been reported toward both Per-emb-33 and L1448 IRS3A by \cite{Tsujimoto2005}.

\section{Conclusions}

We conducted observations of all known Class 0 and I protostars in Perseus molecular cloud in 4.1 cm and 6.4 cm. 8 of 71 detected sources exhibit resolved radio emission. In this paper, we analyzed emission toward the resolved subset of sources with the following conclusions:
\begin{itemize}
\item We observe resolved emission toward 11$\%$ sources from the unbiased radio survey. Their integrated spectral indices are consistent with optically thin free-free emission, however, we find them systematically lower than the median for the whole sample.
\item Toward 5 sources - Per-emb-18, Per-emb-20, Per-emb-30, Per-emb-36, and L1448 IRS3A  -  we detect negative spectral indices in the outflow position indicative of synchrotron emission. Per-emb-36 is the most robust candidate for synchrotron emission and its peak of emission is shifted away from the protostar, indicative of a strong shock. This is one of the few detections of synchrotron emission toward a low-mass protostar. This result suggests that production of cosmic rays might be frequent, however transient, in the shocks of low-mass protostars, which could have significant repercussions for disk and planet formation.
\item Six sources have position angles matching directions of the large-scale molecular outflow, showing that in most cases radio emission can be interpreted as an ionized base of the jet.
\item Two sources, Per-emb-8 and L1448 IRS 3A, show significant misalignment of the resolved radio emission. We suggest that this emission not necessarily comes from the jet but more likely from the ionized outflow cavity wall or even the upper layer of the disk.
\end{itemize}
Since the upgrade of the VLA, we are able to study ionized component of the protostellar outflow with high resolution for an unprecedented number of protostars. This study showed that analysis of the spectral index and alignment of the thermal radio jet can be a useful tool to examine the most embedded protostars. With the James Webb Space Telescope, we will be able to resolve emission from ions and atoms in the direct vicinity of the protostar, which will allow us to disentangle different processes. Future cm-wave facilities will provide even higher resolution and sensitivity to show if negative spectral indices toward jets of low-mass protostars are common phenomena.

\textit{Acknowledgements}: The authors thank the anonymous referee for comments that imporved the clarity of the paper. ŁT is supported by Leiden/ESA Astrophysics Program for Summer Students (LEAPS). Astrochemistry in Leiden is supported by the Netherlands Research School for Astronomy (NOVA), by a Royal Netherlands Academy of Arts and Sciences (KNAW) professor prize, and by the European Union A-ERC grant 291141 CHEMPLAN. AK acknowledges support from the Foundation for Polish Science (FNP) and the Polish National Science Center grants 2013/11/N/ST9/00400 and 2016/21/D/ST9/01098.
ŁT and AK acknowledge support from the HECOLS International Associated Laboratory, supported in part by the Polish NCN grant DEC-2013/08/M/ST9/00664.
ZYL is supported in part by NASA NNX 14AB38G, and NSF AST-1313083, AST-1716259. The National Radio Astronomy Observatory is a facility of the National Science Foundation operated under cooperative agreement by Associated Universities, Inc.
This research made use of: Astropy, a community-developed core Python package for Astronomy (Astropy Collaboration, 2013, http://astropy.org); APLpy, an open-source plotting package for Python hosted at http://aplpy.github.com; Matplotlib library \citep{Hunter2007}; and NASA's Astrophysics Data System.

\bibliography{apj.bib}

\begin{thebibliography}{90}
\expandafter\ifx\csname natexlab\endcsname\relax\def\natexlab#1{#1}\fi

\bibitem[{Aikawa {et~al.}(1997)Aikawa, Umebayashi, Nakano, \&
  Miyama}]{Aikawa1997}
Aikawa, Y., Umebayashi, T., Nakano, T., \& Miyama, S.~M. 1997, \apjl, 486, L51

\bibitem[{{Ainsworth} {et~al.}(2014){Ainsworth}, {Scaife}, {Ray}, {Taylor},
  {Green}, \& {Buckle}}]{Ainsworth2014}
{Ainsworth}, R.~E., {Scaife}, A.~M.~M., {Ray}, T.~P., {Taylor}, A.~M., {Green},
  D.~A., \& {Buckle}, J.~V. 2014, \apjl, 792, L18

\bibitem[{{AMI Consortium: Scaife} {et~al.}(2012{\natexlab{a}}){AMI Consortium:
  Scaife}, {Buckle}, {Ainsworth}, {Davies}, {Franzen}, {Grainge}, {Hobson},
  {Hurley-Walker}, {Lasenby}, {Olamaie}, {Perrott}, {Pooley}, {Ray}, {Richer},
  {Rodr{\'{\i}}guez-Gonz{\'a}lvez}, {Saunders}, {Schammel}, {Scott},
  {Shimwell}, {Titterington}, \& {Waldram}}]{Scaife2012}
{AMI Consortium: Scaife}, A.~M.~M., {Buckle}, J.~V., {Ainsworth}, R.~E.,
  {Davies}, M., {Franzen}, T.~M.~O., {Grainge}, K.~J.~B., {Hobson}, M.~P.,
  {Hurley-Walker}, N., {Lasenby}, A.~N., {Olamaie}, M., {Perrott}, Y.~C.,
  {Pooley}, G.~G., {Ray}, T.~P., {Richer}, J.~S.,
  {Rodr{\'{\i}}guez-Gonz{\'a}lvez}, C., {Saunders}, R.~D.~E., {Schammel},
  M.~P., {Scott}, P.~F., {Shimwell}, T., {Titterington}, D., \& {Waldram}, E.
  2012{\natexlab{a}}, Monthly Notices of the Royal Astronomical Society, 420,
  3334

\bibitem[{{AMI Consortium: Scaife} {et~al.}(2011{\natexlab{a}}){AMI Consortium:
  Scaife}, {Curtis}, {Davies}, {Franzen}, {Grainge}, {Hobson}, {Hurley-Walker},
  {Lasenby}, {Olamaie}, {Pooley}, {Rodr{\'{\i}}guez-Gonz{\'a}lvez}, {Saunders},
  {Schammel}, {Scott}, {Shimwell}, {Titterington}, {Waldram}, \&
  {Zwart}}]{Scaife2011a}
{AMI Consortium: Scaife}, A.~M.~M., {Curtis}, E.~I., {Davies}, M., {Franzen},
  T.~M.~O., {Grainge}, K.~J.~B., {Hobson}, M.~P., {Hurley-Walker}, N.,
  {Lasenby}, A.~N., {Olamaie}, M., {Pooley}, G.~G.,
  {Rodr{\'{\i}}guez-Gonz{\'a}lvez}, C., {Saunders}, R.~D.~E., {Schammel}, M.,
  {Scott}, P.~F., {Shimwell}, T., {Titterington}, D., {Waldram}, E., \&
  {Zwart}, J.~T.~L. 2011{\natexlab{a}}, Monthly Notices of the Royal
  Astronomical Society, 410, 2662

\bibitem[{{AMI Consortium: Scaife} {et~al.}(2011{\natexlab{b}}){AMI Consortium:
  Scaife}, {Hatchell}, {Davies}, {Franzen}, {Grainge}, {Hobson},
  {Hurley-Walker}, {Lasenby}, {Olamaie}, {Perrott}, {Pooley},
  {Rodr{\'{\i}}guez-Gonz{\'a}lvez}, {Saunders}, {Schammel}, {Scott},
  {Shimwell}, {Titterington}, \& {Waldram}}]{Scaife2011}
{AMI Consortium: Scaife}, A.~M.~M., {Hatchell}, J., {Davies}, M., {Franzen},
  T.~M.~O., {Grainge}, K.~J.~B., {Hobson}, M.~P., {Hurley-Walker}, N.,
  {Lasenby}, A.~N., {Olamaie}, M., {Perrott}, Y.~C., {Pooley}, G.~G.,
  {Rodr{\'{\i}}guez-Gonz{\'a}lvez}, C., {Saunders}, R.~D., {Schammel}, M.~P.,
  {Scott}, P.~F., {Shimwell}, T., {Titterington}, D., \& {Waldram}, E.
  2011{\natexlab{b}}, Monthly Notices of the Royal Astronomical Society, 415,
  893

\bibitem[{{AMI Consortium: Scaife} {et~al.}(2012{\natexlab{b}}){AMI Consortium:
  Scaife}, {Hatchell}, {Davies}, {Franzen}, {Grainge}, {Hobson},
  {Hurley-Walker}, {Lasenby}, {Olamaie}, {Perrott}, {Pooley},
  {Rodr{\'{\i}}guez-Gonz{\'a}lvez}, {Saunders}, {Schammel}, {Scott},
  {Shimwell}, {Titterington}, \& {Waldram}}]{Scaife2012a}
---. 2012{\natexlab{b}}, Monthly Notices of the Royal Astronomical Society,
  420, 1019

\bibitem[{{Anglada}(1995)}]{Anglada1995}
{Anglada}, G. 1995, in Revista Mexicana de Astronomia y Astrofisica Conference
  Series, Vol.~1, Revista Mexicana de Astronomia y Astrofisica Conference
  Series, ed. S.~{Lizano} \& J.~M. {Torrelles}, 67

\bibitem[{{Anglada} {et~al.}(1989){Anglada}, {Rodriguez}, {Torrelles},
  {Estalella}, {Ho}, {Canto}, {Lopez}, \& {Verdes-Montenegro}}]{Anglada1989}
{Anglada}, G., {Rodriguez}, L.~F., {Torrelles}, J.~M., {Estalella}, R., {Ho},
  P.~T.~P., {Canto}, J., {Lopez}, R., \& {Verdes-Montenegro}, L. 1989, \apj,
  341, 208

\bibitem[{{Anglada} {et~al.}(1998){Anglada}, {Villuendas}, {Estalella},
  {Beltr{\'a}n}, {Rodr{\'{\i}}guez}, {Torrelles}, \& {Curiel}}]{Anglada1998}
{Anglada}, G., {Villuendas}, E., {Estalella}, R., {Beltr{\'a}n}, M.~T.,
  {Rodr{\'{\i}}guez}, L.~F., {Torrelles}, J.~M., \& {Curiel}, S. 1998, The
  Astronomical Journal, 116, 2953

\bibitem[{Bachiller {et~al.}(1990)Bachiller, Martin-Pintado, Tafalla,
  Cernicharo, \& Lazareff}]{Bachiller1990}
Bachiller, R., Martin-Pintado, J., Tafalla, M., Cernicharo, J., \& Lazareff, B.
  1990, \aap, 231, 174

\bibitem[{{Bally} {et~al.}(1996){Bally}, {Devine}, \& {Reipurth}}]{Bally1996}
{Bally}, J., {Devine}, D., \& {Reipurth}, B. 1996, \apjl, 473, L49

\bibitem[{{Bontemps} {et~al.}(1996){Bontemps}, {Andre}, {Terebey}, \&
  {Cabrit}}]{Bontemps1996}
{Bontemps}, S., {Andre}, P., {Terebey}, S., \& {Cabrit}, S. 1996, \aap, 311,
  858

\bibitem[{{Cabrit} \& {Bertout}(1992)}]{Cabrit1992}
{Cabrit}, S. \& {Bertout}, C. 1992, \aap, 261, 274

\bibitem[{{Caratti o Garatti} {et~al.}(2012){Caratti o Garatti}, {Garcia
  Lopez}, {Antoniucci}, {Nisini}, {Giannini}, {Eisl{\"o}ffel}, {Ray},
  {Lorenzetti}, \& {Cabrit}}]{CarattioGaratti2012}
{Caratti o Garatti}, A., {Garcia Lopez}, R., {Antoniucci}, S., {Nisini}, B.,
  {Giannini}, T., {Eisl{\"o}ffel}, J., {Ray}, T.~P., {Lorenzetti}, D., \&
  {Cabrit}, S. 2012, Astronomy and Astophysics, 538, A64

\bibitem[{{Carrasco-Gonz{\'a}lez} {et~al.}(2008){Carrasco-Gonz{\'a}lez},
  {Anglada}, {Rodr{\'{\i}}guez}, {Torrelles}, \&
  {Osorio}}]{Carrasco-Gonzalez2008}
{Carrasco-Gonz{\'a}lez}, C., {Anglada}, G., {Rodr{\'{\i}}guez}, L.~F.,
  {Torrelles}, J.~M., \& {Osorio}, M. 2008, \apj, 136, 2238

\bibitem[{{Carrasco-Gonz{\'a}lez} {et~al.}(2010){Carrasco-Gonz{\'a}lez},
  {Rodr{\'{\i}}guez}, {Anglada}, {Mart{\'{\i}}}, {Torrelles}, \&
  {Osorio}}]{Carrasco-Gonzalez2010}
{Carrasco-Gonz{\'a}lez}, C., {Rodr{\'{\i}}guez}, L.~F., {Anglada}, G.,
  {Mart{\'{\i}}}, J., {Torrelles}, J.~M., \& {Osorio}, M. 2010, Science, 330,
  1209

\bibitem[{{Chen} {et~al.}(1995){Chen}, {Myers}, {Ladd}, \& {Wood}}]{Chen1995}
{Chen}, H., {Myers}, P.~C., {Ladd}, E.~F., \& {Wood}, D.~O.~S. 1995, \apj, 445,
  377

\bibitem[{{Chiang} {et~al.}(2012){Chiang}, {Looney}, \& {Tobin}}]{Chiang2012}
{Chiang}, H.-F., {Looney}, L.~W., \& {Tobin}, J.~J. 2012, \apj, 756, 168

\bibitem[{{Cleeves} {et~al.}(2013){Cleeves}, {Adams}, \&
  {Bergin}}]{Cleeves2013}
{Cleeves}, L.~I., {Adams}, F.~C., \& {Bergin}, E.~A. 2013, \apj, 772, 5

\bibitem[{Condon {et~al.}(1998)Condon, Cotton, Greisen, Yin, Perley, Taylor, \&
  Broderick}]{Condon1998}
Condon, J.~J., Cotton, W.~D., Greisen, E.~W., Yin, Q.~F., Perley, R.~A.,
  Taylor, G.~B., \& Broderick, J.~J. 1998, \aj, 115, 1693

\bibitem[{{Cunningham} {et~al.}(2009){Cunningham}, {Moeckel}, \&
  {Bally}}]{Cunningham2009}
{Cunningham}, N.~J., {Moeckel}, N., \& {Bally}, J. 2009, \apj, 692, 943

\bibitem[{{Curiel} {et~al.}(1990){Curiel}, {Raymond}, {Moran}, {Rodriguez}, \&
  {Canto}}]{Curiel1990}
{Curiel}, S., {Raymond}, J.~C., {Moran}, J.~M., {Rodriguez}, L.~F., \& {Canto},
  J. 1990, \apjl, 365, L85

\bibitem[{{Curiel} {et~al.}(1993){Curiel}, {Rodriguez}, {Moran}, \&
  {Canto}}]{Curiel1993}
{Curiel}, S., {Rodriguez}, L.~F., {Moran}, J.~M., \& {Canto}, J. 1993, \apj,
  415, 191

\bibitem[{{Davis} {et~al.}(2008){Davis}, {Scholz}, {Lucas}, {Smith}, \&
  {Adamson}}]{Davis2008}
{Davis}, C.~J., {Scholz}, P., {Lucas}, P., {Smith}, M.~D., \& {Adamson}, A.
  2008, Monthly Notices of the Royal Astronomical Society, 387, 954

\bibitem[{Drury(1991)}]{Drury1991}
Drury, L.~O. 1991, \mnras, 251, 340

\bibitem[{{Dzib} {et~al.}(2013){Dzib}, {Loinard}, {Mioduszewski},
  {Rodr{\'{\i}}guez}, {Ortiz-Le{\'o}n}, {Pech}, {Rivera}, {Torres}, {Boden},
  {Hartmann}, {Evans}, {Brice{\~n}o}, \& {Tobin}}]{Dzib2013}
{Dzib}, S.~A., {Loinard}, L., {Mioduszewski}, A.~J., {Rodr{\'{\i}}guez}, L.~F.,
  {Ortiz-Le{\'o}n}, G.~N., {Pech}, G., {Rivera}, J.~L., {Torres}, R.~M.,
  {Boden}, A.~F., {Hartmann}, L., {Evans}, II, N.~J., {Brice{\~n}o}, C., \&
  {Tobin}, J. 2013, \apj, 775, 63

\bibitem[{{Dzib} {et~al.}(2015){Dzib}, {Loinard}, {Rodr{\'{\i}}guez},
  {Mioduszewski}, {Ortiz-Le{\'o}n}, {Kounkel}, {Pech}, {Rivera}, {Torres},
  {Boden}, {Hartmann}, {Evans}, {Brice{\~n}o}, \& {Tobin}}]{Dzib2015}
{Dzib}, S.~A., {Loinard}, L., {Rodr{\'{\i}}guez}, L.~F., {Mioduszewski}, A.~J.,
  {Ortiz-Le{\'o}n}, G.~N., {Kounkel}, M.~A., {Pech}, G., {Rivera}, J.~L.,
  {Torres}, R.~M., {Boden}, A.~F., {Hartmann}, L., {Evans}, II, N.~J.,
  {Brice{\~n}o}, C., \& {Tobin}, J. 2015, \apj, 801, 91

\bibitem[{Eistrup {et~al.}(2016)Eistrup, Walsh, \& Van~Dishoeck}]{Eistrup2016}
Eistrup, C., Walsh, C., \& Van~Dishoeck, E.~F. 2016, Astronomy \& Astrophysics,
  595, A83

\bibitem[{{Enoch} {et~al.}(2009){Enoch}, {Evans}, {Sargent}, \&
  {Glenn}}]{Enoch2009}
{Enoch}, M.~L., {Evans}, II, N.~J., {Sargent}, A.~I., \& {Glenn}, J. 2009,
  \apj, 692, 973

\bibitem[{{Frank} {et~al.}(2014){Frank}, {Ray}, {Cabrit}, {Hartigan}, {Arce},
  {Bacciotti}, {Bally}, {Benisty}, {Eisl{\"o}ffel}, {G{\"u}del}, {Lebedev},
  {Nisini}, \& {Raga}}]{Frank2014}
{Frank}, A., {Ray}, T.~P., {Cabrit}, S., {Hartigan}, P., {Arce}, H.~G.,
  {Bacciotti}, F., {Bally}, J., {Benisty}, M., {Eisl{\"o}ffel}, J.,
  {G{\"u}del}, M., {Lebedev}, S., {Nisini}, B., \& {Raga}, A. 2014, Protostars
  and Planets VI, 451

\bibitem[{{Galv{\'a}n-Madrid} {et~al.}(2015){Galv{\'a}n-Madrid},
  {Rodr{\'{\i}}guez}, {Liu}, {Costigan}, {Palau}, {Zapata}, \&
  {Loinard}}]{galvan-madrid2015}
{Galv{\'a}n-Madrid}, R., {Rodr{\'{\i}}guez}, L.~F., {Liu}, H.~B., {Costigan},
  G., {Palau}, A., {Zapata}, L.~A., \& {Loinard}, L. 2015, \apjl, 806, L32

\bibitem[{Girart {et~al.}(2002)Girart, Curiel, Rodr{\'{\i}}guez, \&
  Cant{\'o}}]{Girart2002}
Girart, J.~M., Curiel, S., Rodr{\'{\i}}guez, L.~F., \& Cant{\'o}, J. 2002,
  Revista Mexicana de Astronomia y Astrofisica, 38, 169

\bibitem[{{G{\"u}del} {et~al.}(2008){G{\"u}del}, {Skinner}, {Audard}, {Briggs},
  \& {Cabrit}}]{Guedel2008}
{G{\"u}del}, M., {Skinner}, S.~L., {Audard}, M., {Briggs}, K.~R., \& {Cabrit},
  S. 2008, Astronomy and Astophysics, 478, 797

\bibitem[{Gueth \& Guilloteau(1999)}]{Gueth1999}
Gueth, F. \& Guilloteau, S. 1999, \aap, 343, 571

\bibitem[{{Gueth} {et~al.}(1996){Gueth}, {Guilloteau}, \&
  {Bachiller}}]{Gueth1996}
{Gueth}, F., {Guilloteau}, S., \& {Bachiller}, R. 1996, \aap, 307, 891

\bibitem[{{Hirota} {et~al.}(2008){Hirota}, {Bushimata}, {Choi}, {Honma},
  {Imai}, {Iwadate}, {Jike}, {Kameya}, {Kamohara}, {Kan-Ya}, {Kawaguchi},
  {Kijima}, {Kobayashi}, {Kuji}, {Kurayama}, {Manabe}, {Miyaji}, {Nagayama},
  {Nakagawa}, {Oh}, {Omodaka}, {Oyama}, {Sakai}, {Sasao}, {Sato}, {Shibata},
  {Tamura}, \& {Yamashita}}]{Hirota2008}
{Hirota}, T., {Bushimata}, T., {Choi}, Y.~K., {Honma}, M., {Imai}, H.,
  {Iwadate}, K., {Jike}, T., {Kameya}, O., {Kamohara}, R., {Kan-Ya}, Y.,
  {Kawaguchi}, N., {Kijima}, M., {Kobayashi}, H., {Kuji}, S., {Kurayama}, T.,
  {Manabe}, S., {Miyaji}, T., {Nagayama}, T., {Nakagawa}, A., {Oh}, C.~S.,
  {Omodaka}, T., {Oyama}, T., {Sakai}, S., {Sasao}, T., {Sato}, K., {Shibata},
  K.~M., {Tamura}, Y., \& {Yamashita}, K. 2008, Publications of the
  Astronomical Society of Japan, 60, 37

\bibitem[{{Hull} {et~al.}(2016){Hull}, {Girart}, {Kristensen}, {Dunham},
  {Rodr{\'{\i}}guez-Kamenetzky}, {Carrasco-Gonz{\'a}lez}, {Cort{\'e}s}, {Li},
  \& {Plambeck}}]{hull2016}
{Hull}, C.~L.~H., {Girart}, J.~M., {Kristensen}, L.~E., {Dunham}, M.~M.,
  {Rodr{\'{\i}}guez-Kamenetzky}, A., {Carrasco-Gonz{\'a}lez}, C., {Cort{\'e}s},
  P.~C., {Li}, Z.-Y., \& {Plambeck}, R.~L. 2016, \apjl, 823, L27

\bibitem[{Hunter(2007)}]{Hunter2007}
Hunter, J.~D. 2007, Computing In Science \& Engineering, 9, 90

\bibitem[{{J{\o}rgensen} {et~al.}(2007){J{\o}rgensen}, {Johnstone}, {Kirk}, \&
  {Myers}}]{Jorgensen2007}
{J{\o}rgensen}, J.~K., {Johnstone}, D., {Kirk}, H., \& {Myers}, P.~C. 2007,
  \apj, 656, 293

\bibitem[{{Karska} {et~al.}(2014){Karska}, {Kristensen}, {van Dishoeck},
  {Drozdovskaya}, {Mottram}, {Herczeg}, {Bruderer}, {Cabrit}, {Evans},
  {Fedele}, {Gusdorf}, {J{\o}rgensen}, {Kaufman}, {Melnick}, {Neufeld},
  {Nisini}, {Santangelo}, {Tafalla}, \& {Wampfler}}]{Karska2014a}
{Karska}, A., {Kristensen}, L.~E., {van Dishoeck}, E.~F., {Drozdovskaya},
  M.~N., {Mottram}, J.~C., {Herczeg}, G.~J., {Bruderer}, S., {Cabrit}, S.,
  {Evans}, N.~J., {Fedele}, D., {Gusdorf}, A., {J{\o}rgensen}, J.~K.,
  {Kaufman}, M.~J., {Melnick}, G.~J., {Neufeld}, D.~A., {Nisini}, B.,
  {Santangelo}, G., {Tafalla}, M., \& {Wampfler}, S.~F. 2014, \aap, 572, A9

\bibitem[{{Kounkel} {et~al.}(2014){Kounkel}, {Hartmann}, {Loinard},
  {Mioduszewski}, {Dzib}, {Ortiz-Le{\'o}n}, {Rodr{\'{\i}}guez}, {Pech},
  {Rivera}, {Torres}, {Boden}, {Evans}, {Brice{\~n}o}, \&
  {Tobin}}]{Kounkel2014}
{Kounkel}, M., {Hartmann}, L., {Loinard}, L., {Mioduszewski}, A.~J., {Dzib},
  S.~A., {Ortiz-Le{\'o}n}, G.~N., {Rodr{\'{\i}}guez}, L.~F., {Pech}, G.,
  {Rivera}, J.~L., {Torres}, R.~M., {Boden}, A.~F., {Evans}, II, N.~J.,
  {Brice{\~n}o}, C., \& {Tobin}, J. 2014, \apj, 790, 49

\bibitem[{{Kristensen} {et~al.}(2012){Kristensen}, {van Dishoeck}, {Bergin},
  {Visser}, {Y{\i}ld{\i}z}, {San Jose-Garcia}, {J{\o}rgensen}, {Herczeg},
  {Johnstone}, {Wampfler}, {Benz}, {Bruderer}, {Cabrit}, {Caselli}, {Doty},
  {Harsono}, {Herpin}, {Hogerheijde}, {Karska}, {van Kempen}, {Liseau},
  {Nisini}, {Tafalla}, {van der Tak}, \& {Wyrowski}}]{Kristensen2012}
{Kristensen}, L.~E., {van Dishoeck}, E.~F., {Bergin}, E.~A., {Visser}, R.,
  {Y{\i}ld{\i}z}, U.~A., {San Jose-Garcia}, I., {J{\o}rgensen}, J.~K.,
  {Herczeg}, G.~J., {Johnstone}, D., {Wampfler}, S.~F., {Benz}, A.~O.,
  {Bruderer}, S., {Cabrit}, S., {Caselli}, P., {Doty}, S.~D., {Harsono}, D.,
  {Herpin}, F., {Hogerheijde}, M.~R., {Karska}, A., {van Kempen}, T.~A.,
  {Liseau}, R., {Nisini}, B., {Tafalla}, M., {van der Tak}, F., \& {Wyrowski},
  F. 2012, \aap, 542, A8

\bibitem[{{Kwon} {et~al.}(2006){Kwon}, {Looney}, {Crutcher}, \&
  {Kirk}}]{Kwon2006}
{Kwon}, W., {Looney}, L.~W., {Crutcher}, R.~M., \& {Kirk}, J.~M. 2006,
  Astrophysical Journal, 653, 1358

\bibitem[{{Lahuis} {et~al.}(2010){Lahuis}, {van Dishoeck}, {J{\o}rgensen},
  {Blake}, \& {Evans}}]{Lahuis2010}
{Lahuis}, F., {van Dishoeck}, E.~F., {J{\o}rgensen}, J.~K., {Blake}, G.~A., \&
  {Evans}, N.~J. 2010, \aap, 519, A3

\bibitem[{{Lee} {et~al.}(2016){Lee}, {Dunham}, {Myers}, {Arce}, {Bourke},
  {Goodman}, {J{\o}rgensen}, {Kristensen}, {Offner}, {Pineda}, {Tobin}, \&
  {Vorobyov}}]{Lee2016}
{Lee}, K.~I., {Dunham}, M.~M., {Myers}, P.~C., {Arce}, H.~G., {Bourke}, T.~L.,
  {Goodman}, A.~A., {J{\o}rgensen}, J.~K., {Kristensen}, L.~E., {Offner},
  S.~S.~R., {Pineda}, J.~E., {Tobin}, J.~J., \& {Vorobyov}, E.~I. 2016, \apjl,
  820, L2

\bibitem[{{Lee} {et~al.}(2015){Lee}, {Dunham}, {Myers}, {Tobin}, {Kristensen},
  {Pineda}, {Vorobyov}, {Offner}, {Arce}, {Li}, {Bourke}, {J{\o}rgensen},
  {Goodman}, {Sadavoy}, {Chandler}, {Harris}, {Kratter}, {Looney}, {Melis},
  {Perez}, \& {Segura-Cox}}]{Lee2015}
{Lee}, K.~I., {Dunham}, M.~M., {Myers}, P.~C., {Tobin}, J.~J., {Kristensen},
  L.~E., {Pineda}, J.~E., {Vorobyov}, E.~I., {Offner}, S.~S.~R., {Arce}, H.~G.,
  {Li}, Z.-Y., {Bourke}, T.~L., {J{\o}rgensen}, J.~K., {Goodman}, A.~A.,
  {Sadavoy}, S.~I., {Chandler}, C.~J., {Harris}, R.~J., {Kratter}, K.,
  {Looney}, L.~W., {Melis}, C., {Perez}, L.~M., \& {Segura-Cox}, D. 2015,
  Astrophysical Journal, 814, 114

\bibitem[{{Lee} {et~al.}(2014){Lee}, {Fern{\'a}ndez-L{\'o}pez}, {Storm},
  {Looney}, {Mundy}, {Segura-Cox}, {Teuben}, {Rosolowsky}, {Arce}, {Ostriker},
  {Shirley}, {Kwon}, {Kauffmann}, {Tobin}, {Plunkett}, {Pound}, {Salter},
  {Volgenau}, {Chen}, {Tassis}, {Isella}, {Crutcher}, {Gammie}, \&
  {Testi}}]{Lee2014}
{Lee}, K.~I., {Fern{\'a}ndez-L{\'o}pez}, M., {Storm}, S., {Looney}, L.~W.,
  {Mundy}, L.~G., {Segura-Cox}, D., {Teuben}, P., {Rosolowsky}, E., {Arce},
  H.~G., {Ostriker}, E.~C., {Shirley}, Y.~L., {Kwon}, W., {Kauffmann}, J.,
  {Tobin}, J.~J., {Plunkett}, A.~L., {Pound}, M.~W., {Salter}, D.~M.,
  {Volgenau}, N.~H., {Chen}, C.-Y., {Tassis}, K., {Isella}, A., {Crutcher},
  R.~M., {Gammie}, C.~F., \& {Testi}, L. 2014, \apj, 797, 76

\bibitem[{{Looney} {et~al.}(2000){Looney}, {Mundy}, \& {Welch}}]{Looney2000}
{Looney}, L.~W., {Mundy}, L.~G., \& {Welch}, W.~J. 2000, \apj, 529, 477

\bibitem[{{Looney} {et~al.}(2007){Looney}, {Tobin}, \& {Kwon}}]{Looney2007}
{Looney}, L.~W., {Tobin}, J.~J., \& {Kwon}, W. 2007, \apjl, 670, L131

\bibitem[{{Machida}(2014)}]{Machida2014}
{Machida}, M.~N. 2014, \apjl, 796, L17

\bibitem[{{Marti} {et~al.}(1993){Marti}, {Rodriguez}, \&
  {Reipurth}}]{Marti1993}
{Marti}, J., {Rodriguez}, L.~F., \& {Reipurth}, B. 1993, \apj, 416, 208

\bibitem[{{Marti} {et~al.}(1995){Marti}, {Rodriguez}, \&
  {Reipurth}}]{Marti1995}
---. 1995, \apj, 449, 184

\bibitem[{{McMullin} {et~al.}(2007){McMullin}, {Waters}, {Schiebel}, {Young},
  \& {Golap}}]{McMullin2007}
{McMullin}, J.~P., {Waters}, B., {Schiebel}, D., {Young}, W., \& {Golap}, K.
  2007, in Astronomical Society of the Pacific Conference Series, Vol. 376,
  Astronomical Data Analysis Software and Systems XVI, ed. R.~A. {Shaw},
  F.~{Hill}, \& D.~J. {Bell}, 127

\bibitem[{{Nisini} {et~al.}(2002){Nisini}, {Giannini}, \&
  {Lorenzetti}}]{Nisini2002}
{Nisini}, B., {Giannini}, T., \& {Lorenzetti}, D. 2002, \apj, 574, 246

\bibitem[{{Nisini} {et~al.}(2015){Nisini}, {Santangelo}, {Giannini},
  {Antoniucci}, {Cabrit}, {Codella}, {Davis}, {Eisl{\"o}ffel}, {Kristensen},
  {Herczeg}, {Neufeld}, \& {van Dishoeck}}]{Nisini2015}
{Nisini}, B., {Santangelo}, G., {Giannini}, T., {Antoniucci}, S., {Cabrit}, S.,
  {Codella}, C., {Davis}, C.~J., {Eisl{\"o}ffel}, J., {Kristensen}, L.,
  {Herczeg}, G., {Neufeld}, D., \& {van Dishoeck}, E.~F. 2015, \apj, 801, 121

\bibitem[{{Ortiz-Le{\'o}n} {et~al.}(2015){Ortiz-Le{\'o}n}, {Loinard},
  {Mioduszewski}, {Dzib}, {Rodr{\'{\i}}guez}, {Pech}, {Rivera}, {Torres},
  {Boden}, {Hartmann}, {Evans}, {Brice{\~n}o}, {Tobin}, {Kounkel}, \&
  {Gonz{\'a}lez-L{\'o}pezlira}}]{Ortiz-Leon2015}
{Ortiz-Le{\'o}n}, G.~N., {Loinard}, L., {Mioduszewski}, A.~J., {Dzib}, S.~A.,
  {Rodr{\'{\i}}guez}, L.~F., {Pech}, G., {Rivera}, J.~L., {Torres}, R.~M.,
  {Boden}, A.~F., {Hartmann}, L., {Evans}, II, N.~J., {Brice{\~n}o}, C.,
  {Tobin}, J., {Kounkel}, M.~A., \& {Gonz{\'a}lez-L{\'o}pezlira}, R.~A. 2015,
  \apj, 805, 9

\bibitem[{{Padovani} {et~al.}(2014){Padovani}, {Galli}, {Hennebelle},
  {Commer{\c c}on}, \& {Joos}}]{Padovani2014}
{Padovani}, M., {Galli}, D., {Hennebelle}, P., {Commer{\c c}on}, B., \& {Joos},
  M. 2014, \aap, 571, A33

\bibitem[{{Padovani} {et~al.}(2013){Padovani}, {Hennebelle}, \&
  {Galli}}]{Padovani2013}
{Padovani}, M., {Hennebelle}, P., \& {Galli}, D. 2013, Astronomy and
  Astophysics, 560, A114

\bibitem[{{Padovani} {et~al.}(2016){Padovani}, {Marcowith}, {Hennebelle}, \&
  {Ferri{\`e}re}}]{Padovani2016}
{Padovani}, M., {Marcowith}, A., {Hennebelle}, P., \& {Ferri{\`e}re}, K. 2016,
  \aap, 590, A8

\bibitem[{{Panagia} \& {Felli}(1975)}]{Panagia1975}
{Panagia}, N. \& {Felli}, M. 1975, \aap, 39, 1

\bibitem[{{Pech} {et~al.}(2016){Pech}, {Loinard}, {Dzib}, {Mioduszewski},
  {Rodr{\'{\i}}guez}, {Ortiz-Le{\'o}n}, {Rivera}, {Torres}, {Boden}, {Hartman},
  {Kounkel}, {Evans}, {Brice{\~n}o}, {Tobin}, \& {Zapata}}]{Pech2016}
{Pech}, G., {Loinard}, L., {Dzib}, S.~A., {Mioduszewski}, A.~J.,
  {Rodr{\'{\i}}guez}, L.~F., {Ortiz-Le{\'o}n}, G.~N., {Rivera}, J.~L.,
  {Torres}, R.~M., {Boden}, A.~F., {Hartman}, L., {Kounkel}, M.~A., {Evans},
  II, N.~J., {Brice{\~n}o}, C., {Tobin}, J., \& {Zapata}, L.~A. 2016, \apj,
  818, 116

\bibitem[{Perley \& Butler(2017)}]{Perley2017}
Perley, R.~A. \& Butler, B.~J. 2017, \apjs, 230, 7

\bibitem[{{Plunkett} {et~al.}(2013){Plunkett}, {Arce}, {Corder}, {Mardones},
  {Sargent}, \& {Schnee}}]{Plunkett2013}
{Plunkett}, A.~L., {Arce}, H.~G., {Corder}, S.~A., {Mardones}, D., {Sargent},
  A.~I., \& {Schnee}, S.~L. 2013, \apj, 774, 22

\bibitem[{{Pravdo} {et~al.}(2004){Pravdo}, {Tsuboi}, \& {Maeda}}]{Pravdo2004}
{Pravdo}, S.~H., {Tsuboi}, Y., \& {Maeda}, Y. 2004, \apj, 605, 259

\bibitem[{{Preibisch}(1997)}]{Preibisch1997}
{Preibisch}, T. 1997, \aap, 324, 690

\bibitem[{{Raga} {et~al.}(2013){Raga}, {Noriega-Crespo}, {Carey}, \&
  {Arce}}]{Raga2013}
{Raga}, A.~C., {Noriega-Crespo}, A., {Carey}, S.~J., \& {Arce}, H.~G. 2013,
  Astronomical Journal, 145, 28

\bibitem[{{Reipurth} \& {Cernicharo}(1995)}]{Reipurth1995}
{Reipurth}, B. \& {Cernicharo}, J. 1995, in Revista Mexicana de Astronomia y
  Astrofisica Conference Series, Vol.~1, Revista Mexicana de Astronomia y
  Astrofisica Conference Series, ed. S.~{Lizano} \& J.~M. {Torrelles}, 43

\bibitem[{{Reipurth} {et~al.}(2002){Reipurth}, {Rodr{\'{\i}}guez}, {Anglada},
  \& {Bally}}]{Reipurth2002}
{Reipurth}, B., {Rodr{\'{\i}}guez}, L.~F., {Anglada}, G., \& {Bally}, J. 2002,
  Astrophysical Journal, 124, 1045

\bibitem[{{Reipurth} {et~al.}(2004){Reipurth}, {Rodr{\'{\i}}guez}, {Anglada},
  \& {Bally}}]{Reipurth2004}
---. 2004, The Astronomical Journal, 127, 1736

\bibitem[{{Reynolds}(1986)}]{Reynolds1986}
{Reynolds}, S.~P. 1986, \apj, 304, 713

\bibitem[{Rodgers-Lee {et~al.}(2017)Rodgers-Lee, Taylor, Ray, \&
  Downes}]{Rodgers-Lee2017}
Rodgers-Lee, D., Taylor, A.~M., Ray, T.~P., \& Downes, T.~P. 2017, \mnras, 472,
  26

\bibitem[{{Rodr{\'{\i}}guez}(1994)}]{Rodriguez1994a}
{Rodr{\'{\i}}guez}, L.~F. 1994, Revista Mexicana de Astronomia y Astrofisica,
  29, 69

\bibitem[{{Rodr{\'{\i}}guez} {et~al.}(1997){Rodr{\'{\i}}guez}, {Anglada}, \&
  {Curiel}}]{Rodriguez1997}
{Rodr{\'{\i}}guez}, L.~F., {Anglada}, G., \& {Curiel}, S. 1997, \apjl, 480,
  L125

\bibitem[{{Rodr{\'{\i}}guez} {et~al.}(1999){Rodr{\'{\i}}guez}, {Anglada}, \&
  {Curiel}}]{Rodriguez1999}
---. 1999, \apj Supplement Series, 125, 427

\bibitem[{{Rodr{\'{\i}}guez} {et~al.}(1986){Rodr{\'{\i}}guez}, {Canto},
  {Torrelles}, \& {Ho}}]{Rodriguez1986}
{Rodr{\'{\i}}guez}, L.~F., {Canto}, J., {Torrelles}, J.~M., \& {Ho}, P.~T.~P.
  1986, \apjl, 301, L25

\bibitem[{{Rodr{\'{\i}}guez} {et~al.}(1989){Rodr{\'{\i}}guez}, {Curiel},
  {Moran}, {Mirabel}, {Roth}, \& {Garay}}]{Rodriguez1989b}
{Rodr{\'{\i}}guez}, L.~F., {Curiel}, S., {Moran}, J.~M., {Mirabel}, I.~F.,
  {Roth}, M., \& {Garay}, G. 1989, \apjl, 346, L85

\bibitem[{{Rodr{\'{\i}}guez} {et~al.}(2000){Rodr{\'{\i}}guez},
  {Delgado-Arellano}, {G{\'o}mez}, {Reipurth}, {Torrelles}, {Noriega-Crespo},
  {Raga}, \& {Cant{\'o}}}]{Rodriguez2000}
{Rodr{\'{\i}}guez}, L.~F., {Delgado-Arellano}, V.~G., {G{\'o}mez}, Y.,
  {Reipurth}, B., {Torrelles}, J.~M., {Noriega-Crespo}, A., {Raga}, A.~C., \&
  {Cant{\'o}}, J. 2000, \aj, 119, 882

\bibitem[{{Rodr{\'{\i}}guez} {et~al.}(1993){Rodr{\'{\i}}guez}, {Marti},
  {Canto}, {Moran}, \& {Curiel}}]{Rodriguez1993}
{Rodr{\'{\i}}guez}, L.~F., {Marti}, J., {Canto}, J., {Moran}, J.~M., \&
  {Curiel}, S. 1993, Revista Mexicana de Astronomia y Astrofisica, 25, 23

\bibitem[{Rodr{\'{\i}}guez-Kamenetzky
  {et~al.}(2016)Rodr{\'{\i}}guez-Kamenetzky, Carrasco-Gonz{\'a}lez, Araudo,
  Torrelles, Anglada, Mart{\'{\i}}, Rodr{\'{\i}}guez, \&
  Valotto}]{Rodriguez-Kamenetzky2016}
Rodr{\'{\i}}guez-Kamenetzky, A., Carrasco-Gonz{\'a}lez, C., Araudo, A.,
  Torrelles, J.~M., Anglada, G., Mart{\'{\i}}, J., Rodr{\'{\i}}guez, L.~F., \&
  Valotto, C. 2016, \apj, 818, 27

\bibitem[{{Sadavoy} {et~al.}(2014){Sadavoy}, {Di Francesco}, {Andr{\'e}},
  {Pezzuto}, {Bernard}, {Maury}, {Men'shchikov}, {Motte}, {NguyLuong},
  {Schneider}, {Arzoumanian}, {Benedettini}, {Bontemps}, {Elia}, {Hennemann},
  {Hill}, {K{\"o}nyves}, {Louvet}, {Peretto}, {Roy}, \& {White}}]{sadavoy2014}
{Sadavoy}, S.~I., {Di Francesco}, J., {Andr{\'e}}, P., {Pezzuto}, S.,
  {Bernard}, J.-P., {Maury}, A., {Men'shchikov}, A., {Motte}, F., {NguyLuong},
  Q., {Schneider}, N., {Arzoumanian}, D., {Benedettini}, M., {Bontemps}, S.,
  {Elia}, D., {Hennemann}, M., {Hill}, T., {K{\"o}nyves}, V., {Louvet}, F.,
  {Peretto}, N., {Roy}, A., \& {White}, G.~J. 2014, \apjl, 787, L18

\bibitem[{{Shirley} {et~al.}(2007){Shirley}, {Claussen}, {Bourke}, {Young}, \&
  {Blake}}]{Shirley2007}
{Shirley}, Y.~L., {Claussen}, M.~J., {Bourke}, T.~L., {Young}, C.~H., \&
  {Blake}, G.~A. 2007, \apj, 667, 329

\bibitem[{Snell \& Bally(1986)}]{Snell1986}
Snell, R.~L. \& Bally, J. 1986, \apj, 303, 683

\bibitem[{{Storm} {et~al.}(2014){Storm}, {Mundy}, {Fern{\'a}ndez-L{\'o}pez},
  {Lee}, {Looney}, {Teuben}, {Rosolowsky}, {Arce}, {Ostriker}, {Segura-Cox},
  {Pound}, {Salter}, {Volgenau}, {Shirley}, {Chen}, {Gong}, {Plunkett},
  {Tobin}, {Kwon}, {Isella}, {Kauffmann}, {Tassis}, {Crutcher}, {Gammie}, \&
  {Testi}}]{Storm2014}
{Storm}, S., {Mundy}, L.~G., {Fern{\'a}ndez-L{\'o}pez}, M., {Lee}, K.~I.,
  {Looney}, L.~W., {Teuben}, P., {Rosolowsky}, E., {Arce}, H.~G., {Ostriker},
  E.~C., {Segura-Cox}, D.~M., {Pound}, M.~W., {Salter}, D.~M., {Volgenau},
  N.~H., {Shirley}, Y.~L., {Chen}, C.-Y., {Gong}, H., {Plunkett}, A.~L.,
  {Tobin}, J.~J., {Kwon}, W., {Isella}, A., {Kauffmann}, J., {Tassis}, K.,
  {Crutcher}, R.~M., {Gammie}, C.~F., \& {Testi}, L. 2014, \apj, 794, 165

\bibitem[{{Tobin} {et~al.}(2015){Tobin}, {Dunham}, {Looney}, {Li}, {Chandler},
  {Segura-Cox}, {Sadavoy}, {Melis}, {Harris}, {Perez}, {Kratter},
  {J{\o}rgensen}, {Plunkett}, \& {Hull}}]{Tobin2015a}
{Tobin}, J.~J., {Dunham}, M.~M., {Looney}, L.~W., {Li}, Z.-Y., {Chandler},
  C.~J., {Segura-Cox}, D., {Sadavoy}, S.~I., {Melis}, C., {Harris}, R.~J.,
  {Perez}, L.~M., {Kratter}, K., {J{\o}rgensen}, J.~K., {Plunkett}, A.~L., \&
  {Hull}, C.~L.~H. 2015, \apj, 798, 61

\bibitem[{{Tobin} {et~al.}(2016{\natexlab{a}}){Tobin}, {Kratter}, {Persson},
  {Looney}, {Dunham}, {Segura-Cox}, {Li}, {Chandler}, {Sadavoy}, {Harris},
  {Melis}, \& {P{\'e}rez}}]{Tobin2016a}
{Tobin}, J.~J., {Kratter}, K.~M., {Persson}, M.~V., {Looney}, L.~W., {Dunham},
  M.~M., {Segura-Cox}, D., {Li}, Z.-Y., {Chandler}, C.~J., {Sadavoy}, S.~I.,
  {Harris}, R.~J., {Melis}, C., \& {P{\'e}rez}, L.~M. 2016{\natexlab{a}},
  Nature, 538, 483

\bibitem[{{Tobin} {et~al.}(2016{\natexlab{b}}){Tobin}, {Looney}, {Li},
  {Chandler}, {Dunham}, {Segura-Cox}, {Sadavoy}, {Melis}, {Harris}, {Kratter},
  \& {Perez}}]{Tobin2016}
{Tobin}, J.~J., {Looney}, L.~W., {Li}, Z.-Y., {Chandler}, C.~J., {Dunham},
  M.~M., {Segura-Cox}, D., {Sadavoy}, S.~I., {Melis}, C., {Harris}, R.~J.,
  {Kratter}, K., \& {Perez}, L. 2016{\natexlab{b}}, \apj, 818, 73

\bibitem[{{Tsujimoto} {et~al.}(2005){Tsujimoto}, {Kobayashi}, \&
  {Tsuboi}}]{Tsujimoto2005}
{Tsujimoto}, M., {Kobayashi}, N., \& {Tsuboi}, Y. 2005, The Astronomical
  Journal, 130, 2212

\bibitem[{Whitney {et~al.}(2003)Whitney, Wood, Bjorkman, \&
  Wolff}]{Whitney2003}
Whitney, B.~A., Wood, K., Bjorkman, J., \& Wolff, M.~J. 2003, \apj, 591, 1049

\bibitem[{Wilner {et~al.}(2005)Wilner, D'Alessio, Calvet, Claussen, \&
  Hartmann}]{Wilner2005}
Wilner, D.~J., D'Alessio, P., Calvet, N., Claussen, M.~J., \& Hartmann, L.
  2005, \apjl, 626, L109

\bibitem[{{Wu} {et~al.}(2004){Wu}, {Wei}, {Zhao}, {Shi}, {Yu}, {Qin}, \&
  {Huang}}]{Wu2004}
{Wu}, Y., {Wei}, Y., {Zhao}, M., {Shi}, Y., {Yu}, W., {Qin}, S., \& {Huang}, M.
  2004, Astronomy and Astophysics, 426, 503

\end{thebibliography}

\begin{deluxetable}{lllllllll}
\tabletypesize{\scriptsize}
\tablecaption{Properties of powering sources of resolved jets}
\tablewidth{0pt}
\tablehead{
\colhead{Source} &
\colhead{Region} &
\colhead{Other names$^{a}$} &
\colhead{Class$^{b}$} &
\colhead{L$_{bol}$$^{b}$} &
\colhead{T$_{bol}$$^{b}$} &
\colhead{PA$^{c}$} &\\
\colhead{} &
\colhead{} &
\colhead{} &
\colhead{} &
\colhead{L$_{\odot}$} &
\colhead{K}&
\colhead{$\degree$} &\\
}
\startdata
    Per-emb-8 &  IC 348 & PER22, IC 348a, IRAS 03415+3152, YSO 48   &    0 & 2.6$\pm$0.5 & 43.0$\pm$6.0 & 135 (5) &\\
    Per-emb-18 &  NGC 1333   & NGC 1333 IRAS7, YSO 24 &  0 & 2.8$\pm$1.7 & 59.0$\pm$12.0 & 159 (2) &\\
    Per-emb-20 &  L1455     & L1455-IRS4   &     0 & 1.4$\pm$0.2 & 65.0$\pm$3.0 & 115 (2)\\
    Per-emb-30 &  B1 & PER19, B1 SMM11, YSO 40  &   0/I & 1.1$\pm$0.0 & 93.0$\pm$6.0 & 109 (2) &\\
    Per-emb-33 & L1448 & PER02, L1448 N(A), L1448 IRS3B, YSO 2 &     0 & 8.3$\pm$0.8 & 57.0$\pm$3.0 & 105 (3) &\\
    Per-emb-36 &  NGC 1333   &  PER06, NGC 1333 IRAS2B, YSO 16  & I & 6.9$\pm$1.0 & 85.0$\pm$12.0 & 24 (1) &\\
    L1448 IRS3A &  L1448 & -  &  I & 9.2$\pm$1.3 & 47.0$\pm$2.0 & 38 (4) &\\
        SVS 13C &  NGC 1333    &  VLA2  & 0 & 1.5$\pm$0.2 & 21.0$\pm$1.0 & 8 (1) &\\
\enddata
\tablenotetext{a}{ Names: YSOXX - \cite{Jorgensen2007}, PERXX - \cite{Karska2014a}, VLAXX - \cite{Rodriguez1997}}
\tablenotetext{b}{ References: \cite{Enoch2009}, \cite{sadavoy2014}}
\tablenotetext{c}{ References: (1) \cite{Plunkett2013}, (2) \cite{Davis2008}, (3) \cite{Kwon2006},
 (4) \cite{Lee2016}, (5) Tobin et al. in prep.}
\label{tab:sourcesinfo}
\end{deluxetable}

\begin{deluxetable}{lcccrrcrrrcrl}
\tabletypesize{\scriptsize}
\rotate
\tablecaption{C-band observation results}
\tablewidth{0pt}
\tablehead{
\colhead{Source} &
\colhead{RA} &
\colhead{DEC} &
\colhead{$F_{\nu,int}$} &
\colhead{$F_{\nu,peak}$} &
\colhead{RMS} &
\colhead{$F_{\nu,int}$} &
\colhead{$F_{\nu,peak}$} &
\colhead{RMS} &
\colhead{Sp. Index$^{a}$} &
\colhead{Sp. Index$^{b}$} &\\
\colhead{} &
\colhead{} &
\colhead{} &
\colhead{(6.4 cm)} &
\colhead{(6.4 cm)} &
\colhead{(6.4 cm)}&
\colhead{(4 cm)} &
\colhead{(4 cm)} &
\colhead{(4 cm)} &
\colhead{Int.} &
\colhead{Peak}&\\
\colhead{} &
\colhead{(J2000)} &
\colhead{(J2000)} &
\colhead{(mJy)} &
\colhead{(mJy bm$^{-1}$)} &
\colhead{(mJy bm$^{-1}$)} &
\colhead{(mJy)} &
\colhead{(mJy bm$^{-1}$)} &
\colhead{(mJy bm$^{-1}$)} &
\colhead{} &
\colhead{}&\\
}
\startdata
     Per-emb-8 & 03:44:43.981 &  +32:01:35.210 & 0.2788$\pm$0.0122 & 0.1323 & 0.0049  & 0.3132$\pm$0.0166 & 0.1240 & 0.0037  & 0.26$\pm$0.15 & -0.14$\pm$0.10 & \\
    Per-emb-18 & 03:29:11.258 &  +31:18:31.072 & 0.1919$\pm$0.0090 & 0.1326 & 0.0058  & 0.1957$\pm$0.0059 & 0.1505 & 0.0043  & 0.04$\pm$0.12 & 0.28$\pm$0.11 & \\
    Per-emb-20 & 03:27:43.276 &  +30:12:28.780 & 0.1477$\pm$0.0083 & 0.1093 & 0.0054  & 0.1384$\pm$0.0124 & 0.1098 & 0.0042  & -0.14$\pm$0.23 & 0.01$\pm$0.14 & \\
    Per-emb-30 & 03:33:27.303 &  +31:07:10.159 & 0.2747$\pm$0.0171 & 0.1542 & 0.0055  & 0.2815$\pm$0.0254 & 0.1681 & 0.0055  & 0.05$\pm$0.24 & 0.19$\pm$0.11 & \\
    Per-emb-33 & 03:25:36.379 &  +30:45:14.727 & 0.1648$\pm$0.0139 & 0.1154 & 0.0053  & 0.1386$\pm$0.0071 & 0.1041 & 0.0042  & -0.38$\pm$0.22 & -0.23$\pm$0.13 & \\
  Per-emb-33-A & 03:25:36.312 &  +30:45:15.153 & 0.1389$\pm$0.0082 & 0.1154 & 0.0053  & 0.1386$\pm$0.0071 & 0.1041 & 0.0042  & -0.00$\pm$0.17 & -0.23$\pm$0.13 & \\
  Per-emb-33-B & 03:25:36.321 &  +30:45:14.913 & \(<\)0.0158$\pm$0.0053 & \(<\)0.0158 & 0.0053  & \(<\)0.0127$\pm$0.0042 & \(<\)0.0127 & 0.0042  & -99.00$\pm$-99.00 & -99.00$\pm$-99.00 & \\
  Per-emb-33-C & 03:25:36.380 &  +30:45:14.722 & 0.0259$\pm$0.0057 & 0.0282 & 0.0053  & \(<\)0.0127$\pm$0.0042 & \(<\)0.0127 & 0.0042  & \(<\)-1.57$\pm$0.88 & \(<\)-1.75$\pm$0.84 & \\
  Per-emb-36-A & 03:28:57.373 &  +31:14:15.764 & 0.2433$\pm$0.0192 & 0.1267 & 0.0051  & 0.2523$\pm$0.0169 & 0.1037 & 0.0044  & 0.08$\pm$0.23 & -0.44$\pm$0.13 & \\
  Per-emb-36-B & 03:28:57.370 &  +31:14:16.072 & \(<\)0.0152$\pm$0.0051 & \(<\)0.0152 & 0.0051  & \(<\)0.0132$\pm$0.0044 & \(<\)0.0132 & 0.0044  & -99.00$\pm$-99.00 & -99.00$\pm$-99.00 & \\
    L1448IRS3A & 03:25:36.499 &  +30:45:21.880 & 0.4922$\pm$0.0149 & 0.3717 & 0.0053  & 0.5074$\pm$0.0149 & 0.3607 & 0.0043  & 0.07$\pm$0.09 & -0.07$\pm$0.04 & \\
        SVS13C & 03:29:01.970 &  +31:15:38.053 & 1.0676$\pm$0.0218 & 0.6812 & 0.0049  & 1.2095$\pm$0.0296 & 0.6753 & 0.0039  & 0.28$\pm$0.07 & -0.02$\pm$0.02 & \\
\enddata
\tablenotetext{a}{ Integrated spectral index - calculated with flux density integrated over the full extent of the source}
\tablenotetext{b}{ Peak spectral index - calcualted with peak value of the flux density.}
\label{tab:cband_res}
\end{deluxetable}

\begin{deluxetable}{lrrrrrrrr}
\tabletypesize{\scriptsize}
\tablecaption{Spectral indices at different positions and position angles of resolved emission}
\tablewidth{0pt}
\tablehead{
\colhead{Source} &
\colhead{PA$^{a}$} &
\colhead{Sp. Index} &
\colhead{Sp. Index} &
\colhead{Sp. Index} &
\colhead{Sp. Index} &\\
\colhead{} &
\colhead{deg} &
\colhead{Central} &
\colhead{Outflow 1} &
\colhead{Outflow 2}&
\colhead{Outflow 3}&\\
}
\startdata
    Per-emb-8  & 13.0$\pm$1.9 & 0.16$\pm$0.12 & -0.25$\pm$0.16 & - \\
    Per-emb-18 & 163.9$\pm$15.6 & 0.37$\pm$0.14 & -0.76$\pm$0.94 & -1.13$\pm$0.42 & - \\
    Per-emb-20 & 137.1$\pm$7.8 & 0.00$\pm$0.18 & -1.38$\pm$0.51 & - \\
    Per-emb-30 & 123.8$\pm$2.9 &  0.28$\pm$0.14 & -0.20$\pm$0.46  &  -1.10$\pm$0.47 & - \\
    Per-emb-33 & 108.0$\pm$3.7 & -0.14$\pm$0.14 & -0.30$\pm$0.74 & -0.27$\pm$0.52 & -0.53$\pm$0.50  \\
    Per-emb-36 & 20.6$\pm$1.7 & 0.67$\pm$0.18 & -0.64$\pm$0.20 & -0.20$\pm$0.28  & - \\
    L1448IRS3A & 79.3$\pm$2.2 &  0.08$\pm$0.05 & -0.61$\pm$0.24 & -0.13$\pm$0.12 & - \\
    SVS13C     & 179.7$\pm$1.6 & 0.39$\pm$0.02 & -0.15$\pm$0.28 & 0.05$\pm$0.36 & - \\
\enddata
\tablenotetext{a}{ Position angle measured from north to east}
\tablecomments{Outflow positions where spectral index was measured are numbered with descending declination. See figures in the Appendix.}
\label{tab:indextab}
\end{deluxetable}

\clearpage
\appendix
\section{Additional plots}

\begin{figure}[H]
  \centering
  \includegraphics[width=0.33\linewidth]{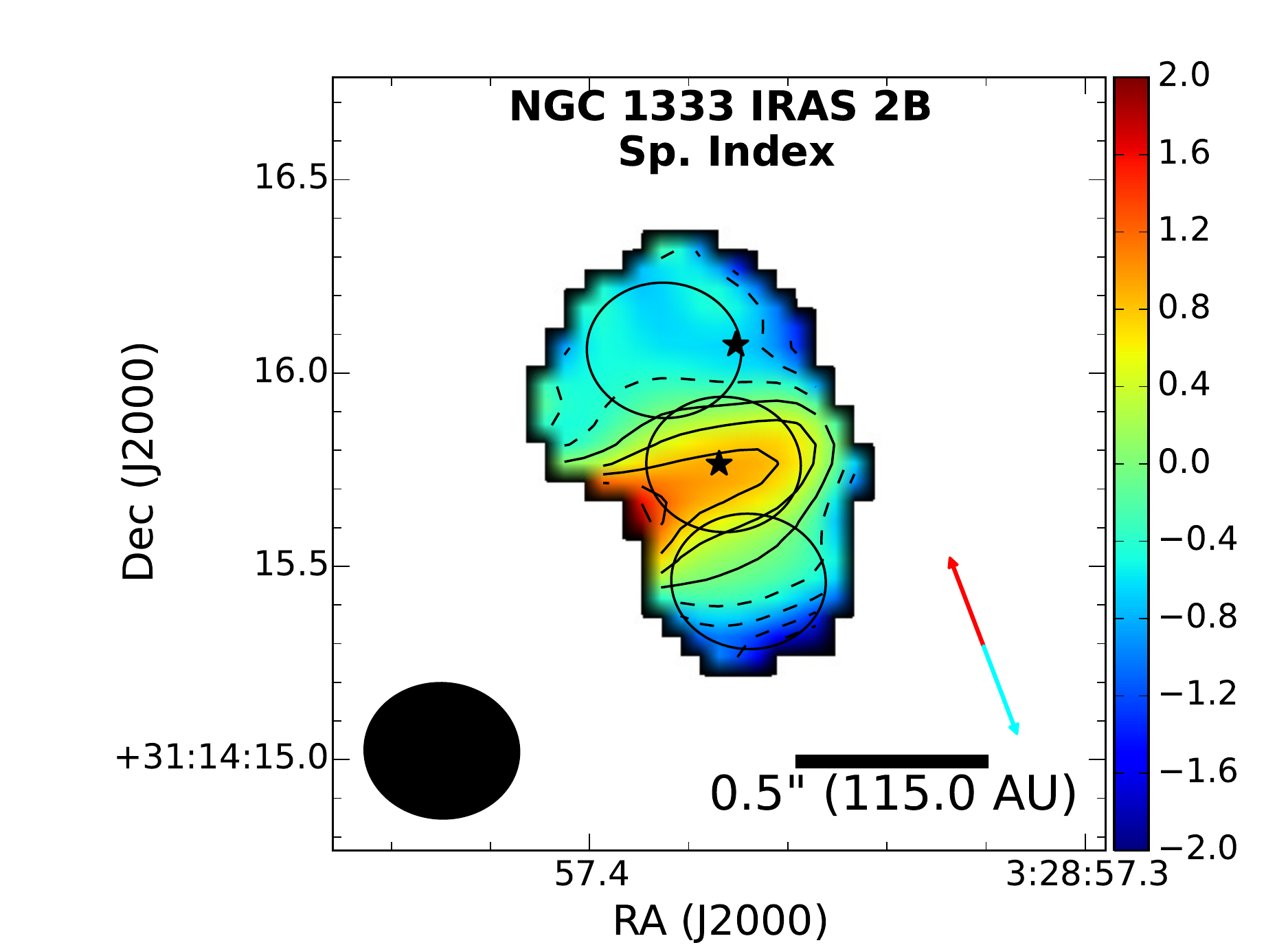}
  \label{fig:lowhist}
 \centering
  \includegraphics[width=0.33\linewidth]{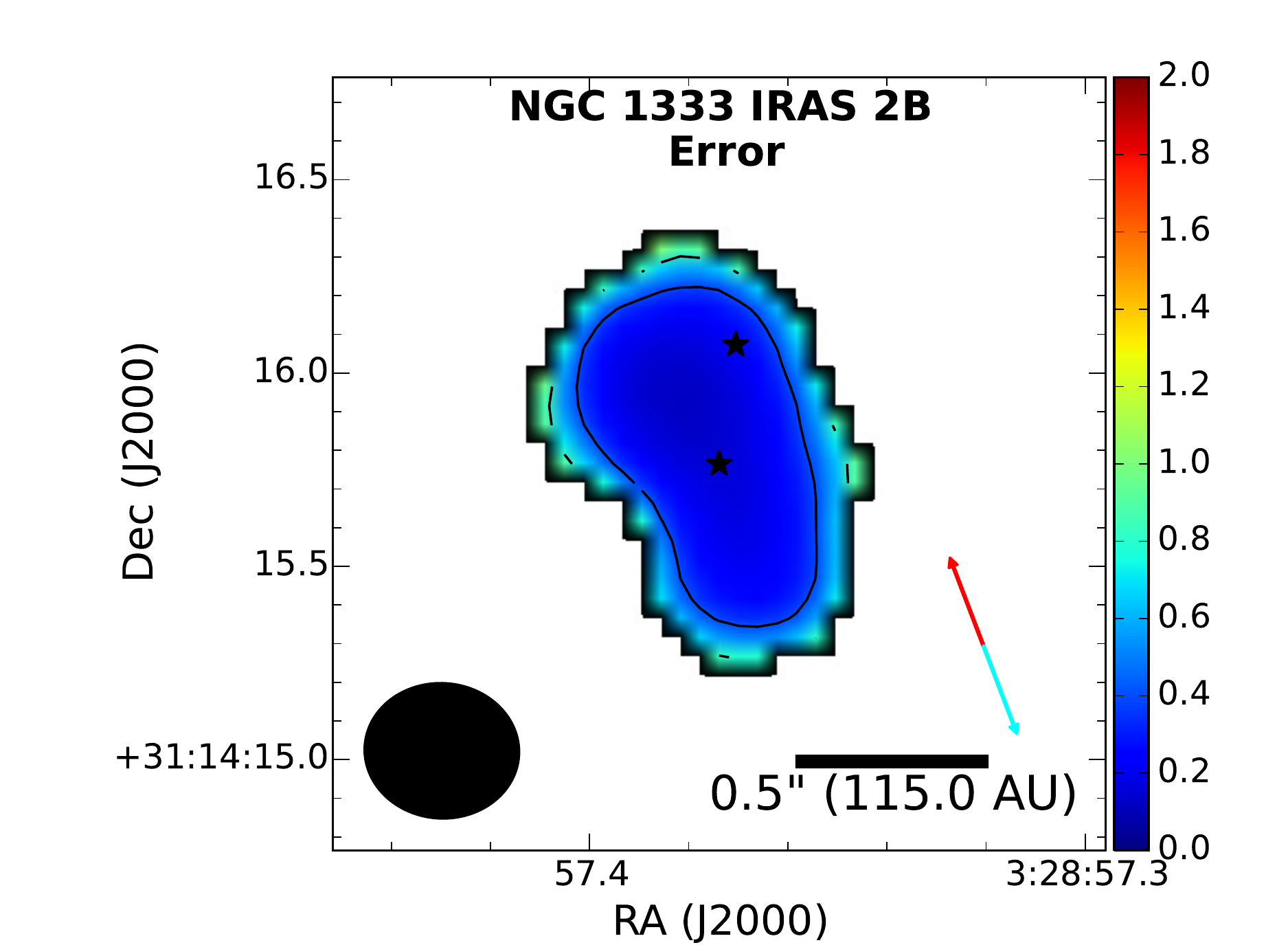}
  \label{fig:lowhist}
  \centering
  \includegraphics[width=0.26\linewidth]{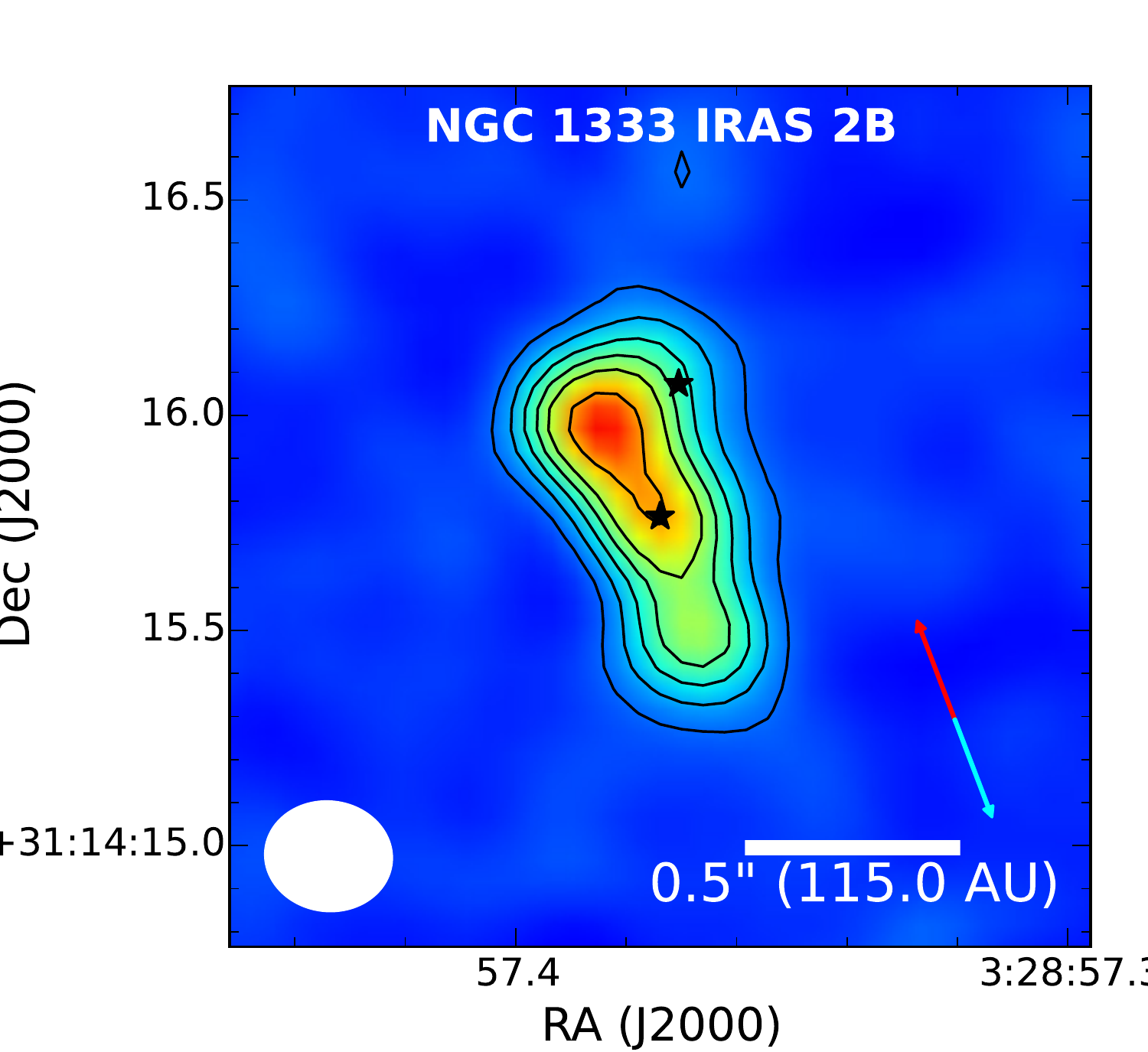}
  \label{fig:highhist}

\caption{Maps of IRAS 2B. Left: Spectral index map with marked positions where median flux was measured to calculate spectral indices. Center: Spectral index error map with contours [ 0.4, 0.8, 1.2, 1.6, 2.0]. Right: 
full bandwidth image centered at 6.05 GHz (5 cm). Contours as in Figure \ref{fig:per36} ($\sigma _{5\ cm}=3.77\ \mu $Jy). Synthesized beam is shown in the left bottom corner (0\farcs29" x 0\farcs25). Stars mark the positions of the protostars based on Ka-band observations \citep{Tobin2016}. Red and blue arrows indicate outflow direction from \cite{Plunkett2013}.}
\label{fig:per36zoom}

\end{figure}

\begin{figure}[H]
  \centering
  \includegraphics[width=0.33\linewidth]{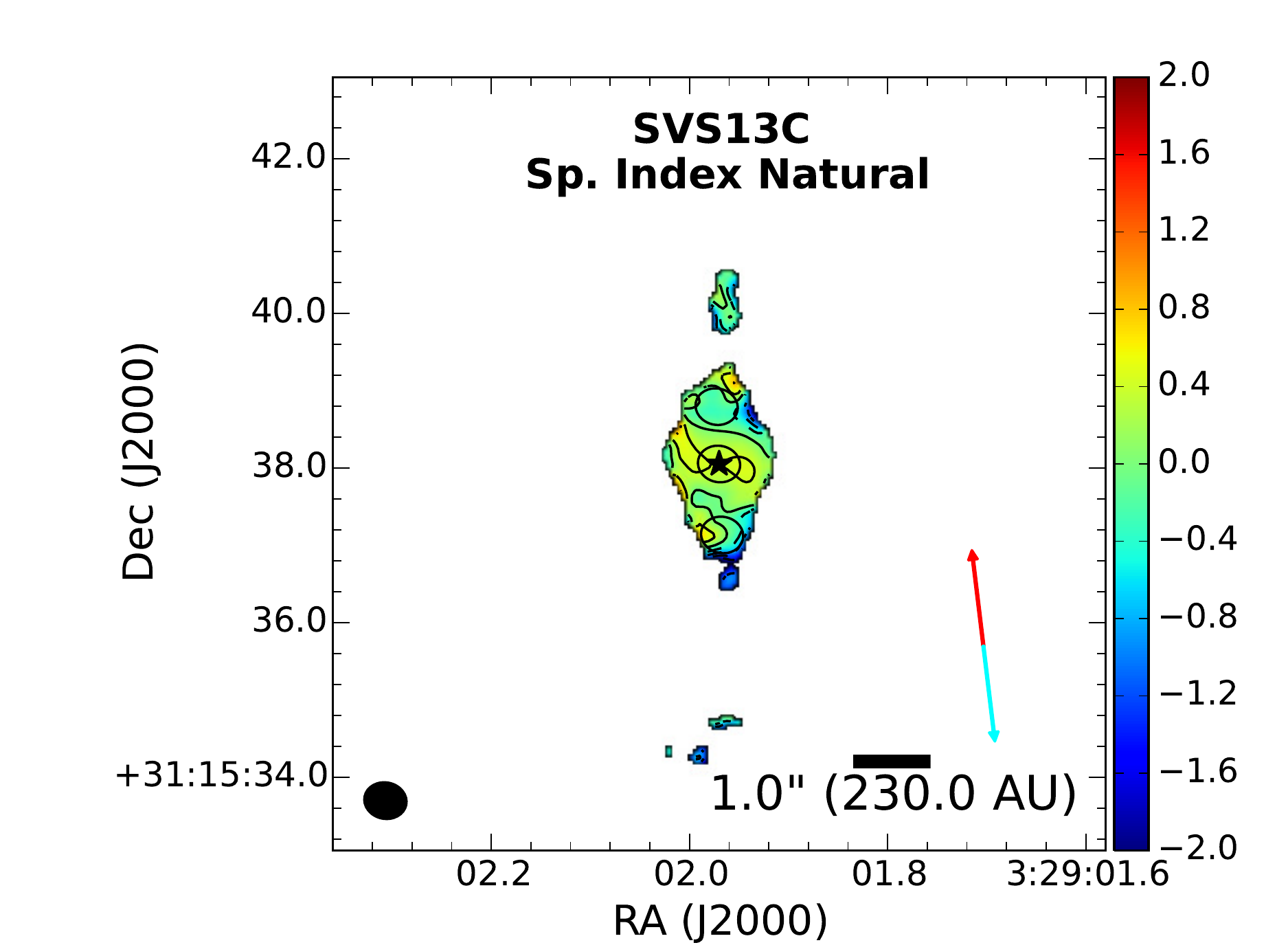}
  \label{fig:lowhist}
 \centering
  \includegraphics[width=0.33\linewidth]{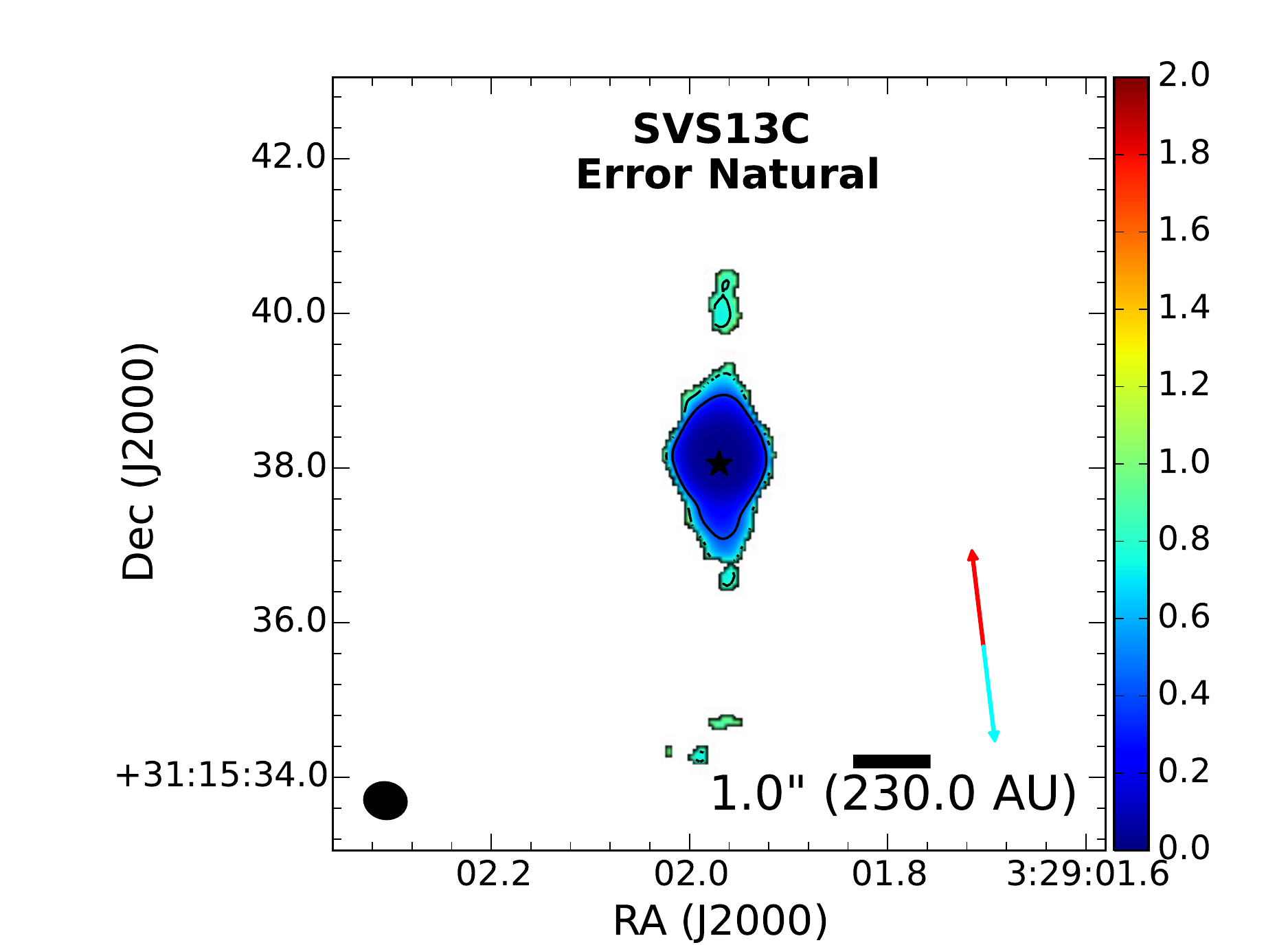}
  \label{fig:lowhist}
  \centering
  \includegraphics[width=0.26\linewidth]{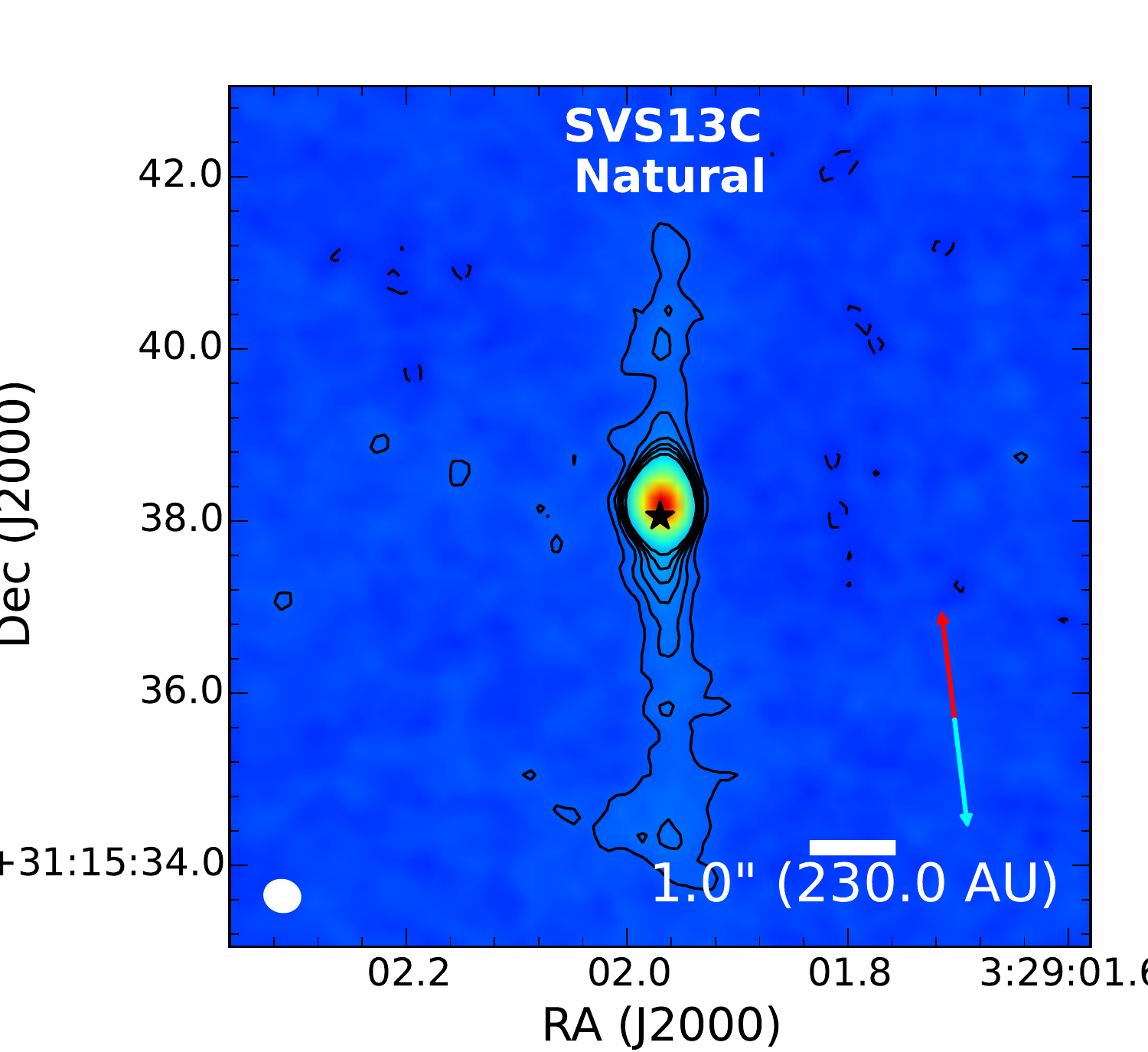}
  \label{fig:highhist}

\caption{ Same as Figure \ref{fig:per36zoom} for  SVS 13C ($\sigma _{5\ cm}=3.06\ \mu $Jy ). 
Synthesized beam of the full bandwidth image is 0\farcs41 x 0\farcs36. Star marks the position of the protostar based on Ka-band observations 
\citep{Tobin2016}. Red and blue arrows indicate outflow direction from \cite{Plunkett2013}.}
\label{fig:per109zoom} 

\end{figure}

\begin{figure}[H]
  \centering
  \includegraphics[width=0.33\linewidth]{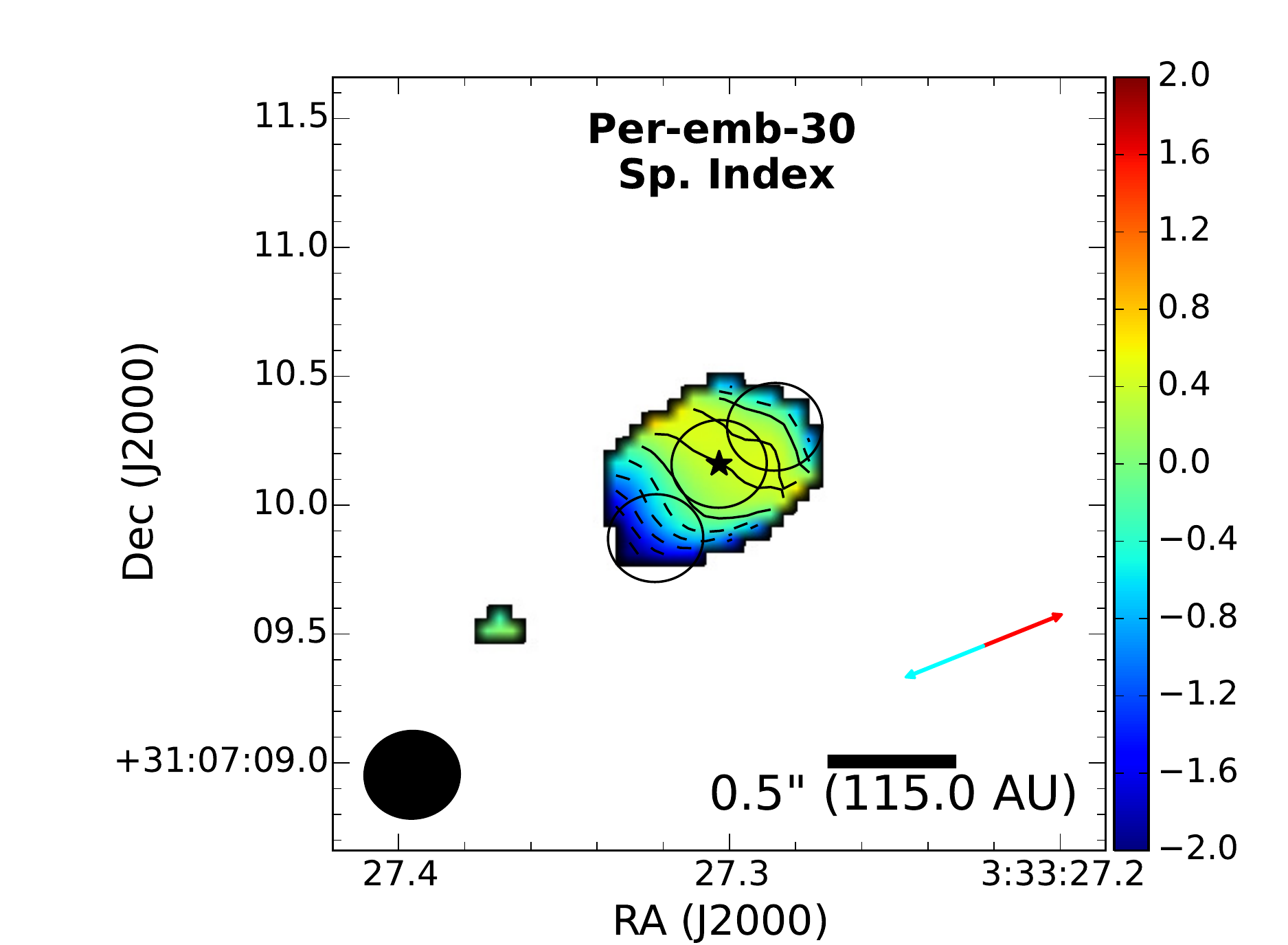}
  \label{fig:lowhist}
 \centering
  \includegraphics[width=0.33\linewidth]{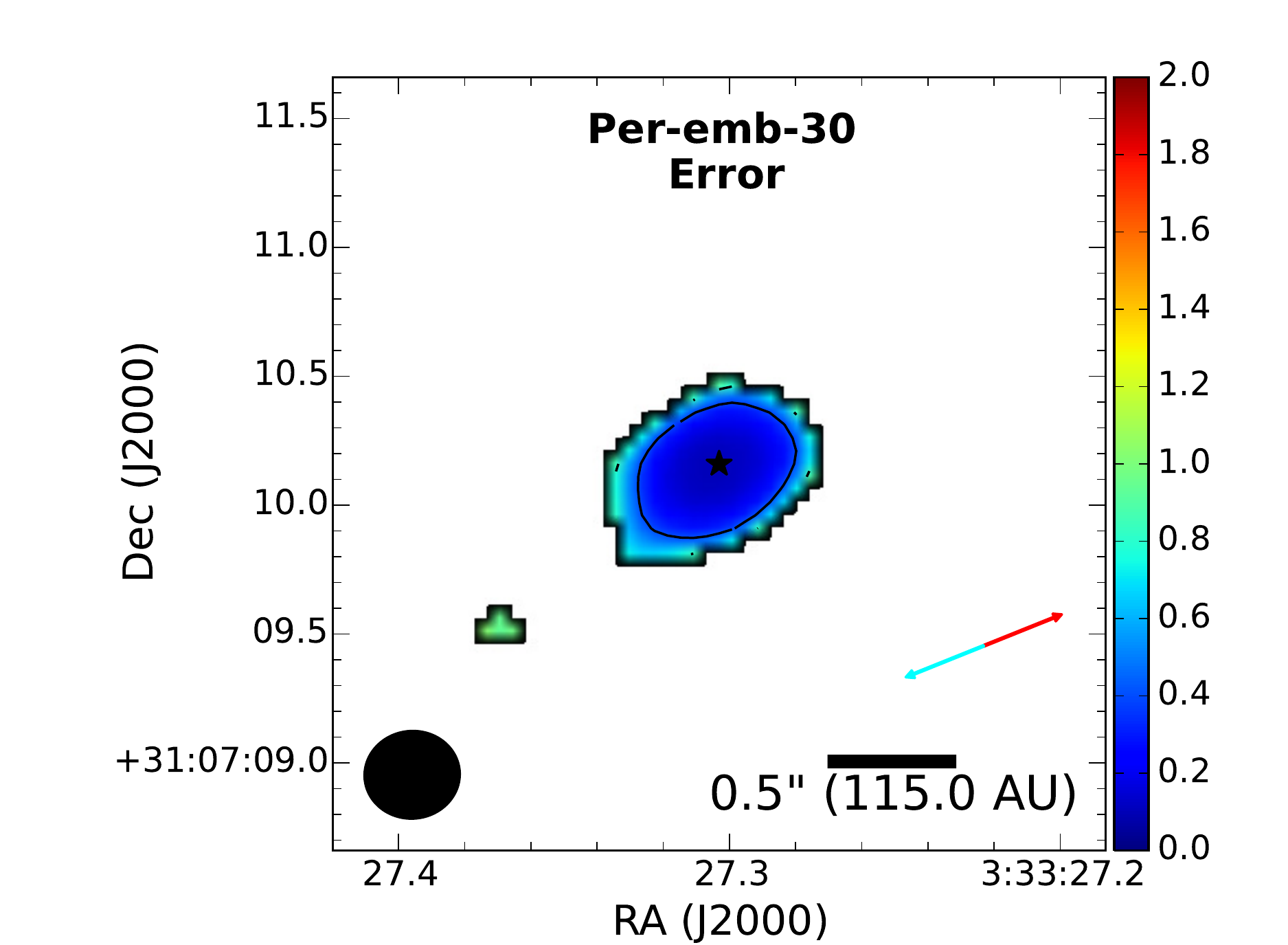}
  \label{fig:lowhist}
  \centering
  \includegraphics[width=0.26\linewidth]{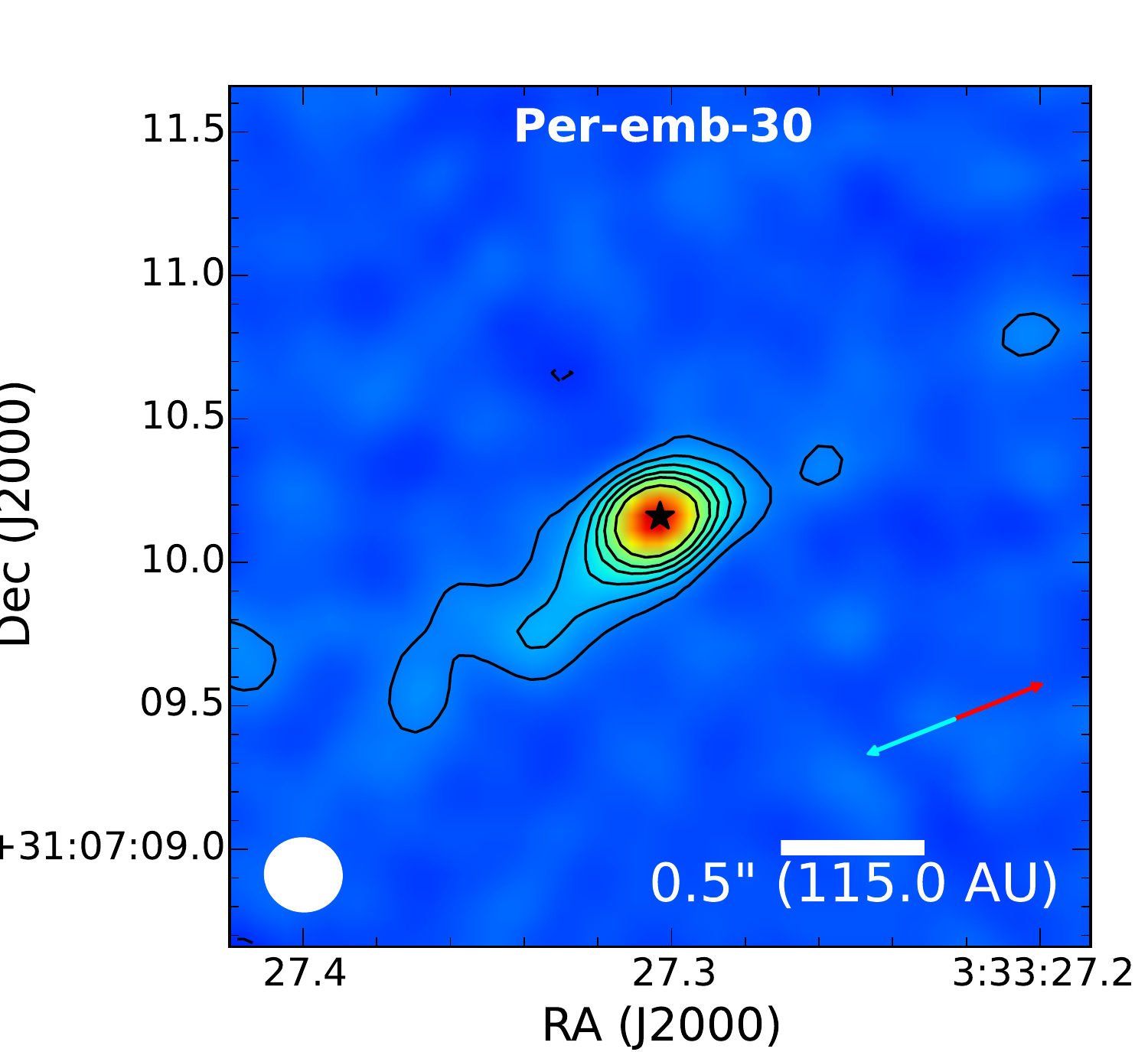}
  \label{fig:highhist}

\caption{ Same as Figure \ref{fig:per36zoom} for Per-emb-30 ($\sigma _{5\ cm}=3.58\ \mu $Jy ). 
Synthesized beam of the full bandwidth image is 0\farcs27 x 0\farcs25).
 Star marks the position of the protostar based on Ka-band observations \citep{Tobin2016}. Red and blue arrows indicate the outflow direction from \cite{Davis2008}.}
\label{fig:per109zoom} 

\end{figure}

\begin{figure}[H]

 \centering
  \includegraphics[width=0.33\linewidth]{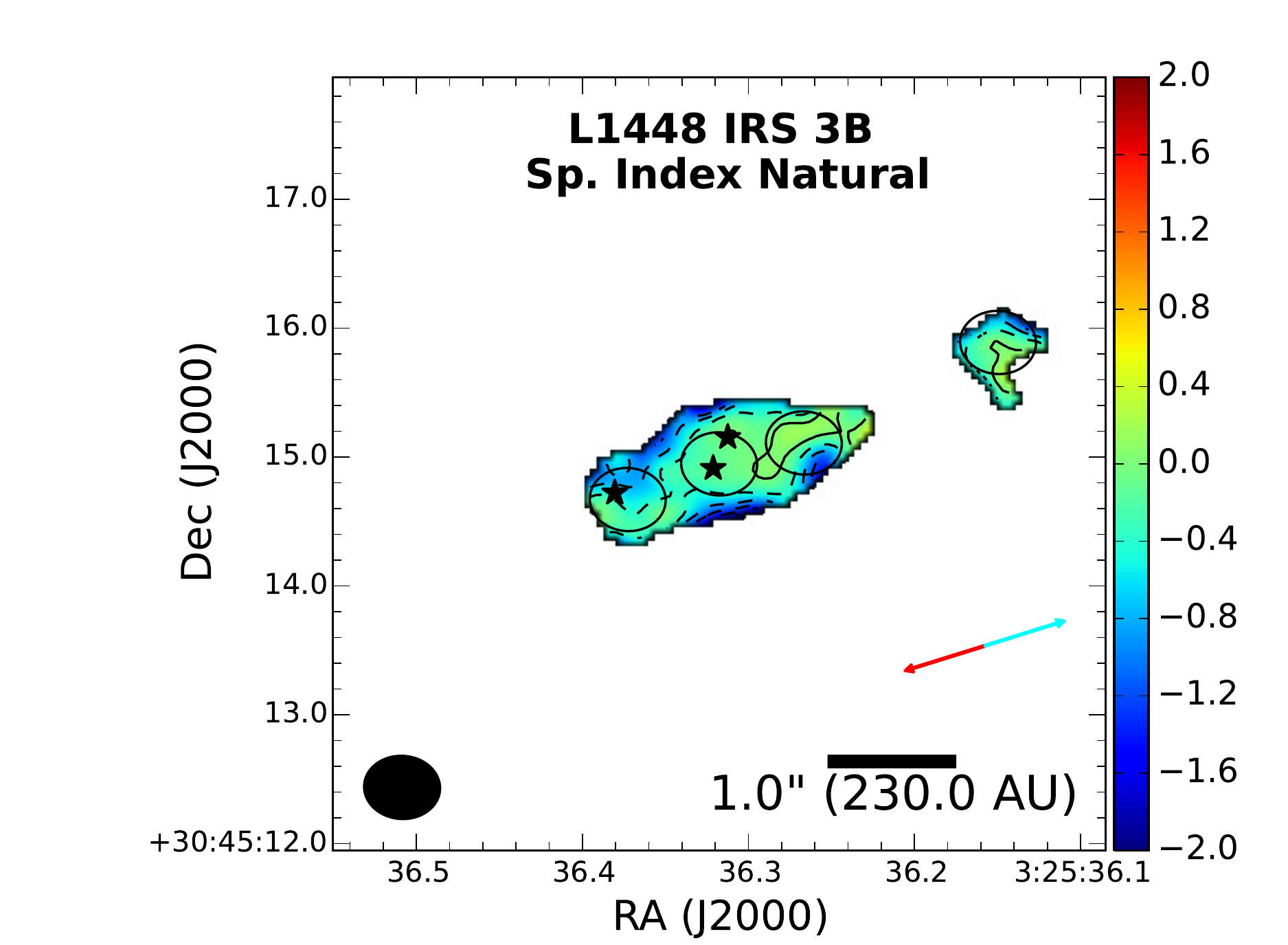}
  \includegraphics[width=0.33\linewidth]{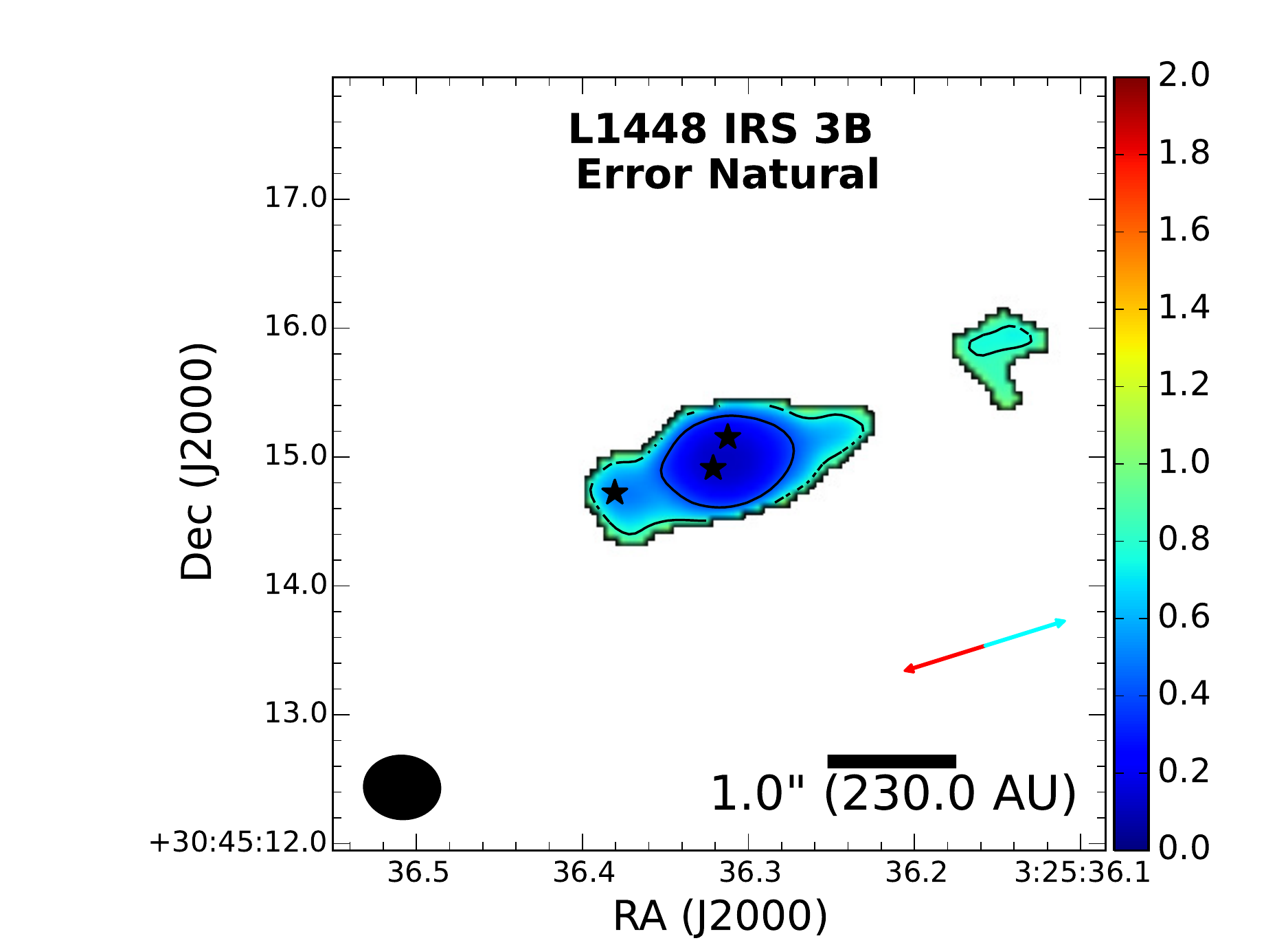}
  \includegraphics[width=0.26\linewidth]{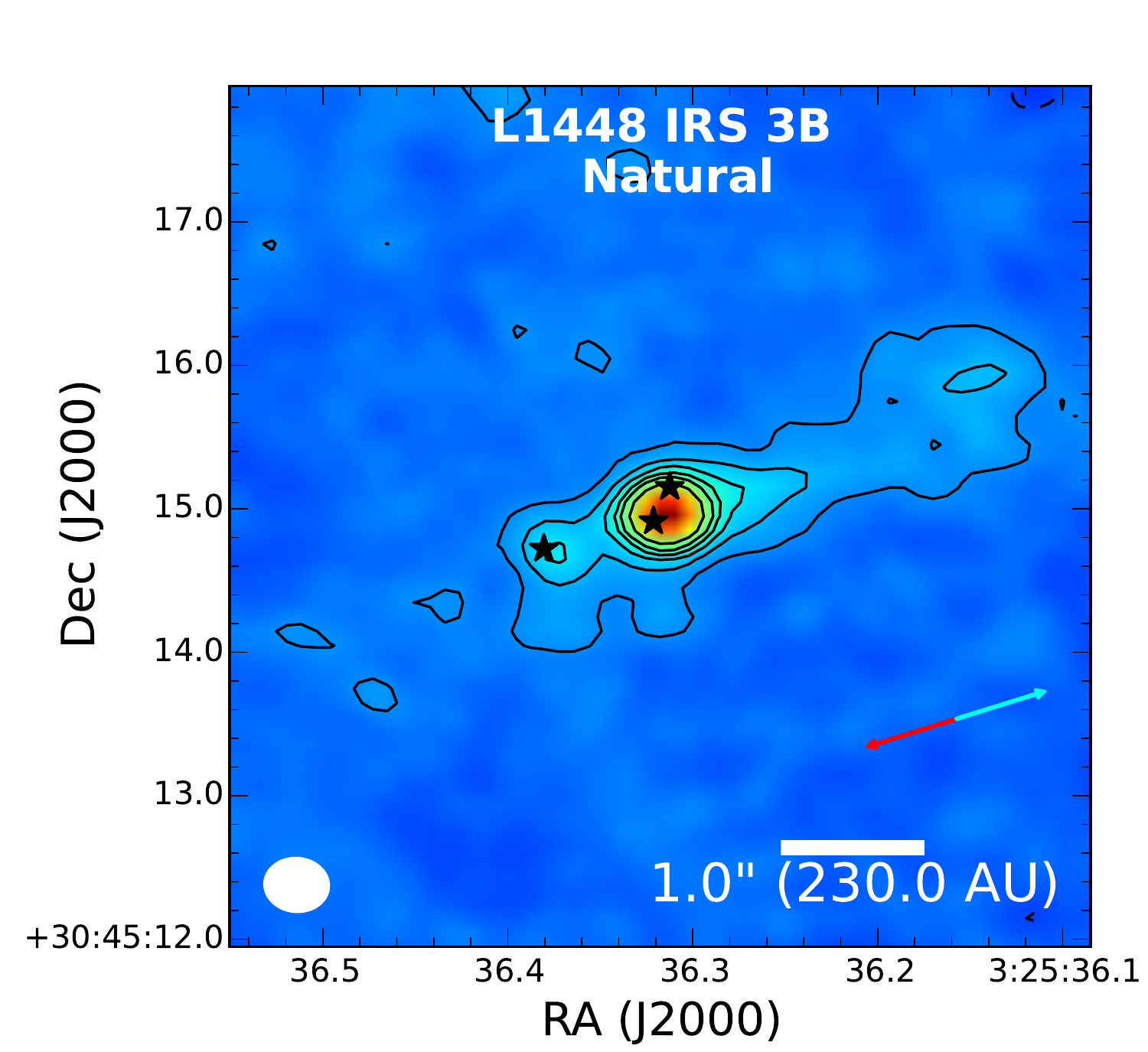}

\caption{ Same as Figure \ref{fig:per36zoom} for L1448 IRS 3B ($\sigma _{5\ cm}=3.35\ \mu $Jy ). 
Synthesized beam of the full bandwidth image is 0\farcs45 x 0\farcs37).
 Star marks the position of the protostar based on Ka-band observations 
\citep{Tobin2016}. Red and blue arrows indicate outflow direction from \cite{Lee2015}.}
\label{fig:per109}
\end{figure}

\begin{figure}[H]
  \centering
  \includegraphics[width=0.33\linewidth]{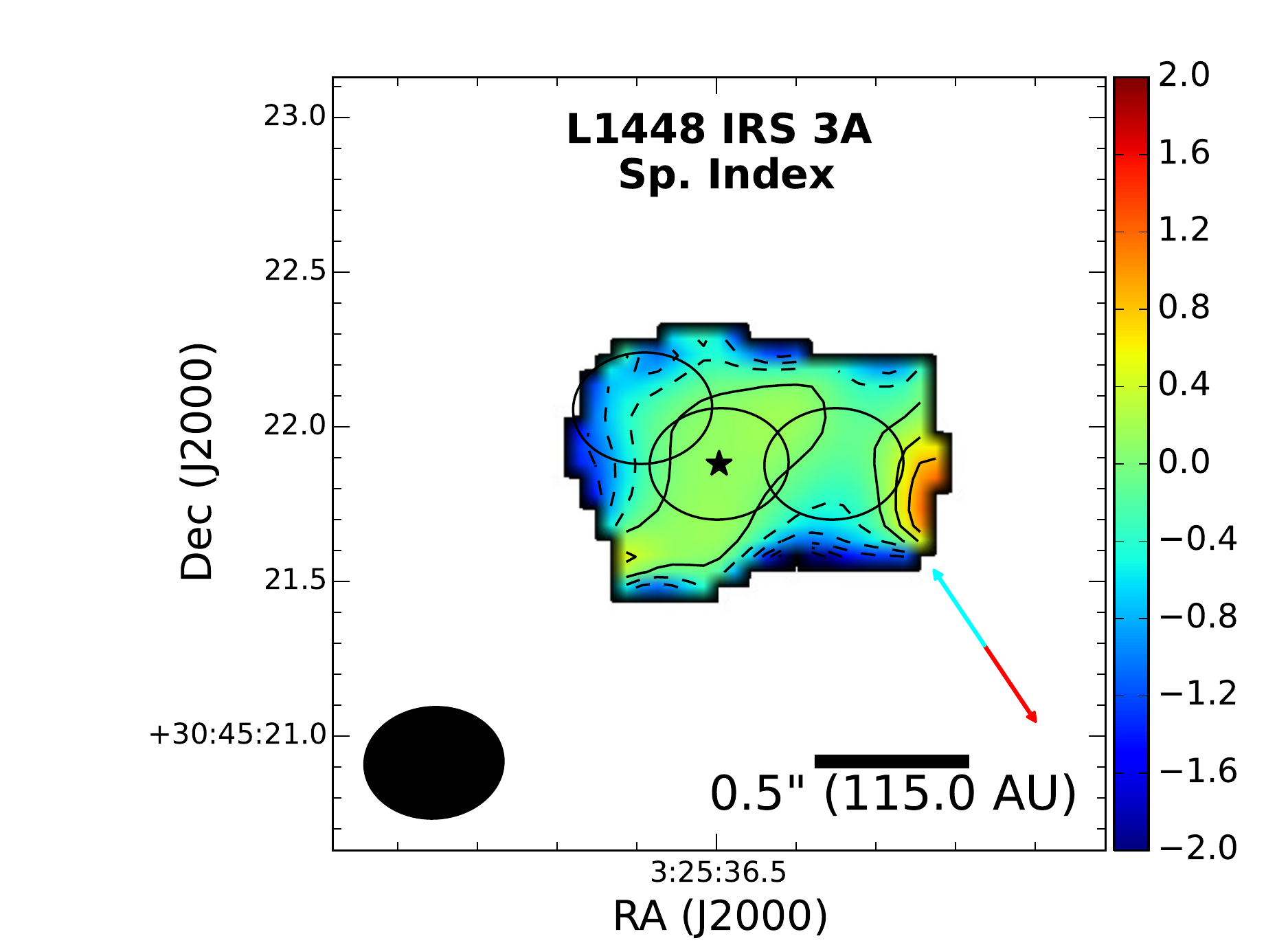}
  \label{fig:lowhist}
 \centering
  \includegraphics[width=0.33\linewidth]{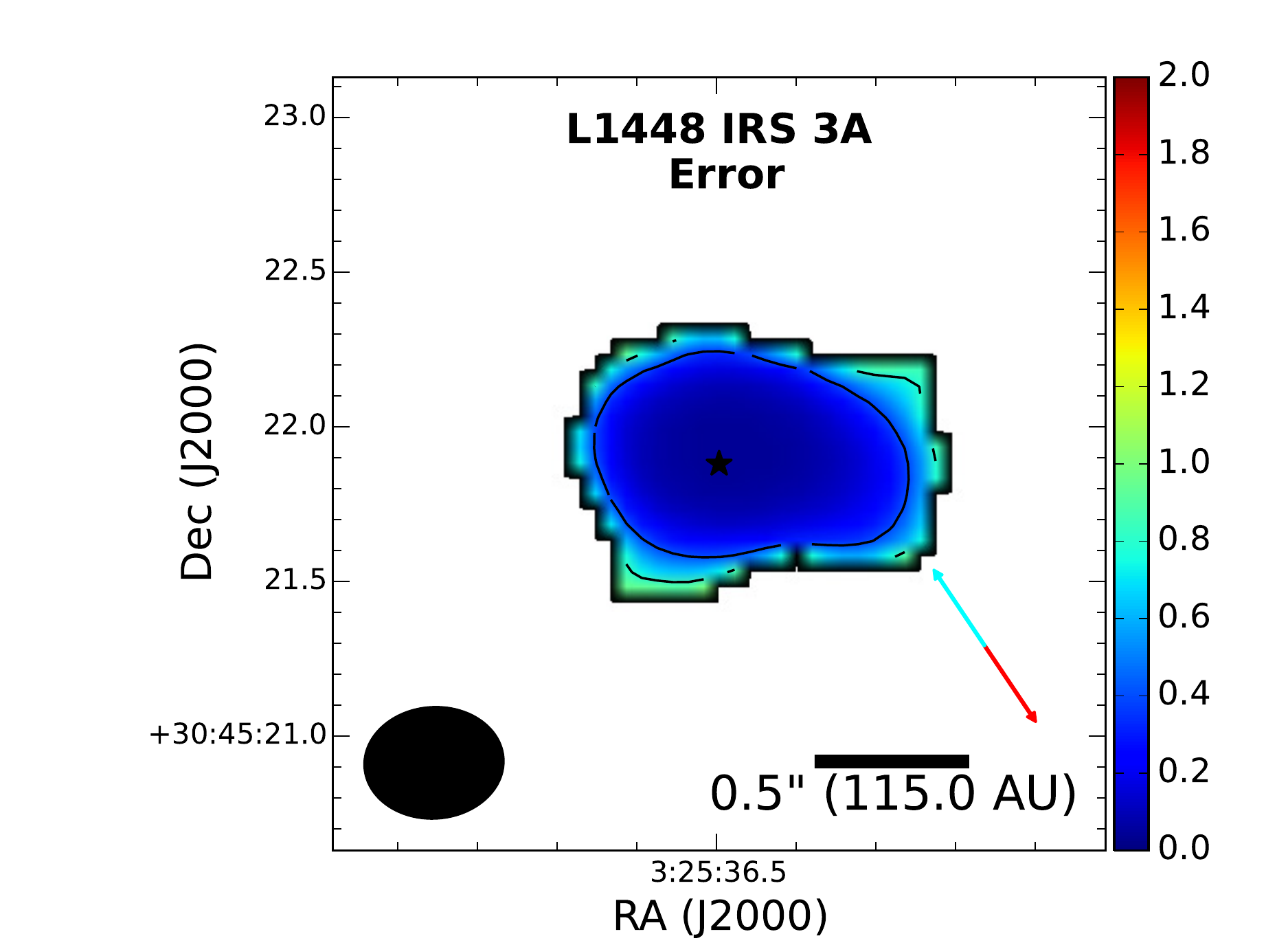}
  \label{fig:lowhist}
  \centering
  \includegraphics[width=0.26\linewidth]{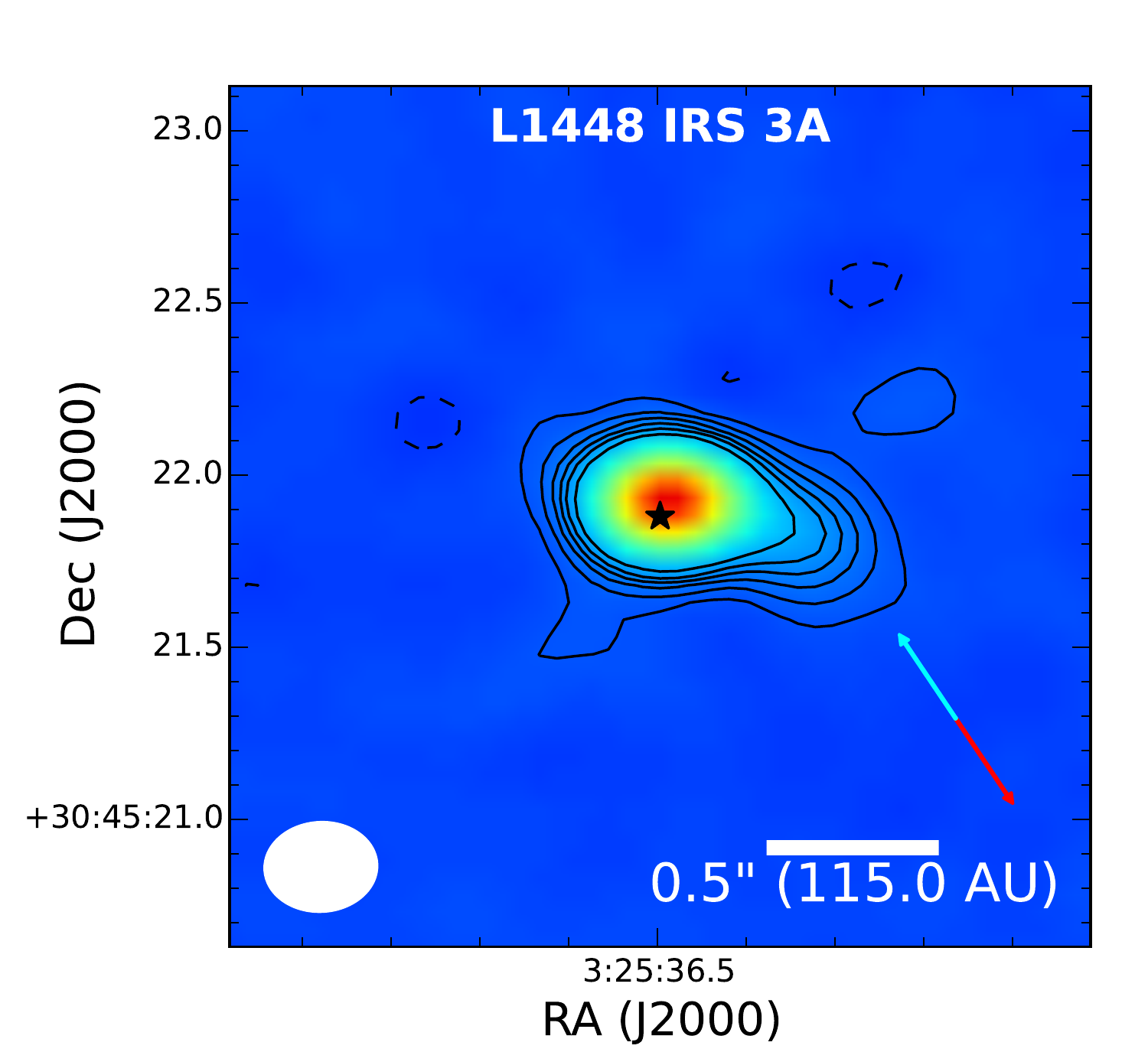}
  \label{fig:highhist}

\caption{  Same as Figure \ref{fig:per36zoom} for L1448 IRS 3A ($\sigma _{5\ cm}=3.91\ \mu $Jy ). 
Synthesized beam of the full bandwidth image is 0\farcs33 x 0\farcs26).
 Star marks the position of the protostar based on Ka-band observations. Red and blue arrows indicate outflow direction from \cite{Lee2015}} 
\label{fig:per109zoom} 
\end{figure}

\begin{figure}[H]
  \centering
  \includegraphics[width=0.33\linewidth]{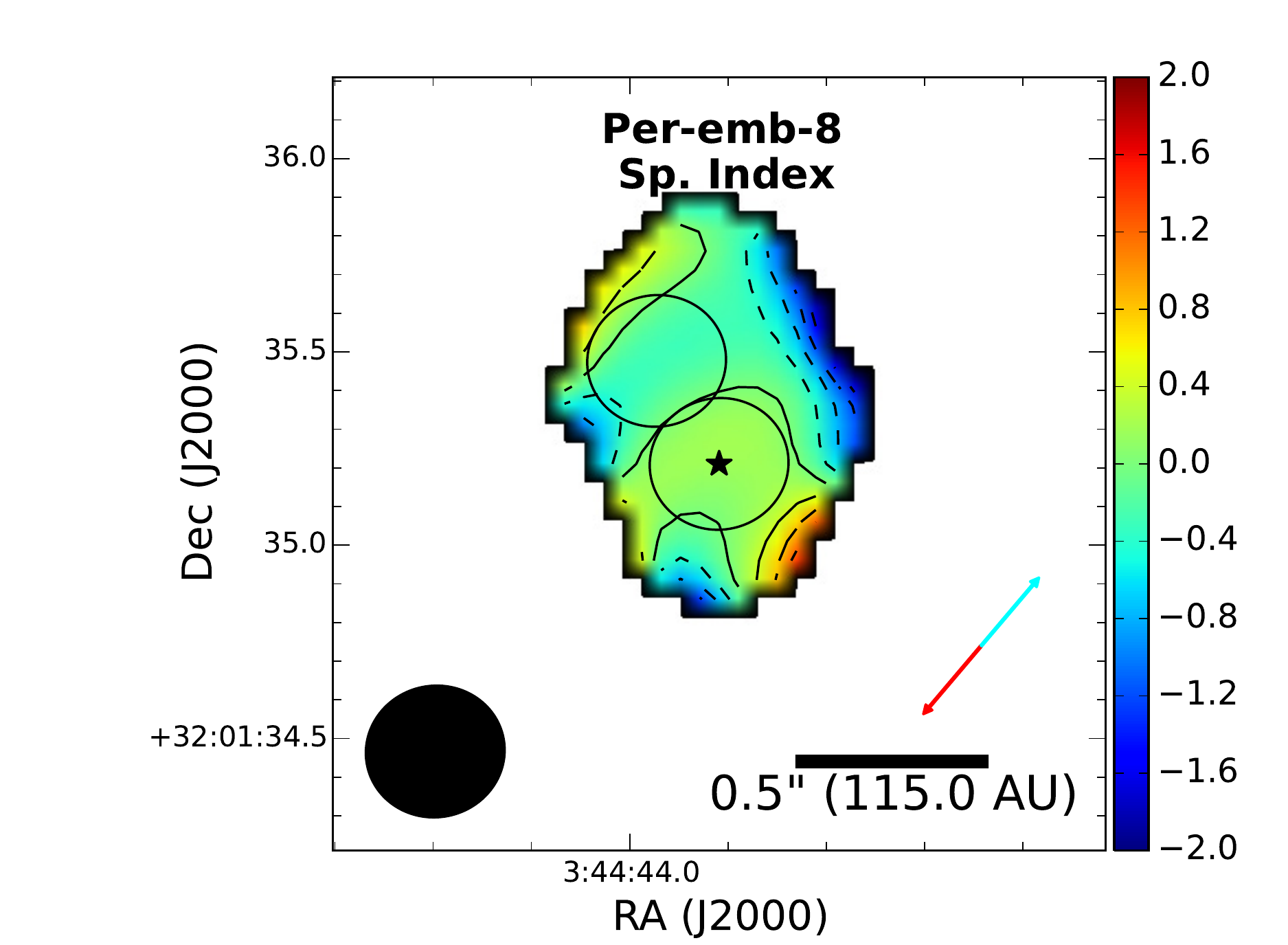}
  \label{fig:lowhist}
 \centering
  \includegraphics[width=0.33\linewidth]{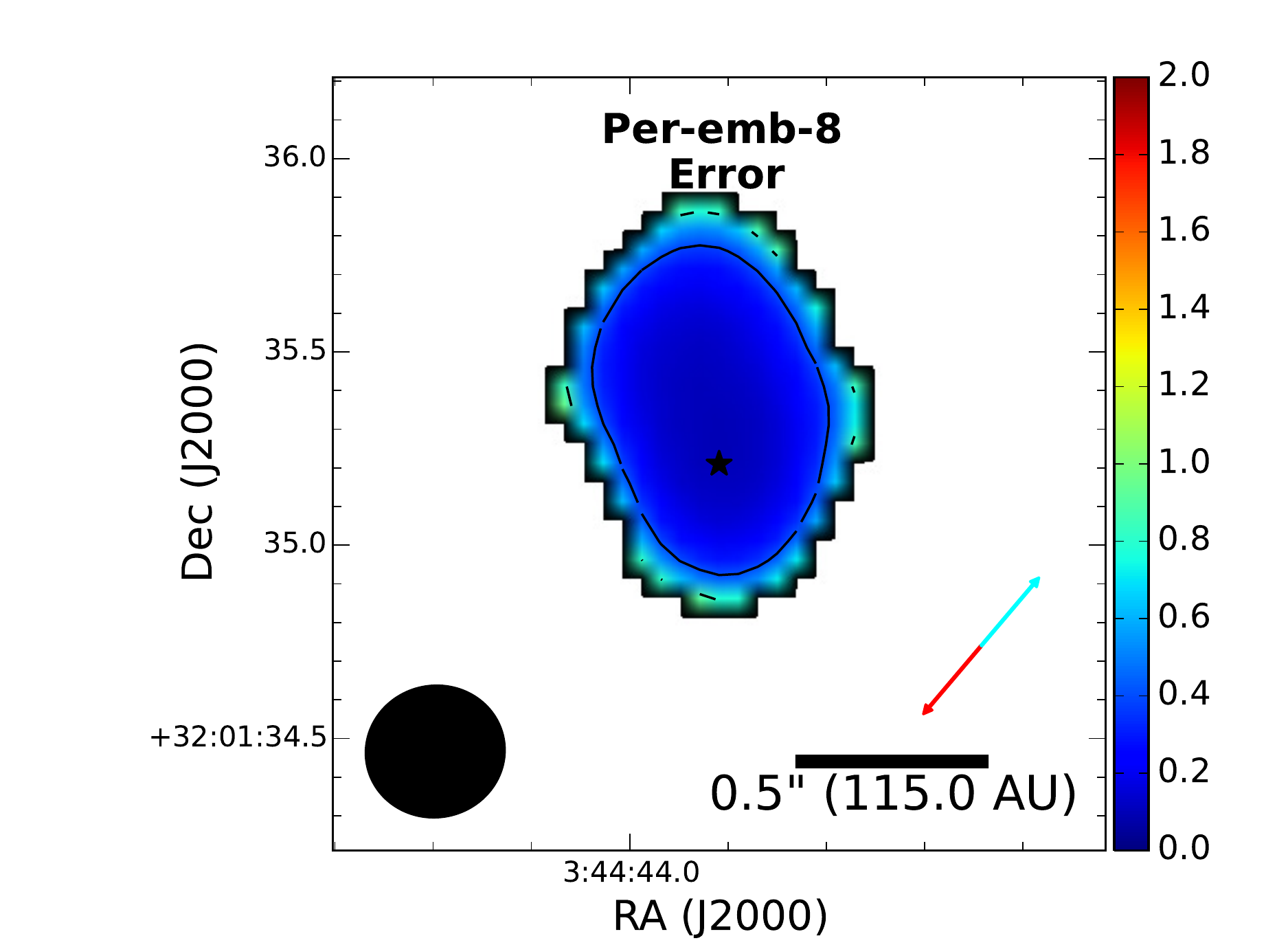}
  \label{fig:lowhist}
  \centering
  \includegraphics[width=0.26\linewidth]{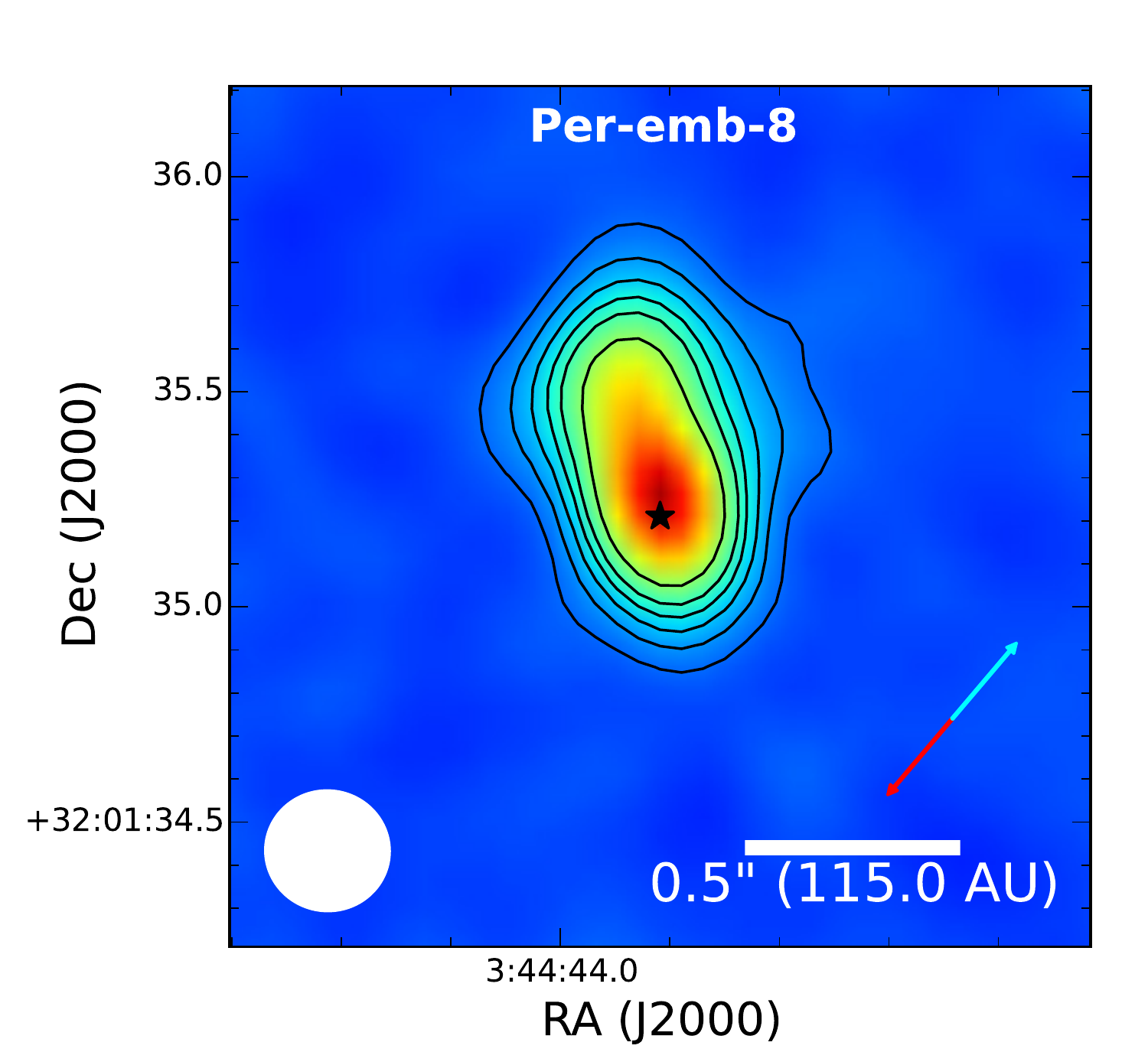}
  \label{fig:highhist}

\caption{Same as Figure \ref{fig:per36zoom} for Per-emb-8. Contours as in Figure \ref{fig:per36} ($\sigma _{5\ cm}=3.34\ \mu $Jy ). 
Synthesized beam of the full bandwidth image is 0\farcs29 x 0\farcs28).
 Star marks the position of the protostar based on Ka-band observations 
\citep{Tobin2016}. Red and blue arrows indicate outflow direction of $135\degree$ based on ALMA observations (outflow is assumed to be perpendicular to the disk (Tobin in prep.).}
\label{fig:per109zoom} 

\end{figure}

\begin{figure}[H]
  \centering
  \includegraphics[width=0.33\linewidth]{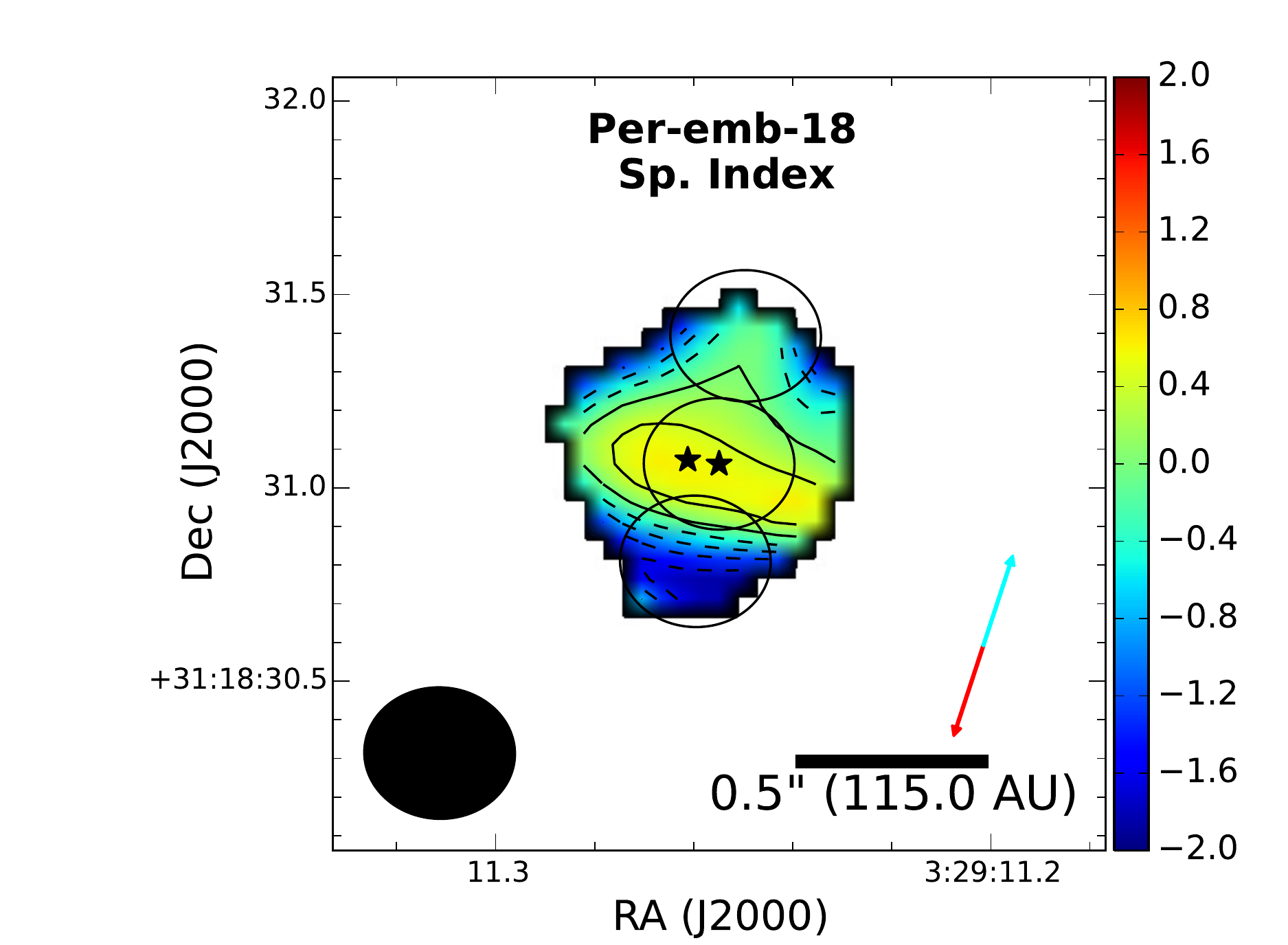}
  \label{fig:lowhist}
 \centering
  \includegraphics[width=0.33\linewidth]{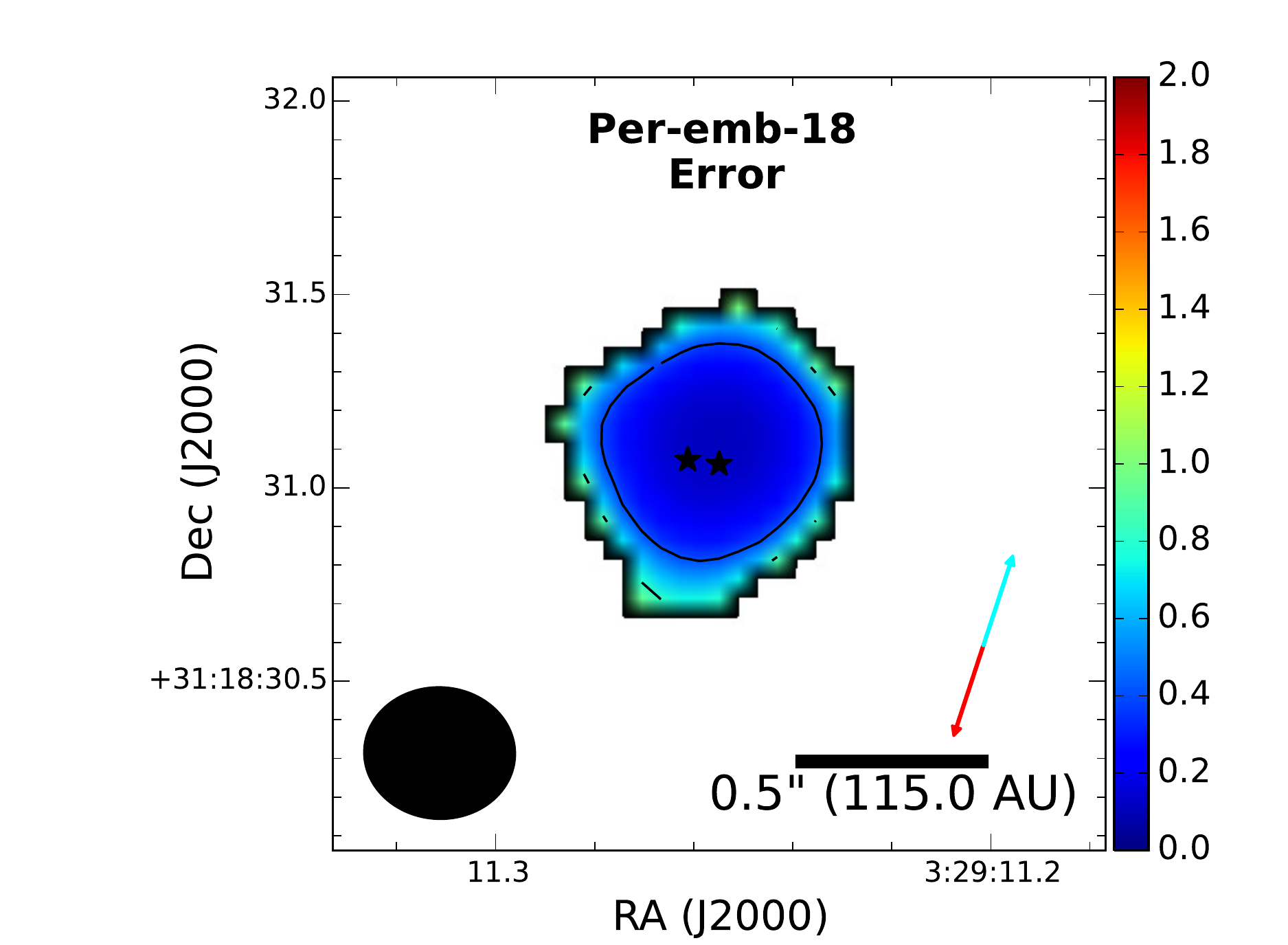}
  \label{fig:lowhist}
  \centering
  \includegraphics[width=0.26\linewidth]{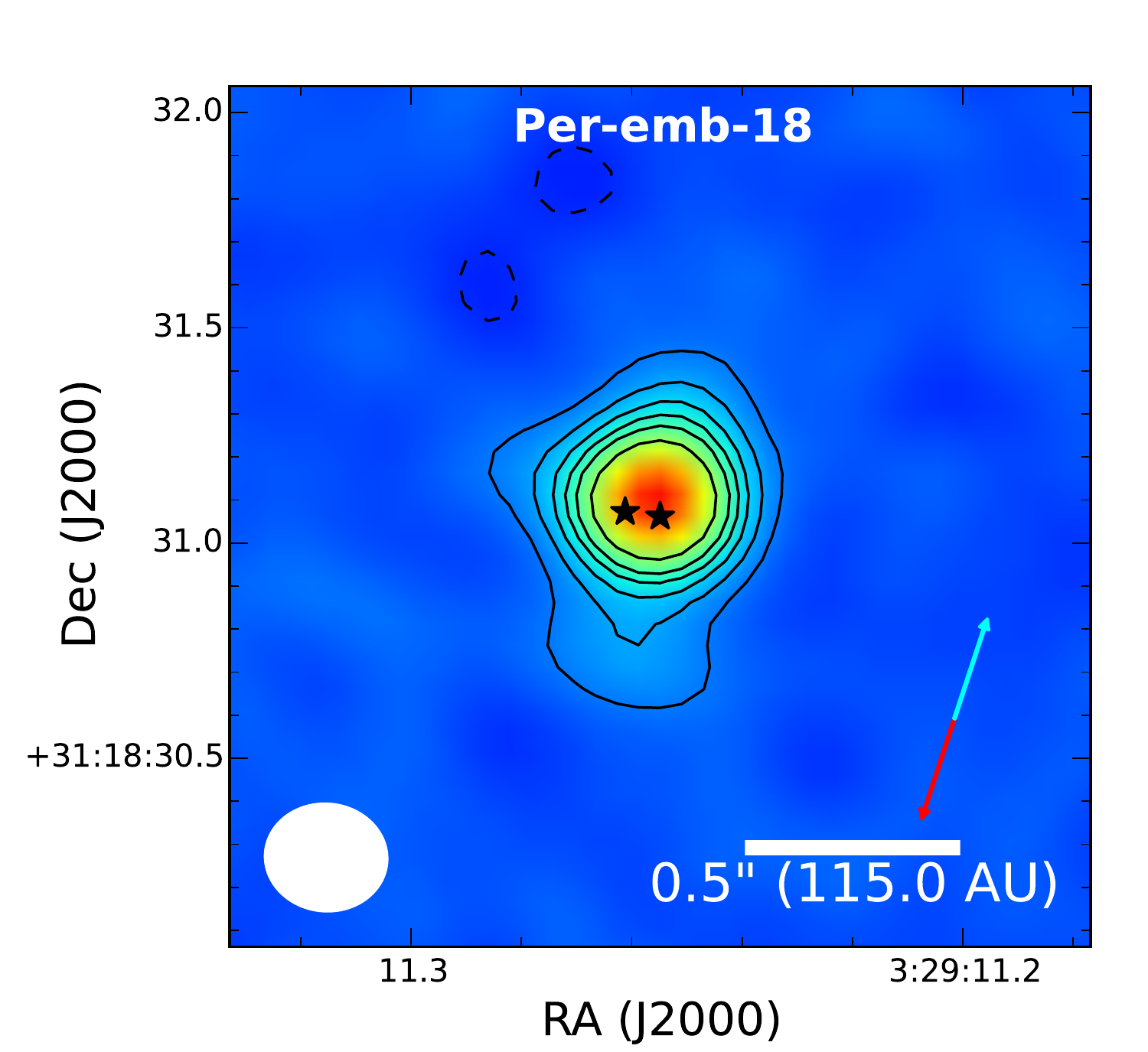}
  \label{fig:highhist}

\caption{Same as Figure \ref{fig:per36zoom} for Per-emb-18. Contours as in Figure \ref{fig:per36} ($\sigma _{5\ cm}=3.34\ \mu $Jy ). 
Synthesized beam of the full bandwidth image is 0\farcs29 x 0\farcs28).
 Stars mark the positions of the protostar based on Ka-band observations 
\citep{Tobin2016}. Red and blue arrows indicate outflow direction from \cite{Davis2008}.}
\label{fig:per109zoom} 

\end{figure}

\begin{figure}[H]
  \centering
  \includegraphics[width=0.33\linewidth]{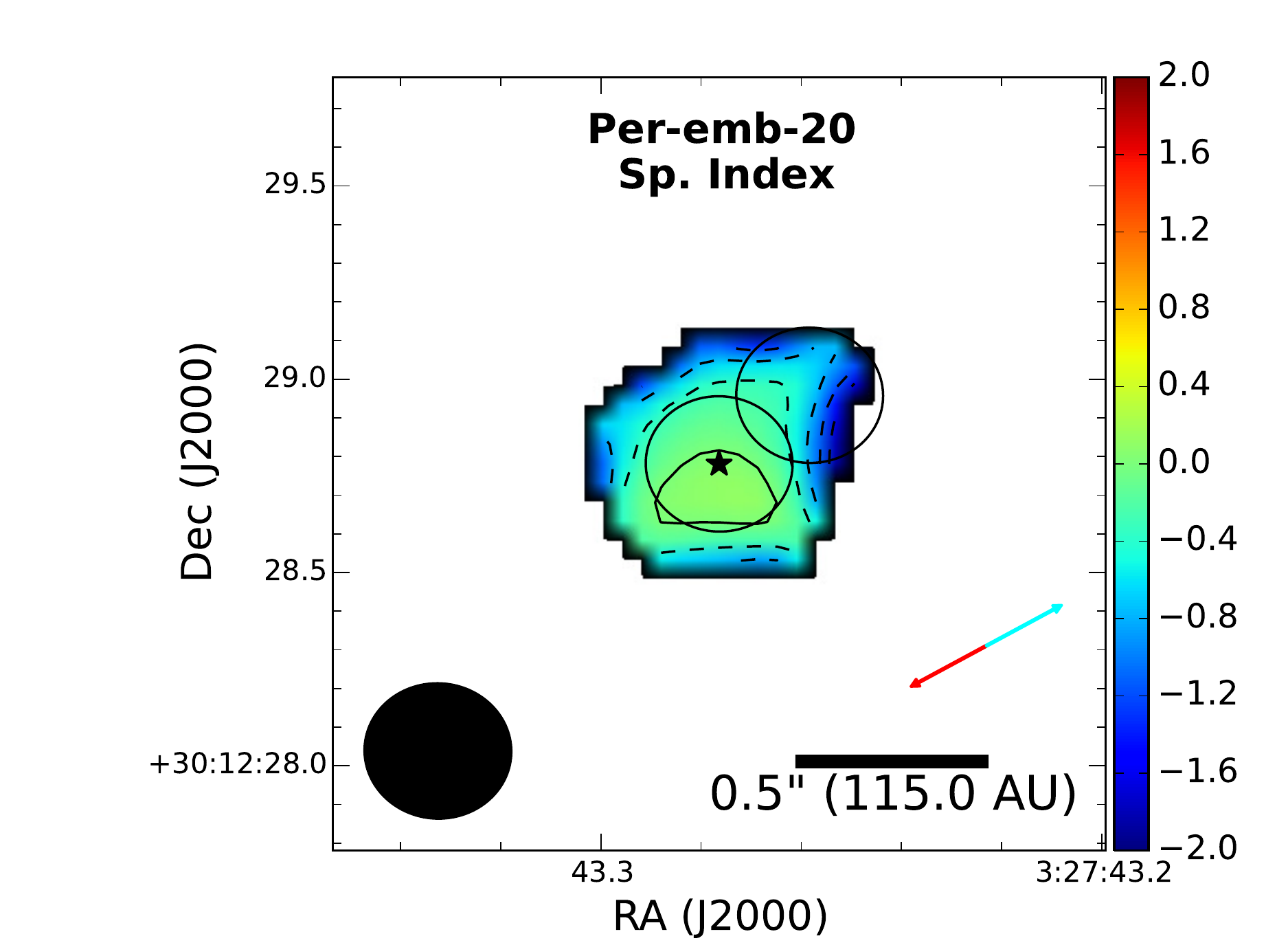}
  \label{fig:lowhist}
 \centering
  \includegraphics[width=0.33\linewidth]{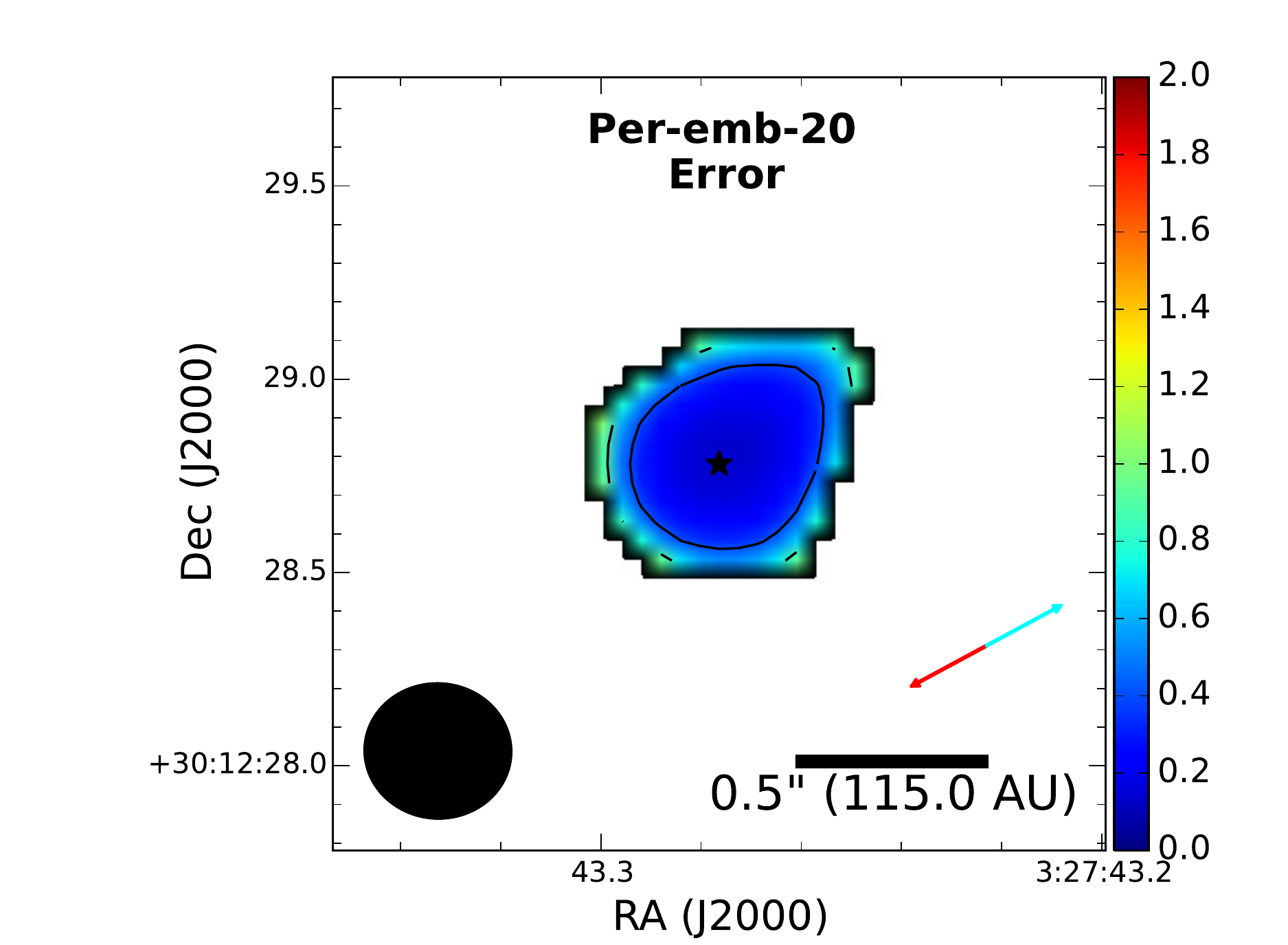}
  \label{fig:lowhist}
  \centering
  \includegraphics[width=0.26\linewidth]{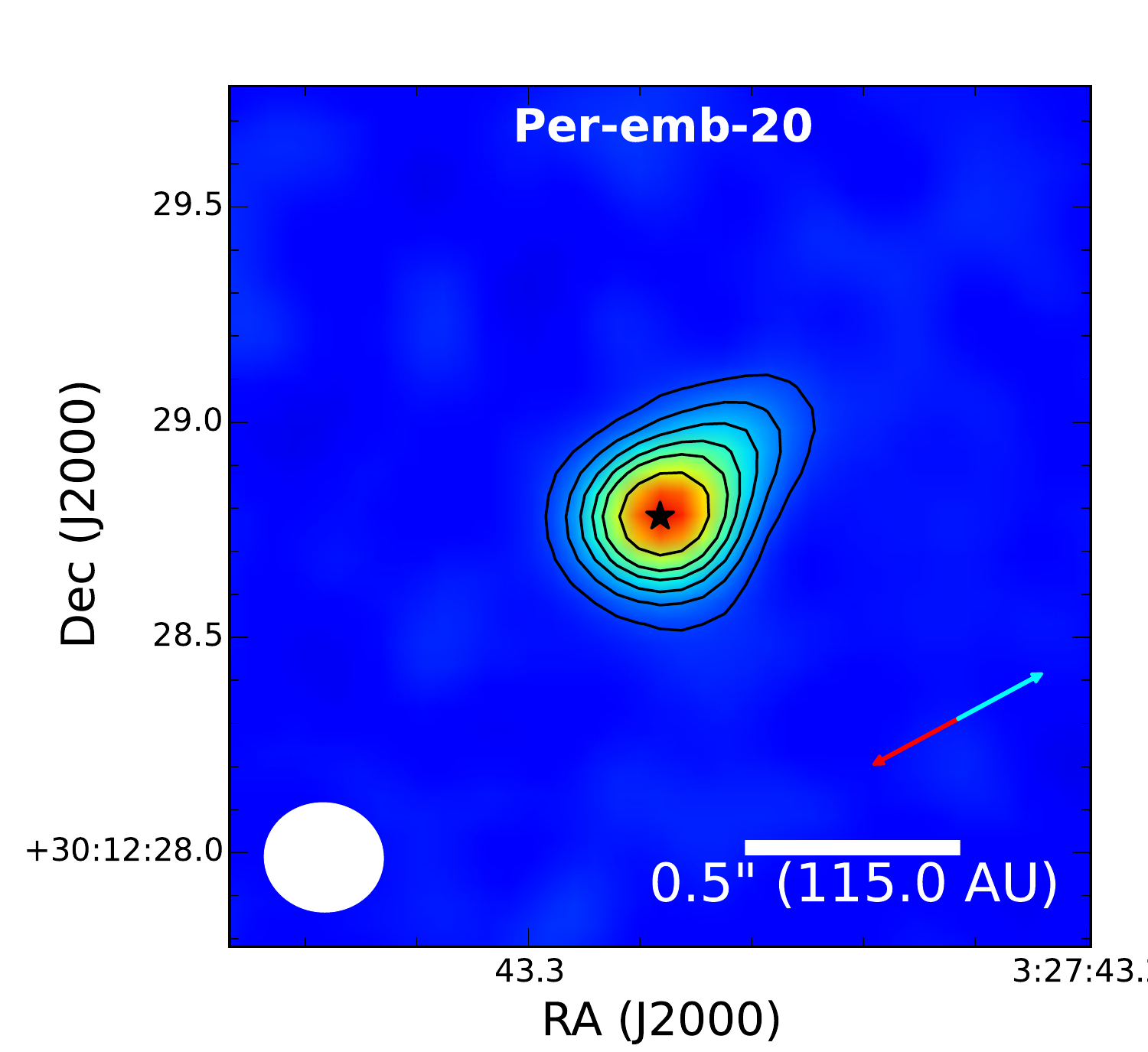}
  \label{fig:highhist}

\caption{ Same as Figure \ref{fig:per36zoom} for Per-emb-20. Contours as in Figure \ref{fig:per36} ($\sigma _{5\ cm}=3.68\ \mu $Jy ). 
Synthesized beam of the full bandwidth image is 0\farcs27 x 0\farcs25).
 Star marks the position of the protostar based on Ka-band observations \citep{Tobin2016}. Red and blue arrows indicate outflow direction from \cite{Davis2008}}

\label{fig:per109zoom} 

\end{figure}

\end{document}